\documentclass[letterpaper, 10pt]{article}
\usepackage{preprint}

\title{On the Joint Estimation of Flow Fields and Particle Properties from Lagrangian Data}

\author{
  Ke Zhou and
  Samuel J. Grauer\thanks{Corresponding author: \href{mailto:sgrauer@psu.edu}{sgrauer@psu.edu}}\\
  {\small Department of Mechanical Engineering, Pennsylvania State University}\vspace*{-1em}\\
}
\date{}

\begin{document}

\maketitle
\setcounter{footnote}{0}
\vspace*{-2em}

% Abstract
\begin{abstract}
\noindent We numerically investigate the feasibility and limits of jointly estimating flow fields and unknown particle properties (e.g., position, size, and density) from Lagrangian particle tracking (LPT) data. LPT offers time-resolved, volumetric measurements of particle trajectories, which are markers of the carrier fluid motion. However, experimental tracks are spatially sparse and potentially noisy, and the problem of reconstructing flow fields may be further complicated by inertial particle transport, such that particle slip velocities must be determined to access the velocity field of the carrier fluid. To address this problem, we develop a data assimilation framework that couples an Eulerian representation of the flow with Lagrangian particle models, enabling the simultaneous inference of carrier fields and particle properties under the governing equations of disperse multiphase flow. We show that flow fields and particle properties can be jointly estimated in three representative regimes: (1)~In a turbulent boundary layer with noisy tracer tracks ($St \to 0$), flow fields and \emph{true} particle positions are jointly estimated, which amounts to a physics-informed particle tracking problem; (2)~in homogeneous isotropic turbulence seeded with inertial particles ($St \sim 1{-}5$), we demonstrate simultaneous recovery of flow states and particle diameters, showing the feasibility of implicit particle characterization; and (3)~in a compressible, shock-dominated flow, we report the first joint reconstructions of velocity, pressure, density, and inertial particle properties (diameter and density), highlighting both the potential and certain limits of joint estimation in supersonic regimes. A systematic sensitivity study reveals how the seeding density, noise level, and Stokes number govern reconstruction accuracy for our method.\par\vspace*{.5em}

\end{abstract}
\vspace*{2em}

%%% Introduction %%%
\section{Introduction}
\label{sec: introduction}
Understanding and modeling turbulent flow in real-world systems \emph{ideally} calls for dense, time-resolved Eulerian fields: velocity, pressure, density, and gradient tensors of the same. These fields allow one to educe coherent structures and superstructures \citep{Schanz2018, Schroder2009, Weiss2023}, quantify energy transfer and dissipation \citep{Schneiders2017, Schroder2022}, determine surface loading from pressure and shear stress \citep{Rival2017, vanGent2017, Jux2020}, analyze buoyancy-driven transport in convection \citep{Muralidhar2002, Wissink2023}, and so forth. In principle, scale-resolving computational fluid dynamics (CFD) methods can deliver such fields, but direct numerical simulation (DNS) becomes intractable when applied to device-scale flows at realistic Reynolds numbers \citep{Slotnick2014, Cary2021}, and large-eddy simulation depends on subgrid-scale closures that require tuning for new physical regimes \citep{Argyropoulos2015, Duraisamy2019}. Even in a small sub-domain, accurate simulations hinge on the fidelity of boundary conditions that are often difficult to specify, such as representative upstream disturbances \citep{Buchta2022, Johnson2023}, wall heat fluxes \citep{Roy2006}, or coherent structures in the inflow \citep{Gorle2015, Xiao2019}. These boundaries are challenging to measure experimentally, and it is uncommon to set up CFD simulations using empirical unsteady boundary conditions.\par

Experiments, by contrast, capture the true behavior of flows but yield measurements that are typically sparse, noisy, and indirectly related to the quantities of interest. Data assimilation (DA) offers a means to combine these complementary capabilities: enforcing the governing equations via CFD methods while matching available measurements, thereby producing physically consistent reconstructions of real flows \citep{He2025, Zaki2025a, Zaki2025b}. In this context, Lagrangian particle tracking (LPT, Schr{\"o}der \& Schanz, \citeyear{Schroder2023}) is an especially attractive diagnostic because it provides time-resolved, volumetric velocity and acceleration information in complex configurations and at high Reynolds numbers. The data, however, are limited to particle trajectories (also known as ``tracks’’) that are spatially sparse, subject to localization and tracking errors, and, in some flows, whose velocities deviate from the underlying fluid velocity due to particle inertia. This leads to the central question motivating our study: Given experimentally realizable Lagrangian data, to what extent can we recover the antecedent sequence of Eulerian flow states and salient particle properties, such as their \emph{true} positions, sizes, and densities, by supplementing the tracks with the equations of motion for the carrier and particle phases in a DA reconstruction?\par

% LPT
\subsection{Capabilities and limitations of Lagrangian particle tracking}
\label{sec: introduction: LPT}
Lagrangian particle tracking has emerged as a leading diagnostic for volumetric velocimetry in both laboratory and field environments \citep{Schroder2024, Bristow2023, Li2024a}. Compared to tomographic particle image velocimetry (PIV, Scarano, \citeyear{Scarano2012}), it achieves higher spatial resolution and nearly ghost-free particle fields, enabling more accurate computation of derivatives and more reliable pressure inference \citep{vanGent2017}. Tracks are obtained by seeding the flow with tracer particles (or by leveraging natural tracers), imaging them with one or more cameras, and reconstructing three-dimensional (3D) particle positions before linking them in time. Multi-camera triangulation provides high-accuracy localization from overlapping views, while single-camera methods such as digital in-line holography (DIH, Toloui et al., \citeyear{Toloui2015}; Mallery et al. \citeyear{Mallery2019}), plenoptic cameras \citep{Fahringer2015, Hall2017}, or defocusing imaging \citep{Guo2019} enable four-dimensional measurements in settings where optical access is limited. Recent advances, including the predictor--corrector approach of Shake-The-Box (STB) tracking \citep{Schanz2016, Schroder2024}, object-aware LPT near boundaries \citep{Wieneke2024}, field-scale deployments in atmospheric turbulence \citep{Bristow2023, Li2024a}, and multi-pulse schemes for high-speed compressible flows \citep{Novara2019, Manovski2021}, have expanded the technique's reach. Hence, LPT is now frequently applied to flows of practical relevance.\par

Two challenges are especially relevant to the question of what can be inferred from LPT data. The first challenge is localization and tracking error. Although present in all LPT variants, these errors are most severe for single-camera systems (i.e., based on plenoptic or DIH imaging), which suffer from strongly anisotropic uncertainties due to depth-of-focus limitations. The largest errors occur along the optical axis \citep{Katz2010, Gao2013}, degrading velocity estimates and, in turn, derived quantities such as vorticity and pressure. The second challenge is inertial transport. Particles with finite response times, characterized by their Stokes number $St = \tau_\mathrm{p}/\tau_\mathrm{f}$, where $\tau_\mathrm{p}$ and $\tau_\mathrm{f}$ are characteristic time scales of the particle and carrier fluid, deviate systematically from the local fluid velocity \citep{Melling1997, Raffel2018}. While small particles, with $St < 0.1$, are used to ensure good ``traceability'' in LPT, inertial particle transport may occur inadvertently in both laboratory and natural settings. Examples include helium-filled soap bubbles in wind tunnels \citep{Wolf2019, Faleiros2021}, solid particles in shocked flows \citep{Ragni2011}, snowflakes in the atmospheric boundary layer \citep{Li2022, Bristow2023}, and sediment transport in waterways \citep{Righetti2004}, with $St$ varying from 0.1 to 10 in these examples. Track errors and particle inertia \emph{seemingly} reduce the information content of LPT data and thus prevent the accurate determination of flow fields from particle tracks in a typical LPT workflow.\par

% DA methods
\subsection{Data assimilation for Lagrangian particle tracking}
\label{sec: introduction: DA}
Early reconstruction approaches for LPT converted Lagrangian tracks into Eulerian fields by interpolation \citep{Malik1995, Agui1987}, but these methods were limited by the particle sampling resolution and prone to amplifying noise. Modern DA methods incorporate the Navier--Stokes equations as constraints, either as \emph{hard} constraints (ensemble Kalman filters, Deng et al., \citeyear{Deng2018}; adjoint--variational methods, Gronskis et al., \citeyear{Gronskis2013}; Foures et al., \citeyear{Foures2014}; He et al., \citeyear{He2024}) or \emph{soft} ones (physics-based interpolation with B-splines, Gesemann et al., \citeyear{Gesemann2016}; radial basis functions, Casa \& Krueger, \citeyear{Casa2013}; vorticity formulations, Jeon et al., \citeyear{Jeon2018}). By enforcing physical consistency, these methods have the potential to reconstruct velocity fields beyond the limits of interpolation per se, recovering information at scales finer than the particle-sampling Nyquist wavenumber (though this has yet to be demonstrated experimentally). Machine learning variants, like those based on physics-informed neural networks (PINNs, Raissi et al., \citeyear{Raissi2019}), offer flexible functional representations and have been successfully applied to a broad range of flows. They are increasingly relied upon for LPT DA, including in the present study, owing to their ease of implementation, robustness in inverse settings, and demonstrated accuracy with sparse or noisy data \citep{Zhou2024}.\par

Two central difficulties remain when processing LPT data, stemming from the challenges introduced above. One is localization and tracking errors, which directly limit the fidelity of reconstructed velocities; these errors can be compounded by biases which are inadvertently introduced during track filtering. The other is the assumption of ideal tracers, which breaks down when particles have non-negligible inertia, creating the additional challenge of jointly inferring particle properties alongside flow fields. In other words, the particle--flow coupling must be modeled to reconstruct the flow, and it is not known a priori whether carrier-phase flow fields and particle properties can be jointly determined from LPT track data. A related theoretical question is whether the same framework can also recover the true particle positions from noisy measurements. Although this is rarely a primary objective in LPT, since particle positions are merely a means to recover flow fields, successful recovery of the true positions from noisy data would indicate that the governing physics and the measurements jointly constrain the track geometries, highlighting the information content of Lagrangian data in turbulence. Addressing these challenges requires a DA framework that treats particle positions as the measured quantity rather than velocities, encodes the coupled carrier-phase and particle-phase dynamics, and operates directly on noisy particle position data. The development and demonstration of such a framework is one focus of this work.\par

% Approach
\subsection{Present approach and investigations of joint estimation}
\label{sec: introduction: approach}
We empirically demonstrate the feasbility of jointly estimating flow fields and particle properties using a framework that we term \emph{neural-implicit particle advection} (NIPA). NIPA couples an Eulerian flow model, parameterized by coordinate neural networks (i.e., PINNs when trained with a physics loss), to individual Lagrangian particle models that embed the particle advection equation as a hard constraint. For inertial particles, their size, density, and other attributes enter the model as trainable parameters, enabling estimation of properties that determine their response times. The governing physics enter via soft constraints on the Navier--Stokes equations for the carrier phase and an extended Maxey--Riley formulation for the particle (or ``disperse'') phase \citep{Subramaniam2022}. By design, NIPA works directly from the raw track positions, avoiding any biases induced by filtering the tracks, and it estimates both flow states and particle properties without requiring direct observations of either.\par

The NIPA framework provides numerical means to assess the factors that govern reconstruction accuracy across a range of flow regimes and measurement conditions. Moreover, it represents an algorithmic advance in LPT DA. We probe the degree to which flow fields can be recovered from Lagrangian data through synthetic test cases that include noisy tracks from ideal tracers in incompressible turbulence, inertial particles in the same, and inertial particles in a compressible flow with a shock wave. In each case, we examine whether the available track data and governing equations suffice to recover the flow field and particle properties, and we the study the sensitivity of reconstruction accuracy to the seeding density, noise level, and Stokes number. These results provide empirical evidence for the existence of an inverse mapping from a Lagrangian measurement manifold to a flow's global attractor in state space. Despite the empirical nature of this work, our findings motivate a formal analysis of the joint observability of flow--particle systems \citep{Zaki2025a}. The remainder of this paper outlines the reconstruction methodology in \S~\ref{sec: method}, describes the flow cases used for numerical testing in \S~\ref{sec: cases} and the implementation details in \S~\ref{sec: implementation}, tests the joint estimation of particle positions for ideal tracers in \S~\ref{sec: tracer}, and extends this treatment to inertial particles for turbulent and compressible flow reconstruction in \S~\ref{sec: inertial}. Lastly, in \S~\ref{sec: sensitivity}, we investigate the sensitivity of reconstruction accuracy to both particle inertia and noise before concluding the manuscript in \S~\ref{sec: conclusions}.\par

%%% Methodology %%%
\section{Methodology: neural-implicit particle advection}
\label{sec: method}
Understanding when flow fields and particle properties can be jointly recovered from Lagrangian data requires a DA algorithm that accounts for the physics of both phases. Existing DA methods either reconstruct the flow but treat the particles as ideal tracers, neglecting particle--fluid interactions, or they infer the particle dynamics given full knowledge of the flow field \citep{Dominguez2025}. Neither approach is adequate to the task at hand. To remedy this shortcoming, we introduce a new framework for LPT DA, which we call \emph{NIPA}. It features dedicated models of both the flow and the particles, which are coupled through the governing equations of disperse multiphase flow, namely, the Navier--Stokes and extended Maxey--Riley equations. By simultaneously optimizing the flow and particle models to satisfy these equations, NIPA recovers not only time-resolved flow fields but also enhanced particle trajectories and otherwise unknown particle properties. This tool thus enables the systematic evaluation of how dataset attributes (e.g., the number and spacing of particles, data fidelity, particle inertia, and compressibility) affect reconstruction accuracy. Below, we present our framework, define the objective loss terms, and describe our flow and particle models.\par

% DA architecture
\subsection{Framework for data assimilation}
\label{sec: method: framework}
\begin{figure}[htb]
    \centering
    \vspace{-0.5em}
    \includegraphics[width= 1\textwidth]{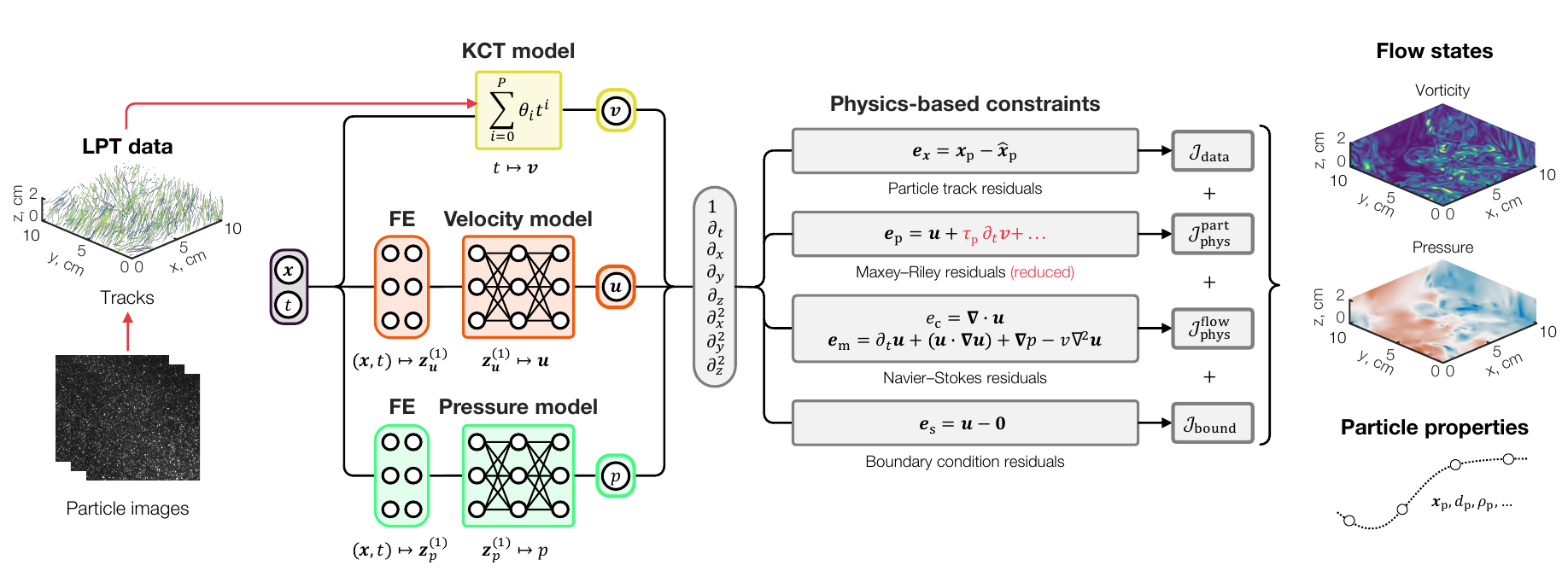}
    \caption{Architecture of the neural DA solver. Eulerian flow fields are represented by one or more neural networks. Fourier encoding (FE) enhances network expressivity and broadband spectral coverage. Each particle is represented by a Lagrangian kinematics-constrained track (KCT) model. Flow fields, particle kinematics, and particle properties (when $St > 0$) are jointly inferred from the data and governing equations.}
    \label{fig: schematic}
\end{figure}

The NIPA framework employs ``neural-implicit flow states'' coupled to a set of particle models, with one model per particle, to reconstruct unsteady flow fields from Lagrangian tracks. A schematic of the approach is shown in \cref{fig: schematic}. The flow field is represented using one or more coordinate neural networks, which take space--time input coordinates and return flow variables at that position and time,
\begin{equation}
    \label{equ: flow states}
    \mathsf{F} : \mathcal{V} \times \mathcal{T} \rightarrow \mathbb{R}^{d+1}, \quad \left(\boldsymbol{x}, t\right) \mapsto \left(\boldsymbol{u}, p\right),
\end{equation}
where $\boldsymbol{x} \in \mathcal{V}$ are the spatial coordinates in the flow domain $\mathcal{V}$, $t \in \mathcal{T}$ is a time within the measurement interval $\mathcal{T}$, $\boldsymbol{u}$ and $p$ are the velocity and pressure fields, respectively, and $d = 2$ or 3 is the number of spatial dimensions. Additional variables can be added to the outputs of $\mathsf{F}$ as needed, such as density or temperature in compressible flows. When coordinate neural networks like $\mathsf{F}$ are trained to minimize residuals to a set of physical equations, they are deemed to be PINNs, i.e., ``physics-informed neural networks.'' Network architectures used in this work are described in more detail in \S~\ref{sec: method: flow model}.\par

We next introduce the concept of our Lagrangian particle model, whose detailed formulation is derived in \S~\ref{sec: method: KCTs}. Measured particle tracks natively comprise sequences of spatial positions determined at the image times. Accordingly, our particle models are defined with respect to these tracked locations, which may be adjusted during training to account for localization uncertainty. For the $k$th particle, a dedicated model $\sdx{\mathsf{P}}[k]$ is built to represent its velocity as a continuous function of time,
\begin{equation}
    \label{equ: track model}
    \sdx{\mathsf{P}}[k] :
    \sdx{\mathcal{T}}[k] \rightarrow
    \mathbb{R}^d, \quad
    t \mapsto \sdx{\boldsymbol{v}}[k] \quad\text{for}\quad k \in \{1, \dots, n_\mathrm{p}\},
\end{equation}
where $\sdx{\boldsymbol{v}}[k]$ is the velocity of particle $k$ at time $t \in \sdx{\mathcal{T}}[k]$, $\sdx{\mathcal{T}}[k] \subseteq \mathcal{T}$ is the time span of that track, and $n_\mathrm{p}$ is the total number of particle tracks. Although $\sdx{\mathsf{P}}[k]$ may take various forms, a physical track model must satisfy both kinematic and dynamic constraints. Kinematic consistency corresponds to an advection equation,
\begin{equation}
    \label{equ: particle advection}
    \sdx{\boldsymbol{x}_{j}}[k] - \sdx{\boldsymbol{x}_{j-1}}[k] =
    \int_{t_{j-1}}^{t_{j}} \sdx{\boldsymbol{v}}[k]\mathopen{} \left(t\right) \mathrm{d}t \quad\text{for}\quad j \in \{1, \dots, n_k-1\}
\end{equation}
where $\sdx{\boldsymbol{x}_j}[k]$ is the position of particle $k$ at the $j$th time step $t_j$, and $n_k$ is the number of measurement points along that track. We embed this advection equation \eqref{equ: particle advection} as a hard constraint into $\sdx{\mathsf{P}}[k]$, as elaborated in \S~\ref{sec: method: KCTs}. The dynamic consistency of $\sdx{\mathsf{P}}[k]$ is weakly enforced by the physics-based optimization described in \S~\ref{sec: method: loss}.\par

Our formulation of $\sdx{\mathsf{P}}[k]$ provides a general framework for modeling the kinematics of both ideal tracers and inertial particles. In the latter case, properties of inertial particles like their size, density, and shape determine $\tau_\mathrm{p}$ and govern the particles' coupling to the carrier fluid \citep{Subramaniam2022}. These properties, and hence any slip velocities, are nearly always unknown in practice, complicating the reconstruction problem. To address this, we treat the relevant particle properties as model parameters to be learned, enabling the implicit characterization of particles from LPT data in conjunction with the equations of motion for both phases. In well-characterized experiments, by contrast, where the pertinent particle properties are known prior to reconstruction, they may be fixed in the model.\par

% Loss functions
\subsection{Composite objective function}
\label{sec: method: loss}
Flow and particle states ought to satisfy the governing equations for both phases, they must match known boundary conditions, and they should be consistent with the observed LPT data. To achieve these objectives, the flow model $\mathsf{F}$ and particle models $\sdx{\mathsf{P}}[k]$ are trained in tandem by minimizing a composite loss,
\begin{equation}
    \label{equ: objective loss: total}
    \mathscr{J}_\mathrm{total} =
    \chi_1 \mathscr{J}_\mathrm{data}^{\boldsymbol{x}} +
    \chi_2 \,\mathscr{J}_\mathrm{phys}^\mathrm{flow} +
    \chi_3 \,\mathscr{J}_\mathrm{phys}^\mathrm{part} +
    \chi_4 \,\mathscr{J}_\mathrm{bound},
\end{equation}
where $\chi_i$ are weighting coefficients that balance the relative contributions of each term. The four loss terms, detailed below, are the \emph{position-based} data fidelity term $\mathscr{J}_\mathrm{data}^{\boldsymbol{x}}$ (cf. the velocity-based term in \S~\ref{sec: method: loss: baseline}), the flow physics term $\mathscr{J}_\mathrm{phys}^\mathrm{flow}$, the particle physics term $\mathscr{J}_\mathrm{phys}^\mathrm{part}$, and the boundary condition term $\mathscr{J}_\mathrm{bound}$.\par

% Data loss
\subsubsection{Data fidelity loss}
\label{sec: method: loss: data}
Particle positions are treated as trainable parameters, meaning that the estimated position of each particle at each measurement time can be adjusted during reconstruction. Without additional constraints, however, allowing the particles to move risks unmooring the models from the measurements. To anchor reconstructions to the experiment, therefore, we introduce a data fidelity term that penalizes discrepancies between measured and estimated positions, weighted by the localization uncertainty,
\begin{equation}
    \label{equ: objective loss: data loss}
    \mathscr{J}_\mathrm{data}^{\boldsymbol{x}} =
    \left( \frac{1}{d n_\mathrm{p}} \sum_{k=1}^{n_\mathrm{p}} \frac{1}{n_k} \sum_{i=1}^{n_k} \smash{\underbrace{\left\lVert \sdx{\boldsymbol{x}_{i}}[k] - \sdx{\widehat{\boldsymbol{x}}_{i}}[k] \right\rVert_{\mathsfbi{L}}^2}_\text{chi-squared statistic}} - 1 \right)^2.
    \vphantom{\left(\sum_1^{1_1} \underbrace{1}_1\right)}
\end{equation}
Here, $\sdx{\boldsymbol{x}_i}[k]$ and $\sdx{\widehat{\boldsymbol{x}}_i}[k]$ are the measured and estimated (i.e., trainable) positions of the $k$th particle at time $t_i$, with $n_\mathrm{p}$ total tracks and $n_k$ samples in the $k$th track. The norm is the matrix-weighted Mahalanobis norm,
\begin{equation*}
    \left\lVert \Delta\boldsymbol{x} \right\rVert_{\mathsfbi{L}}^2 = \Delta\boldsymbol{x}^\top \mathsfbi{L}^\top \mathsfbi{L} \Delta\boldsymbol{x}, \quad\text{where}\quad \mathsfbi{L}^\top \mathsfbi{L} = \boldsymbol{\Gamma}^{-1}
\end{equation*}
and $\boldsymbol{\Gamma}$ is the covariance matrix for the localization errors. For independent, centered Gaussian errors, the weighted norm follows a chi-squared distribution with $d$ degrees of freedom, where $d$ is the dimension of $\boldsymbol{x}$ and also the expected value of the statistic. Normalizing the chi-squared statistic by $d$, subtracting 1, and squaring the result, as in \eqref{equ: objective loss: data loss}, encourages consistency of the residuals $\lVert\sdx{\boldsymbol{x}_{i}}[k] - \sdx{\widehat{\boldsymbol{x}}_{i}}[k]\rVert_2$ with the expected statistics of the localization errors. Values below $d$ are indicative of overfitting to noise, while values above $d$ suggest underfitting. In isolation, \eqref{equ: objective loss: data loss} acts as a maximum likelihood criterion for the particle positions. Notably, there exists an infinite set of minimizers for \eqref{equ: objective loss: data loss} for any set of LPT data. When coupled with the physics and boundary losses, however, the data loss enables a physics-informed particle tracking method. Both uses of \eqref{equ: objective loss: data loss} are demonstrated in \S~\ref{sec: tracer}.\par

% Physics loss
\subsubsection{Physics-based loss terms}
\label{sec: method: loss: physics}
The flow physics loss penalizes residuals of the carrier-phase governing equations,
\begin{equation}
    \label{equ: flow physics loss}    \mathscr{J}_\mathrm{phys}^\mathrm{flow} = \frac{\dim(\boldsymbol{e}_\mathrm{f})^{-1}}{\left|\mathcal{V} \times \mathcal{T}\right|} \int_\mathcal{T} \int_\mathcal{V} \left\lVert\boldsymbol {e}_\mathrm{f} \right\rVert_2^2 \mathrm{d}\boldsymbol{x} \,\mathrm{d}t,
\end{equation}
where $\boldsymbol{e}_\mathrm{f}$ is the residual vector associated with the governing flow equations within $\mathcal{V} \times \mathcal{T}$. For domains with transient boundaries, the space--time domain is not the simple Cartesian product of $\mathcal{V}$ and $\mathcal{T}$; see \citet{Tang2025} for details on handling such situations. While most LPT DA algorithms are restricted to incompressible flows, the present framework can be extended to compressible configurations with minimal modification. Appendix~\ref{app: carrier} summarizes the governing equations used in this study.\par

The particle physics loss is based on residuals from an equation of motion for the disperse phase,
\begin{equation}
    \label{equ: particle physics loss}
    \mathscr{J}_\mathrm{phys}^\mathrm{part} = \frac{\dim\mathopen{} \left( \smash{\boldsymbol{e}_\mathrm{p}} \right)^{-1}}{n_\mathrm{p}} \sum_{k=1}^{n_\mathrm{p}} \frac{1}{\left|\sdx{\mathcal{T}}[k]\right|} \int_{\sdx{\mathcal{T}}[k]} \left\lVert \sdx{\boldsymbol{e}_\mathrm{p}}[k]\right\rVert_2^2 \mathrm{d}t,
\end{equation}
where $\sdx{\boldsymbol{e}_\mathrm{p}}[k]$ is the residual of the $k$th particle's governing equation. For small spherical particles with a vanishing particle Reynolds number, the dynamics are described by the Maxey--Riley equation \citep{Maxey1983, Capecelatro2023}. The specific form adopted in this work, including its adaptation for compressible flows, is summarized in appendix~\ref{app: disperse}. In cases with inertial particle transport, information about particle properties is obtained through the residuals in \eqref{equ: particle physics loss}, since the equation of motion (and therefore $\sdx{\boldsymbol{e}_\mathrm{p}}[k]$) generally depends on the flow velocity $\boldsymbol{u}$, the particle velocity $\sdx{\boldsymbol{v}}[k]$, and particle characteristics such as their sizes and densities.\par

In many LPT experiments, the particles are carefully selected to behave as ideal tracers with negligible slip velocities. In the limit $St \to 0$, the particle and carrier motions are equal. Hence, we specify velocity and acceleration residuals for ideal tracers:
\begin{equation}
    \label{equ: particle residual}
    \sdx{\boldsymbol{e}_{\mathrm{p}, \boldsymbol{u}}}[k] =
    \boldsymbol{u} - \sdx{\boldsymbol{v}}[k]
    \quad\text{and}\quad
    \sdx{\boldsymbol{e}_{\mathrm{p}, \boldsymbol{a}}}[k] =
    \frac{\mathrm{D}\boldsymbol{u}}{\mathrm{D}t} - \sdx{\boldsymbol{a}}[k],
\end{equation}
where $\mathrm{D}/\mathrm{D}t$ is the material derivative operator, $\sdx{\boldsymbol{a}}[k]$ is the acceleration of a particle, and $\boldsymbol{e}_\mathrm{p}$ is the concatenation of $\boldsymbol{e}_{\mathrm{p}, \boldsymbol{u}}$ and $\boldsymbol{e}_{\mathrm{p}, \boldsymbol{a}}$.\par

% Boundary loss
\subsubsection{Boundary condition loss}
\label{sec: method: loss: boundary}
Although neural state estimation does not explicitly require boundary conditions, incorporating known constraints can improve reconstruction accuracy. In turbulent boundary layers (TBLs), for instance, enforcing a no-slip condition at the wall enhances near-wall resolution, where positional uncertainties are worsened by optical reflections and where the flow scales are smallest. This combination of large relative uncertainty and fine-scale dynamics makes boundary conditions especially valuable near walls. The no-slip boundary loss is
\begin{equation}
    \label{equ: boundary loss}
    \mathscr{J}_\mathrm{bound} = \frac{\dim(\boldsymbol{u})^{-1}}{\left|\mathcal{A} \times \mathcal{T}\right|} \int_\mathcal{T} \int_\mathcal{A} \left\lVert\boldsymbol{u} \right\rVert_2^2 \mathrm{d}\boldsymbol{x} \,\mathrm{d}t,
\end{equation}
where $\mathcal{A} \subseteq \partial\mathcal{V}$ denotes the no-slip portion of the domain boundary $\partial \mathcal{V}$. This formulation can be extended to moving-wall conditions \citep{Tang2025}, enabling the inference of wall motion, or to hybrid constraints that feature multiple variables (e.g., a constant free-stream density, adiabatic walls, or known pressures at tap locations) and boundary types (Dirichlet, Neumann, Robin), depending on the pertinent physics.\par

% Baseline reconstruction
\subsubsection{Baseline flow-only reconstruction}
\label{sec: method: loss: baseline}
To evaluate the added value of joint flow--particle reconstruction, we implement a simplified baseline that neglects the particle models and instead relies on velocity information derived from the particle tracks through an intermediate step, as is common in LPT DA \citep{DiLeoni2023, Shin2025}. In this baseline, the framework reduces to a flow network $\mathsf{F}$ that is trained to reproduce the track-based velocity estimates $\widehat{\boldsymbol{v}}$ while satisfying the governing equations and boundary conditions. The corresponding loss is
\begin{equation}
    \label{equ: baseline: total loss}
    \mathscr{J}_\mathrm{total} =
    \chi_1 \mathscr{J}_\mathrm{data}^{\boldsymbol{u}} +
    \chi_2 \mathscr{J}_\mathrm{phys}^\mathrm{flow} +
    \chi_3 \mathscr{J}_\mathrm{bound},
\end{equation}
where $\mathscr{J}_\mathrm{data}^{\boldsymbol{u}}$ enforces agreement between the reconstructed flow velocity and particle velocities inferred from the tracks. The velocity-based data fidelity term is
\begin{equation}
    \label{equ: baseline: data loss}
    \mathscr{J}_\mathrm{data}^{\boldsymbol{u}} 
    = \frac{1}{n_\mathrm{p}} \sum_{k=1}^{n_\mathrm{p}}
    \frac{1}{n_k} \sum_{i=1}^{n_k}
    \left\lVert \boldsymbol{u} - \sdx{\widehat{\boldsymbol{v}}_{i}}[k] \right\rVert_2^2,
\end{equation}
where $\boldsymbol{u}$ is the velocity field from $\mathsf{F}$ and $\sdx{\widehat{\boldsymbol{v}}_{i}}[k]$ is the velocity estimate for the $k$th particle at time $t_i$. Such estimates may be obtained by differentiating the tracked positions, but finite differencing amplifies noise, so smoothing or interpolation schemes like those based on B-splines, polynomial fits, or kernel convolution are often applied. Here, we demonstrate ``baseline reconstructions'' using both a second-order finite difference scheme (with single-sided differences at the track ends) and a quasi-optimal B-spline filter, detailed in appendix~\ref{app: B-spline}. While these methods can suppress the effects of noise in $\sdx{\widehat{\boldsymbol{v}}_{i}}[k]$, they also embed errors in $\mathscr{J}_\mathrm{data}^{\boldsymbol{u}}$ through the heuristic choice of filter parameters. By contrast, our position-based formulation in \eqref{equ: objective loss: data loss} and the associated particle models bypass the intermediate estimation of velocity. Instead, we anchor the reconstruction to the measured positions, subject to the localization error statistics.\par

We wish to emphasize that this ``velocity data loss'' formulation is used solely as a baseline. It represents the performance attainable when particle--fluid coupling and particle properties are ignored, thereby highlighting the added value of joint estimation.\par

% Flow Model
\subsection{Neural flow model}
\label{sec: method: flow model}
The flow model consists of one or more neural networks, here denoted by $\mathsf{F}$, comprising an input layer, output layer, and $n_\mathrm{L}$ hidden layers,
\begin{subequations}
    \label{equ: NN architecture}
    \begin{equation}
        \sdx{\boldsymbol{z}}[n_\mathrm{L}+1] = \mathsf{F}\mathopen{} \left(\sdx{\boldsymbol{z}}[0]\right) = \sdx{\mathsfbi{W}}[n_\mathrm{L}+1] \left[\sdx{\mathsf{L}}[n_\mathrm{L}] \circ \sdx{\mathsf{L}}[n_\mathrm{L}-1] \circ \dots \circ \sdx{\mathsf{L}}[2] \circ \mathsf{G}\mathopen{}\left(\sdx{\boldsymbol{z}}[0]\right)\right] + \sdx{\boldsymbol{b}}[n_\mathrm{L}+1],
    \end{equation}
    with
    \begin{equation}
        \sdx{\boldsymbol{z}}[L] = \mathsf{L}^L\mathopen{}\left(\sdx{\boldsymbol{z}}[L-1]\right) = \text{swish}\mathopen{} \left(\sdx{\mathsfbi{W}}[L] \sdx{\boldsymbol{z}}[L-1] + \sdx{\boldsymbol{b}}[L]\right), \quad\text{for}\quad L \in \{2, \dots, n_\mathrm{L}\}.
    \end{equation}
\end{subequations}
In this expression, $\sdx{\boldsymbol{z}}[L]$ are the outputs at layer $L$ and $\mathsf{L}$ represents the operator that maps between consecutive hidden layers. This operator consists of an affine transform followed by a nonlinear activation that is applied element-wise. The matrix $\sdx{\mathsfbi{W}}[L]$ and the vectors $\sdx{\boldsymbol{b}}[L]$ are the weights and biases of the affine transform, respectively; the latter provides an additive offset at each layer (not to be confused with measurement or estimation biases). We use the swish activation,
\begin{equation}
    \label{equ: swish}
    \text{swish}(z_i) = \frac{z_i \,\exp(z_i)}{1 + \exp(z_i)},
\end{equation}
which has been shown to mitigate issues associated with vanishing gradients and is also smooth, making it well suited for representing flow fields that must be at least twice differentiable \citep{Jagtap2023}. Furthermore, to mitigate the low-frequency bias that commonly occurs in gradient-based training \citep{Wang2021}, the first hidden layer $\sdx{\mathsf{L}}[1]$ is replaced by a Fourier encoding $\mathsf{G}$ \citep{Tancik2020},
\begin{equation}
    \label{equ: FE}
    \sdx{\boldsymbol{z}}[1] = \mathsf{G}\mathopen{}\left(\sdx{\boldsymbol{z}}[0]\right) = \left[\sin\mathopen{}\left(2\pi \boldsymbol{f}_1 \boldsymbol{\cdot} \sdx{\boldsymbol{z}}[0]\right), \,\cos\mathopen{}\left(2\pi \boldsymbol{f}_1 \boldsymbol{\cdot} \sdx{\boldsymbol{z}}[0]\right), \dots, \,\sin\mathopen{}\left(2\pi \boldsymbol{f}_{n_\mathrm{f}} \boldsymbol{\cdot} \sdx{\boldsymbol{z}}[0]\right), \,\cos\mathopen{}\left(2\pi \boldsymbol{f}_{n_\mathrm{f}} \boldsymbol{\cdot} \sdx{\boldsymbol{z}}[0]\right)\right],
\end{equation}
where $\boldsymbol{f}_i$ are random frequency vectors, sampled once upon initialization and fixed during training, and $n_\mathrm{f}$ is the number of features. For an incompressible flow represented by a single network, the input is $\sdx{\boldsymbol{z}}[0] = (\boldsymbol{x}, t)$ and the output is $\sdx{\boldsymbol{z}}[n_\mathrm{L}+1] = (\boldsymbol{u}, p)$.\par

Two additional strategies are used to improve reconstruction accuracy.  
First, some flow variables (e.g., density, temperature, total energy in compressible flows) must be strictly positive. Enforcing positivity stabilizes the inverse problem, and we achieve this by reparameterizing the corresponding outputs with a softplus function,
\begin{equation*}
    \label{equ: softplus}
    \text{softplus}(z_i) = \log\mathopen{} \left[1 + \exp(z_i)\right],
\end{equation*}
which smoothly maps $\mathbb{R} \to (0, \infty)$. Second, different flow variables often have distinct spectral characteristics, which can hinder the performance of a shared network. For example, in homogeneous isotropic turbulence (HIT), velocity and pressure follow $\kappa^{-5/3}$ and $\kappa^{-7/3}$ scalings, respectively \citep{Kolmogorov1941, Obukhov1949, Corrsin1951}. In such cases, it is advantageous to assign dedicated subnetworks, e.g., $\mathsf{F}_{\boldsymbol{u}} : (\boldsymbol{x}, t) \mapsto \boldsymbol{u}$ and $\mathsf{F}_p : (\boldsymbol{x}, t) \mapsto p$. Both strategies are employed in this work where appropriate and noted, accordingly.\par

% KCTs
\subsection{Kinematics-constrained particle model}
\label{sec: method: KCTs}
\begin{figure}[htb]
    \vspace{-0.5em}
    \centering
    \includegraphics[width=0.4\linewidth]{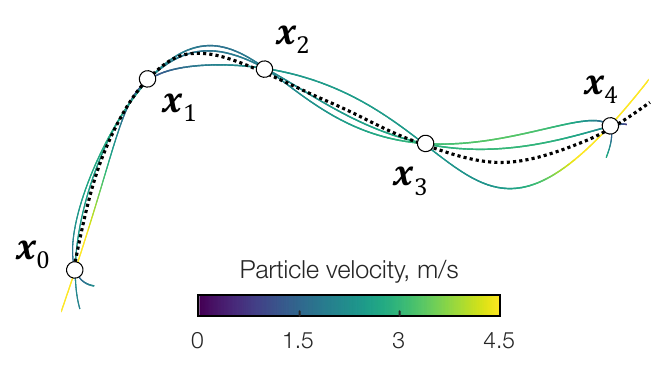}
    \caption{Illustration of a KCT (\emph{kinematics-constrained track}) model. Colored trajectories show particle histories outputted by the KCT, with color indicating particle speed. Dotted lines indicate the ground truth. Embedding \eqref{equ: particle advection} as a hard constraint ensures that tracks pass through the specified positions (dots), regardless of adjustments to velocity via $\boldsymbol{\theta}$. Accordingly, $\boldsymbol{\theta}$ remains unconstrained.}
    \label{fig: KCT}
\end{figure}

Particles are represented by models that we term ``kinematics-constrained tracks'' (KCTs), which embed the advection equation \eqref{equ: particle advection} as a hard constraint. As a result, trajectories given by KCTs always integrate to the positions specified by the model parameters. Measured positions are incorporated directly into the models. When high-fidelity LPT data are available, the specified particle positions can be fixed. In cases with appreciable localization uncertainty, however, they are treated as trainable variables and refined through optimization of the position-based data fidelity term $\mathscr{J}_\mathrm{data}^{\boldsymbol{x}}$ in conjunction with the physics and boundary losses. Velocities, accelerations, and intermediate positions are determined by free parameters of the model, but these parameters are themselves unconstrained. Instead, kinematic consistency is enforced by the model formulation, rather than by explicit nonlinear restrictions in parameter space. This ensures that the model outputs meet a baseline of physical fidelity without complicating gradient-based training. \Cref{fig: KCT} illustrates a representative KCT. In the example, tracked positions are fixed while other model parameters are adjusted, yielding multiple velocity histories that integrate to the same set of data.\par

In what follows, the KCT formulation is presented for a single velocity component $v$, with the particle index $k$ omitted for clarity. Extension to the full vector $\sdx{\boldsymbol{v}}[k]$ is straightforward. The construction draws on the theory of functional connections \citep{Leake2022}, which provides a systematic framework for converting constrained optimization problems into unconstrained ones. Within this framework, each velocity component can be written as a function of time,
\begin{equation}
    \label{equ: TFC formulation}
    v\mathopen{}\left(t\right) =
    g\mathopen{} \left(t\right) + \sum_{j=1}^{n_\mathrm{c}} \eta_j \,\varphi_j\mathopen{}\left(t\right),
\end{equation}
where $g$ is a user-chosen ``free function,'' $\eta_j$ is a projection coefficient that enforces the integral constraint over the interval $[t_{j-1}, t_j]$, $\varphi_j$ is a switch function that activates the projections over time, and $n_\mathrm{c} = n_k - 1$ is the number of constraints. The choice of $g$ determines the form of the projection coefficients. Here, $g$ is taken as a $P$th-order polynomial in time,
\begin{equation}
    \label{equ: free function}
    g\mathopen{}\left(t\right) = \sum_{i = 0}^{P} \theta_i \,t^i,
\end{equation}
with the coefficients $\theta_i$ being trainable parameters. To enforce the advection constraint, the projection coefficients are set to
\begin{equation}
    \label{equ: projection coefficients}
    \eta_j = \left(x_j - x_{j-1}\right) - \int_{t_{j-1}}^{t_j} g\mathopen{}\left(t\right) \mathrm{d}t,
\end{equation}
where $x_j$ is the tracked position at time $t_j$. The switch functions are linear combinations of a set of so-called ``support functions,''
\begin{equation}
    \label{equ: switch function}
    \varphi_j\mathopen{}\left(t\right) = \sum_{i = 1}^{n_\mathrm{c}} s_i\mathopen{}\left(t\right) A_{ij},
\end{equation}
where $s_i$ is the $i$th support function and $A_{ij}$ is a weighting coefficient. The switch functions must successively activate the integral constraints over the corresponding intervals, which amounts to the condition
\begin{equation}
    \label{equ: condition}
    \int_{t_{i-1}}^{t_i} \varphi_j\mathopen{} \left(t\right) \mathrm{d}t =
    \begin{cases} 1, & i = j \\ 0, & i \neq j \end{cases}
    \quad\text{for}\quad i, j \in \{1, \dots, n_\mathrm{c}\}.
\end{equation}
Enforcing this condition is equivalent to solving the linear system
\begin{equation}
    \label{equ: linear constraints}
    \mathsfbi{S}\mathsfbi{A} = \mathsfbi{I},
\end{equation}
where $\mathsfbi{S}$ is an $n_\mathrm{c}\times n_\mathrm{c}$ matrix with entries
\begin{equation}
    \label{equ: matrix S} S_{ij} = \int_{t_{i-1}}^{t_i} s_j\mathopen{} \left(t\right) \mathrm{d}t
\end{equation}
and $\mathsfbi{I}$ is the identity matrix. The support functions $s_j$ must yield a nonsingular matrix $\mathsfbi{S}$ but are otherwise flexible. While \citet{Leake2022} suggested monomials (e.g., $s_j(t) = t^{j-1}$), the resulting KCT model becomes ill-conditioned for long tracks. Instead, we use radial basis functions,
\begin{equation}
    \label{equ: support function}
    s_j\mathopen{} \left(t\right) = \exp\mathopen{} \left[-\beta_j^2 \left(t - t_{0,j}\right)^2\right],
\end{equation}
with $\beta_j = (t_j - t_{j-1})^{-1}$ and where $t_{0,j} = (t_{j-1}+t_j)/2$ is the midpoint for the $j$th interval.\par

Given a set of support functions, the coefficient matrix $\mathsfbi{A}$ is determined by solving \eqref{equ: linear constraints}, which may be done independently of the free function $g$ and its parameters. Conversely, the projection coefficients $\eta_j$ explicitly depend upon the polynomial parameters $\theta_i$ through \eqref{equ: projection coefficients}. Substituting the polynomial form of $g$ from \eqref{equ: free function} yields a closed-form expression for the projection coefficients,
\begin{equation}
    \eta_j = \left(x_j - x_{j-1}\right) -
    \sum_{i=0}^P \frac{\theta_i}{i + 1} \left(t_{j}^{i+1} - t_{j-1}^{i+1}\right),
\end{equation}
where $t_j^{i+1}$ is time $t_j$ raised to the $(i+1)$th power.\par

For ease of implementation, it is convenient to recast \eqref{equ: TFC formulation} in matrix form. We thus define a series of entities: a parameter vector $\boldsymbol{\theta} = \{\theta_0, \dots, \theta_P\}$, time vector $\boldsymbol\tau_\mathrm{v}(t) = \{t^j \mid j=0, \dots, P\}$, displacement vector $\boldsymbol{\delta} = \{x_j - x_{j-1} \mid j = 1, \dots, n_\mathrm{c}\}$, support-function vector $\boldsymbol{s}_\mathrm{v}(t) = \{s_j(t) \mid j = 1, \dots, n_\mathrm{c}\}$, and $(P + 1)\times n_\mathrm{c}$ support matrix $\mathsfbi{C}$, with entries
\begin{equation}
    \label{equ: support matrix}
    C_{ij} = i^{-1} \left(t_j^i - t_{j-1}^i\right).
\end{equation}
Using these definitions, the KCT velocity model can be written compactly as
\begin{equation}
    \label{equ: track velocity matrix}
    v(t) = \boldsymbol{\theta}^\top \boldsymbol\tau_\mathrm{v}(t) + \left(\boldsymbol{\delta}^\top - \boldsymbol{\theta}^\top \mathsfbi{C}\right) \mathsfbi{A}^\top \boldsymbol{s}_\mathrm{v}(t).
\end{equation}
This representation, which is one component of $\mathsf{P}$ for a single particle, provides a continuous, differentiable velocity profile that inherently satisfies the advection constraint. The displacement vector $\boldsymbol{\delta}$ encodes the measured particle positions. For noisy tracks, however, the positions in $\boldsymbol{\delta}$ can be adjusted during training alongside $\boldsymbol{\theta}$, enabling refinement of particle positions at measurement times.\par

Integration and differentiation of the velocity model yield continuous expressions for the particle's position and acceleration. The position is
\begin{equation}
    \label{equ: track position matrix}
    x(t) = \boldsymbol{\theta}^\top \boldsymbol\tau_\mathrm{x}(t) + \left(\boldsymbol{\delta}^\top - \boldsymbol{\theta}^\top \mathsfbi{C}\right) \mathsfbi{A}^\top \boldsymbol{s}_\mathrm{x}(t) + x_0,
\end{equation}
where $x_0$ is the initial position of the chosen track component. The acceleration is
\begin{equation}
    \label{equ: track acceleration matrix}
    a(t) = \boldsymbol{\theta}^\top \boldsymbol\tau_\mathrm{a}(t) + \left(\boldsymbol{\delta}^\top - \boldsymbol{\theta}^\top \mathsfbi{C}\right) \mathsfbi{A}^\top \boldsymbol{s}_\mathrm{a}(t).
\end{equation}
As in \eqref{equ: track velocity matrix}, all time dependencies are carried by the $\boldsymbol{\tau}$ and $\boldsymbol{s}$ vectors. For position, these are
\begin{subequations}
    \begin{align}
        \boldsymbol\tau_\mathrm{x}(t) &= \left\{t^{j+1}/(j+1) \;\middle|\; j = 0, \dots, P\right\} \quad\text{and} \\
        \boldsymbol{s}_\mathrm{x}(t) &= \left\{\frac{\sqrt{\pi}}{2 \, \beta_j} \,\mathrm{erf}\mathopen{} \left[\beta_j \left(t - t_{0,j}\right) \right] \;\middle|\; j = 1, \dots, n_\mathrm{c} \right\},
    \end{align}
    and for acceleration, they are
    \begin{align}
        \boldsymbol\tau_\mathrm{a}(t) &= \left\{j \,t^{j-1}  \;\middle|\; j = 0, \dots, P\right\} \quad\text{and} \\
        \boldsymbol{s}_\mathrm{a}(t) &= \left\{-2 \,\beta_j^2(t - t_{0,j}) \,s_j(t) \;\middle|\; j = 1, \dots, n_\mathrm{c} \right\}.
    \end{align}
\end{subequations}\par

It should be emphasized that while the KCT model enforces particle kinematics as a hard constraint, the dynamics emerge through the joint optimization of $\mathsf{F}$ and $\sdx{\mathsf{P}}[k]$ by minimization of the particle physics loss in \eqref{equ: objective loss: total}. For any prescribed set of positions, the KCT admits many velocity histories, as illustrated in \cref{fig: KCT}. Accordingly, the free parameters $\boldsymbol{\theta}$ and $\boldsymbol{\delta}$ must be identified from the interplay of physics and data losses. In cases involving inertial transport with particles of unknown size, density, or other properties, the kinematic parameters alone are insufficient. Additional training variables associated with the relevant particle dynamics (diameter $d_\mathrm{p}$, density $\rho_\mathrm{p}$, etc.) must be included in the optimization.\par

% Comparison
\subsubsection{Comparison to alternative track models}
\label{sec: method: KCTs: comparison}
Track processing for LPT generally falls into two categories: filtering and optimization. In filtering methods, the particle position time series is differentiated and smoothed by convolving the raw trajectories with a discrete kernel. Velocities and accelerations are estimated at the measurement times. The simplest example is finite differencing, which amplifies noise. For this reason, low-pass filtering is commonly introduced, with Gaussian kernels being a standard choice \citep{Mordant2004, Berk2021, Will2021, Brandt2022}. These methods are simple and computationally efficient, but they suffer from large endpoint errors due to the asymmetric support of the stencil. \citet{Berk2024a} found that Gaussian smoothing requires kernels with 20 points or more to perform well, which effectively removes 10 points from each end of a track. Short tracks may then be discarded entirely, introducing a potential selection bias, since tracks associated with larger velocities and accelerations are often the shortest ones \citep{Lawson2018}.\par

In optimization methods, including our KCT approach, an analytical function is fitted to the raw trajectories, thereby providing a continuous representation in time and avoiding the worst of the endpoint issue. Early techniques in this class represented tracks using low-order polynomials \citep{Malik1993, Voth2002, DiLeoni2023}; these models were typically under-parameterized and optimized by least squares \citep{Lynch2013}. Such models inherently suppress noise, but they can also smooth out bona fide fluctuations. More recently, spline-based models have become prominent, with TrackFit as a notable example \citep{Gesemann2016, Gesemann2021, Buchwald2025}. In that method, each trajectory is represented by an over-parameterized B-spline. Regularization is supplied by a third-order smoothness penalty, and the weighting parameter is selected so as to balance realistic spectral characteristics against the suppression of noise. See \citet{Gesemann2021} for a full description.\par

In a similar spirit, KCTs also over-parameterize each track and yield a continuous, differentiable representation of the trajectory. We do not claim that KCTs confer a theoretical advantage over spline-based methods. Rather, we have simply found that the KCT formulation performs well across all of our tests, being easy to initialize and sample and well conditioned in the joint optimization. More broadly, the main point of the present work is not the use of KCTs per se, but the use of parameterized tracks within a physics-constrained reconstruction, which amounts to a form of physics-informed particle tracking. A similar approach could be implemented using the B-splines of TrackFit or other smooth track models.\par

%%% Flow Cases %%%
\section{Flow cases and datasets}
\label{sec: cases}
We investigate the joint estimation problem using three flow configurations: a turbulent boundary layer, forced homogeneous isotropic turbulence, and supersonic flow over a cone--cylinder body. These are referred to as the \emph{TBL}, \emph{HIT}, and \emph{cone} cases, respectively, and they are introduced below in the order in which they appear in the paper. First, in the TBL case, we estimate true tracer trajectories from noisy data, corresponding to physics-informed tracking in the $St \to 0$ limit. Second, in the HIT and cone--cylinder cases, we estimate inertial particle properties for particles with finite $St$ and $Re_\mathrm{p}$. Third, in the HIT case, we jointly estimate particle positions and sizes from noisy tracks to assess the interaction between noise and inertia. Each dataset mimics realistic LPT conditions in terms of particle densities and tracking accuracy.\par

% TBL
\subsection{Turbulent boundary layer with noisy particle tracks}
\label{sec: cases: TBL}
The first case is drawn from a DNS of channel flow with a favorable pressure gradient, obtained from the Johns Hopkins Turbulence Database (JHTDB, Perlman et al., \citeyear{Perlman2007}). An imposed pressure gradient drives the development of TBLs along the top and bottom walls, with friction Reynolds numbers up to $Re_\tau \approx 1000$. The DNS domain spans $8\pi \times 3\pi \times 2$ and is discretized with $2048 \times 1536 \times 512$ voxels in the streamwise, spanwise, and wall-normal directions, and the data are stored at a dimensionless temporal resolution of 0.0065. Our study focuses on a $126 \times 54 \times 80$-voxel sub-volume at the bottom wall, corresponding to a physical region of $56 \times 12 \times 4.25$~mm$^3$ (air, $\nu = 15$~mm$^2$/s). In viscous units, this region spans $1546 \times 331 \times 117$ and covers the buffer layer. The viscous length and time scales are $\ell_\nu = 0.036$~mm and $\tau_\nu = 0.09$~ms, respectively.\par

Synthetic LPT data are generated by advecting 70~000 ideal tracers, yielding a particle image density of 0.07~particles-per-pixel (ppp) for a 1~MP camera. This lies squarely within the capability of standard multi-camera STB workflows (which extend to about 0.1~ppp, Schanz et al., \citeyear{Schanz2016}) and slightly above the typical range reported for single-camera techniques such as DIH and plenoptic LPT (0.03--0.06~ppp, Shao et al., \citeyear{Shao2020}; Yamakaitis et al., \citeyear{Yamakaitis2025a}). Advection is performed with a fourth-order Runge--Kutta scheme and periodic boundary conditions. To mitigate boundary-related artifacts, the tracks are first computed in an extended outer domain and then cropped to an inner region that is 15\% smaller in each direction. Tracks are ``recorded'' for 51 consecutive frames at a temporal resolution of 0.057~ms. The mean particle spacing is about $9\ell_\nu$, which is sufficient to resolve most of the flow's energy content using PINNs. The particle field is downsampled by factors of 8 and 64 to produce datasets with lower seeding densities, having mean inter-particle spacings of $18\ell_\nu$ and $36\ell_\nu$.\par

To simulate experimental localization errors, the ground truth particle positions are corrupted with random noise drawn from the error probability density function (PDF) reported for iterative particle reconstruction at 0.1~ppp \citep{Wieneke2012, Schanz2022}. A range of localization uncertainties is considered, with standard deviations of $\sigma_x = \sigma_y \in \{0.05, 0.15, 0.25, 0.35, 0.45, 0.5\}$ in pixel units in the $x$--$y$ plane. In viscous units, this corresponds to $0.09$--$0.9\ell_\nu$. To reflect the anisotropy commonly observed in experiments, the wall-normal noise is taken to be twice as large as the other components, with $\sigma_z/\sigma_x = 2$, under the assumption that the $z$-axis coincides with the optical axis. This ratio is broadly consistent with STB and with \emph{advanced} DIH and plenoptic techniques. For example, \citet{Shao2020} developed a deep learning method for DIH LPT that reduced the positional uncertainty ratio to $\sigma_z/\sigma_x = 2$--4, and \citet{Moaven2024} proposed a ray-bundling method that achieved a ratio of about 3 for plenoptic LPT. By contrast, when processed using conventional numerical refocusing, low-aperture LPT methods can exhibit much larger anisotropy ratios, often up to 10. In previous work, we showed that DA can compensate for such issues \citep{Zhou2023, Zhou2024}, but a comprehensive study of noise distributions is outside the scope of the present paper.\par

The selected localization noise levels span a broad range of measurement conditions. Low noise levels, with $\sigma_x \leq 0.15$~px, roughly correspond to multi-camera STB \citep{Schanz2016}, whereas high noise levels, with $\sigma_x \geq 0.35$~px, are consistent with low-aperture single-camera systems \citep{Mallery2019, Moaven2024, Yamakaitis2024a}. At the high end of this range, recent campaigns using DIH and plenoptic LPT to measure turbulent flows have reported particle image densities and extraction rates similar to those considered here. For instance, deep learning methods for DIH-LPT can achieve extraction rates up to 95\% at seeding densities reaching 0.06~ppp \citep{Shao2020}. Likewise, the plenoptic LPT algorithm of \citet{Moaven2024} extracts more than 90\% of particles at seeding densities up to 0.06 particles per microlens (analogous to ppp for plenoptic imaging), although a larger sensor would be required to recover the same number of particles in a plenoptic configuration. Other sources of error, such as ghost particles and broken tracks, are not considered here because modern processing methods have been shown to suppress most such artifacts \citep{Schanz2016, Mallery2020}.\par

The TBL case is specifically designed to reproduce the conditions of a recent LPT experiment conducted by \citet{Schroder2024, Schroder2025}. In it, a TBL with $Re_\tau \approx 995$ was measured in a near-wall region of size $60 \times 12 \times 2$~mm$^3$. Particle images were captured by five high-speed cameras at a temporal resolution of 0.042~ms, with a particle image density of 0.06--0.07~ppp for a $2048 \times 512$ pixel sensor. The flow properties and acquisition parameters in this experiment closely match those of our synthetic case, which features a slightly larger domain ($56 \times 12 \times 4.25$~mm$^3$) and lower temporal resolution (0.057~ms), thereby posing a somewhat harder reconstruction problem. Similar near-wall measurements have also been performed using low-aperture imaging systems \citep{Yamakaitis2024b, Yamakaitis2025b}. Given this experimental backdrop, we use the TBL case to address our first question: can noisy tracks, when coupled with flow physics constraints and a localization error PDF, support accurate reconstruction of both the flow field and the \emph{true} tracer trajectories?\par

% HIT 1
\subsection{Homogeneous isotropic turbulence with bidisperse inertial particles}
\label{sec: cases: HIT}
The second case is based on a DNS of forced incompressible HIT, also taken from the JHTDB. Inertial particle transport in HIT has been studied extensively, both numerically \citep{Eaton2009, Rosa2016} and experimentally \citep{Petersen2019, Ferran2023}, making it a suitable numerical testbed for evaluating the joint estimation problem. The selected JHTDB dataset has a Taylor-scale Reynolds number of $Re_\lambda = 433$ and is simulated in a periodic domain of size $2\pi \times 2\pi \times 2\pi$, which is discretized into $1024^3$ voxels. We focus on the central $128^3$-voxel sub-volume, spanning 100 frames at a dimensionless temporal resolution of 0.002. To mimic laboratory conditions, the data are dimensionalized assuming air as the carrier fluid, giving a physical volume of $10^3$~cm$^3$, a measurement duration of 0.04~s, and a sampling rate of 2500~Hz. The Taylor and Kolmogorov length scales are $\ell_\lambda = 14.38$~mm and $\ell_\eta = 0.35$~mm, with corresponding time scales of $\tau_\lambda = 31.9$~ms and $\tau_\eta = 8.2$~ms.\par

Particle tracks are generated by simulating 50~000 spherical soda-lime glass beads of density $\rho_\mathrm{p} = 2500$~kg/m$^3$. Diameters are drawn from two Gaussian distributions, one with mean 32~$\upmu$m and standard deviation 2~$\upmu$m, and the other having mean 73~$\upmu$m and standard deviation 4~$\upmu$m, with equal sampling from each distribution. The resulting particle field has a volume fraction of $3.4 \times 10^{-6}$ and a mass loading of $6.9 \times 10^{-3}$, placing it near the upper boundary of the one-way-coupled regime \citep{Elghobashi1994, Brandt2022}. The corresponding Stokes numbers, based on $\tau_\eta$ and $\tau_\mathrm{p}$, are approximately 1 and 5, indicating substantial inertial lag and clustering \citep{Samimy1991}. Particle dynamics are governed by the Maxey--Riley equation \citep{Maxey1997}, which applies to small spherical particles in the one-way-coupled regime, together with a Schiller--Naumann drag correction that is valid for $Re_\mathrm{p} < 800$ \citep{Schiller1933}. Here, the maximum particle Reynolds number is about 6.5. Following \citet{Eaton2009} and \citet{Ling2013}, the Basset history force is neglected. Trajectories are integrated using a second-order Runge--Kutta scheme,\footnote{We use a second-order Runge--Kutta scheme to preserve the native temporal resolution of the DNS data; at the time steps used here, it yields nearly identical particle positions to a fourth-order scheme.} with periodic boundaries applied at the periphery of an extended domain (150 voxels). The simulation is run for 201 frames to minimize initialization and boundary artifacts. Reconstructions are performed for the $128^3$-voxel sub-volume over the final 100 frames at a temporal resolution of 0.4~ms. On average, 33~000 particles occupy the probe volume at any instant, corresponding to 0.033~ppp for a 1~MP camera. Once again, this seeding level lies comfortably within the capability of leading LPT techniques \citep{Schroder2023}. The mean inter-particle spacing is approximately $8.9 \ell_\eta$ or $0.216 \ell_\lambda$, which is low enough for joint reconstruction (see \S~\ref{sec: sensitivity: seeding}).\par 

The configuration of this case is motivated by extensive experimental studies of inertial particle dynamics in HIT. A pertinent example is the work of \citet{Petersen2019}, who investigated particle clustering and settling dynamics in HIT at $Re_\lambda \approx 200$--500. In that study, sub-Kolmogorov inertial particles ($d_\mathrm{p} = 30$--90~$\upmu$m) exhibited Stokes numbers of order 1 to 10; the carrier fluid was simultaneously seeded with 1--2~$\upmu$m DEHS (di-ethyl-hexyl-sebacate) tracers having negligible Stokes numbers. A single imaging system was used to perform simultaneous \emph{planar} PIV and LPT, with intensity-based image processing used to segregate the tracers and inertial particles. This enabled detailed analysis of particle--flow interactions, including preferential sampling of high-strain\slash low-vorticity regions and turbulence modulation. The same facility was later used at comparable $Re_\lambda$ and $St$ to study conditional Lagrangian statistics of inertial particles \citep{Berk2021} and slip velocities \citep{Berk2024b}. Although the flow and particle parameters in these studies are broadly comparable to the present case, the planar nature of the measurements can bias trajectory sampling \citep{Berk2021}. Extending such measurements to 3D would alleviate this issue, but doing so whilst tracking both tracer and inertial particles would restrict the admissible concentration of the latter. Moreover, optical segregation of tracers and inertial particles becomes quite difficult for smaller particles, e.g., toward $St \sim 1$, where phenomena such as inertial clustering \citep{Wang1993, Petersen2019} and gravitational settling \citep{Aliseda2002, Rosa2016} are especially pronounced.\par

Such limitations could, in principle, be addressed by joint estimation, in which time-resolved 3D flow fields and particle characteristics are inferred solely from inertial track data. Ideally, no additional seeding of tracers would be required to measure the carrier phase, nor would image-based phase segregation be needed, thereby permitting measurements of denser inertial particle fields. The present HIT case therefore allows us to assess whether inertial particle trajectories, themselves, contain enough information to support joint recovery of the flow field and unknown particle properties.\par

% Cone
\subsection{Supersonic cone--cylinder flow with inertial particle transport}
\label{sec: cases: cone}
The third case involves a steady compressible axisymmetric flow at Mach~2 over a 15$^\circ$ half-angle cone--cylinder body, generating an oblique shock wave at the nose and an expansion fan over the shoulder. The inflow density and temperature are 0.55~kg/m$^3$ and 166.7~K, and the cylinder radius is 20~mm, consistent with the experiments of \citet{Venkatakrishnan2004}. The flow is simulated with the compressible, axisymmetric Navier--Stokes solver in SU2~7.3.0. The computational domain spans a radius of 0.15~m and a length of 0.25~m, with $\gamma = 1.4$. For reference, the freestream velocity is $U_\infty \approx 520$~m/s and the cone length is 40~mm, leading to a characteristic Reynolds number of $Re \sim 10^6$. Viscous scales for this flow are on the order of a micron in length and tens of nanoseconds in time, and the physical shock thickness is likewise sub-micron, i.e., well below both the resolution of the CFD grid and that of present LPT scenario. Nevertheless, particle tracks throughout the domain provide physical anchors for DA reconstruction. Further details on the mesh, solver settings, and experimental validation are reported by \citet{Molnar2023}.\par

Particle tracks are simulated for 2000 solid spherical particles, modeled as agglomerated \ce{TiO2} seed. Diameters are drawn from a Gaussian distribution with mean $d_\mathrm{p} = 2~\upmu$m and standard deviation 0.5~$\upmu$m, and densities from a Gaussian distribution with mean $\rho_\mathrm{p} = 950$~kg/m$^3$ and standard deviation 100~kg/m$^3$; these values were obtained from the calibration measurements of \citet{Williams2015}. The resulting mean response time is $\tau_\mathrm{p} \approx 20~\upmu$s, which implies appreciable lag through the shock wave and expansion fan. The particle volume fraction and mass loading for this scenario are $10^{-9}$ and $10^{-6}$, respectively, lying well within the one-way coupled regime \citep{Brandt2022}. Particle transport is computed using a compressible drag law (see appendix~\ref{app: disperse: compressible drag}); the particles are injected at the entrance of the computational domain at the freestream velocity and advected downstream with periodic reinjection.\par

This case is intentionally idealized with respect to the measurement configuration and conservative with respect to the particle dynamics. The synthetic tracks are generated for eight frames at an implied imaging rate of 0.5~MHz. This frame rate, and the eight-pulse LPT configuration, lie near the boundary of what is currently possible in high-speed aerodynamic measurements, for which double-frame PIV remains the standard approach \citep{Beresh2021}. Nevertheless, recent work has begun to extend multi-pulse LPT into this regime \citep{Novara2019, Manovski2021}, and such developments may ultimately enable LPT-based interrogation of high-speed flows. The particle properties, by contrast, correspond to a relatively unfavorable seeding condition. The resulting test therefore asks whether joint estimation remains useful when high-speed LPT data are available but particle filtering remains significant. Although we consider only the mean flow in this case, the characteristic time scales of turbulent fluctuations are much shorter than those associated with mean velocity gradients. As a result, effective Stokes numbers in real experiments on unsteady high-speed flows may be comparable to those tested here, even for tracers with much faster response times.\par

As mentioned above, shorter response times than those reported by \citet{Williams2015} have indeed been achieved for \ce{TiO2} seed. For example, \citet{Urban2001} reported response times of 3--4~$\upmu$s for \ce{TiO2} particles from oblique-shock tests in supersonic shear layers, and \citet{Scarano2003} determined relaxation times below 2~$\upmu$s. \citet{Ragni2011} further developed specialized seeding and calibration procedures and reported typical response times of 2.5--3.25~$\upmu$s for \ce{TiO2}, reducing the response time to as little as 0.5~$\upmu$s through dehydration, a custom cyclone seeder, and a 1~$\upmu$m particle filter. Such measures may be necessary to ensure adequate traceability, although the particle filter in particular was reported to substantially reduce the number of particles delivered to the probe volume. In light of these studies, the calibration data of \citet{Williams2015} are likely associated with agglomerated seed. The present cone case thus provides a conservative test of whether joint estimation can compensate for non-ideal tracer behavior without a priori knowledge of the particle response. We note again that the filtering effect associated with $\tau_\mathrm{p} \approx 20~\upmu$s tracers in steady flow may be viewed as a proxy for effect of very agile tracers when resolving turbulent fluctuations in high-speed flows, for which effective values of $St$ can be high. For completeness, we also consider a lower-inertia case with $\tau_\mathrm{p} = 3.1~\upmu$s in appendix~\ref{app: low-inertia particles}.\par

% HIT 2
\subsection{Homogeneous isotropic turbulence with varying inertial particles and noise}
\label{sec: cases: sensitivity}
Lastly, we extend the HIT case from \S~\ref{sec: cases: HIT} to establish how localization uncertainty and inertial effects interact when jointly determining flow fields and particle properties. We use the central $64^3$-voxel sub-domain of the HIT dataset while keeping the temporal resolution, number of frames, and advection schemes unchanged. A dense field of 6600 particles is first simulated and then downsampled by factors of $2^N$, for $N \in \{1, \dots, 6\}$, producing seeding densities with inter-particle spacings from $7.3\ell_\eta$ to $29.3\ell_\eta$. The particle density is increased from 2500~kg/m$^3$ in \S~\ref{sec: cases: HIT} to 6000~kg/m$^3$, and diameters are drawn from a Gaussian distribution with mean 33~$\upmu$m and standard deviation 4~$\upmu$m. This ensures one-way coupling, even at the densest seeding condition. The mean Stokes number is 3, and no localization error is applied at this stage.\par

We then vary the localization error and Stokes number independently. A total of 3300 particles are simulated, giving a mean spacing of $9.2\ell_\eta$, sufficient for a scale-resolving reconstruction in the tracer limit. Particle densities are then adjusted to set $St \in \{1, \dots, 5\}$ via $\rho_\mathrm{p} = 2000 St$~kg/m$^3$; diameters are fixed to avoid the two-way coupled regime. The clean tracks are corrupted with additive Gaussian noise, with $\sigma_x = \sigma_y \in \{0.1, \dots, 0.5 \}$ in pixel units, assuming a 1~MP camera. Similar to \S~\ref{sec: cases: TBL}, the lowest level ($\sigma_x = 0.1$~px) corresponds to STB accuracy in ideal laboratory conditions \citep{Schanz2016}. Once again, to mimic anisotropy, the magnitude of wall-normal noise is doubled ($\sigma_z/\sigma_x = 2$). These settings yield 25 cases spanning a matrix of localization errors and Stokes numbers.\par

%%% Numerical Details %%%
\section{Implementation of the method and evaluation metrics}
\label{sec: implementation}

% Architecture
\subsection{Neural network architectures}
\label{sec: implementation: architecture}
We implement NIPA in TensorFlow~2.10. Exact partial derivatives of the model outputs with respect to $\boldsymbol{x}$ and $t$, and derivatives of the loss function with respect to the model parameters, are computed using automatic differentiation. The integrals over $\mathcal{V}$, $\mathcal{A}$, and $\mathcal{T}$ in \eqref{equ: flow physics loss}--\eqref{equ: boundary loss} are approximated by Monte Carlo sampling. At each training step, the probe volume is sampled with a batch of 5000 points drawn uniformly from $\mathcal{V}$, and the boundary loss is evaluated using 1000 points drawn uniformly from $\mathcal{A}$. For the particles, 5000 tracks are sampled at each iteration, and eight random times are drawn for each track; the first and last times are also included to suppress boundary inflections, culminating in a particle batch size of 50~000 points.\par

\begin{table}[htb]
    \renewcommand{\arraystretch}{1.25}
    \caption{Network architectures used for the flow models (velocity, pressure, or full primitive state) in the three test cases.}
    \centering\vspace*{.3em}
    \begin{tabular}{c l c}
        \hline\hline
        \multicolumn{1}{c}{\bf Case} &
        \multicolumn{1}{c}{\bf Flow Model} &
        \multicolumn{1}{c}{\bf Layers $\boldsymbol\times$ Neurons} \\
        \hline
        \multirow{2}{*}{TBL} & $\mathsf{F}_{\boldsymbol{u}} : \left(\boldsymbol{x}, t\right) \mapsto \boldsymbol{u}$ & $10 \times 300$\\
        & $\mathsf{F}_{p} : \left(\boldsymbol{x}, t\right) \mapsto p $ & $10 \times 150$\\[.5em]
        \multirow{2}{*}{HIT} & $\mathsf{F}_{\boldsymbol{u}} : \left(\boldsymbol{x}, t\right) \mapsto \boldsymbol{u}$ & $15 \times 300$ \\
        & $\mathsf{F}_{p} : \left(\boldsymbol{x}, t\right) \mapsto p $ & $15 \times 150$\\[.5em]
        cone & $\mathsf{F}_{(\rho, \boldsymbol{u}, T)} : \boldsymbol{x} \mapsto \left(\rho, \boldsymbol{u}, T \right) $ & $10 \times 250$ \\[.25em]
        \hline\hline
    \end{tabular}
    \label{tab: architecture}
\end{table}

The network architectures are tailored to the complexity of each flow case, with the corresponding parameters listed in \cref{tab: architecture}. Flow models and sub-models are denoted by $\mathsf{F}$, with subscripts indicating the target fields, e.g., $\mathsf{F}_{\boldsymbol{u}}$ and $\mathsf{F}_{p}$ represent neural models of the velocity and pressure fields, respectively. All the networks employ a Fourier encoding layer (see \S~\ref{sec: method: flow model}), with frequency vectors $\boldsymbol{f}_i$ drawn from a standard Gaussian distribution for spatial features and from a zero-mean Gaussian distribution with a standard deviation of 0.2 for temporal features. The number of features is fixed at 1024. This choice provides sufficient expressivity for all the flow cases. We tested networks with 128, 256, 512, and 1024 Fourier features and found that the accuracy of the learned fields saturated before 1024 features, with little additional computational cost incurred between 512 and 1024. An assessment of the the networks' representational power is provided in appendix~\ref{app: expressivity}.\par

% Initialization
\subsection{Model initialization strategy}
\label{sec: implementation: initialization}
Initialization of the flow model is straightforward: the network weights are drawn from a standard normal distribution and the biases are set to zero. Initialization of the KCTs proceeds in three steps. First, long tracks are split into shorter segments to reduce computational costs and keep the associated matrix operations manageable; guidelines for selecting an appropriate segment length are provided in appendix~\ref{app: KCT: split}. Second, for ideal tracers, the particle displacements $\boldsymbol{\delta}$ are directly initialized from the raw (noisy) track data, while the polynomial coefficients in $\boldsymbol{\theta}$ are set to zero. We refer to this as a \emph{cold start} since it requires no prior information. Conversely, inertial tracks are \emph{warm-started} using filtered track data to aid convergence. This is necessary because additional unknowns, such as particle diameters and densities, are inferred in the inertial cases, making the inverse problem more ill-posed. Warm-started KCTs thus inherit both the regularity and the bias of the chosen filter, which is a necessary trade-off for stable optimization. Finally, all trainable parameters are normalized to be of order unity. This allows for particle quantities with different units and scales to be trained together using a single learning rate. Appendix~\ref{app: KCT: warm start} provides details of the warm-start procedure, and appendix~\ref{app: KCT: transform} describes the normalization and parameter selection strategy.\par

% Training
\subsection{Training procedure}
\label{sec: implementation: training}
Flow and KCT models are trained together by minimizing $\mathscr{J}_\mathrm{total}$. The weighting coefficients for each loss term $\chi_i$ are chosen through a simple parameter sweep, in which reconstructions are performed for different values of $\chi_i$. We found that accuracy is relatively insensitive to $\chi_i$ near the optimum. Training is performed using the Adam optimizer, with the learning rate for flow networks fixed at $10^{-3}$. For particle models, the learning rate is annealed from $10^{-4}$ to $10^{-5}$ and finally to $10^{-6}$ to improve precision. All cases are trained to convergence, typically requiring about 2000 epochs per learning rate. Computations are performed on an NVIDIA RTX A6000 GPU with 48~GB of onboard memory. The total wall-clock training time is approximately 25~hours for the TBL cases, 15~hours for the HIT cases, and 3~hours for the cone case.\par

More advanced neural architectures and optimization strategies may further improve DA performance. Examples include nonlinear residual layers such as PirateNet \citep{wang2024a}, multi-stage networks designed to better capture multi-scale features \citep{wang2024b}, quasi-second-order optimizers like SOAP \citep{Wang2025}, and automatic loss-weighting methods \citep{Wang2022d}. Incorporating such techniques could improve both reconstruction accuracy and training efficiency in future studies.\par

% Spatial Errors
\subsection{Error metrics and spectral resolution}
\label{sec: implementation: metrics}
Reconstruction accuracy is evaluated using global and spectral error metrics. For a field variable $\varphi$, the normalized root-mean-square error (NRMSE) is
\begin{equation}
    \label{equ: NRMSE}
    e_\varphi = \left(\frac{\left\langle \left\lVert \varphi - \varphi_\mathrm{exact} \right\rVert_2^2 \right\rangle} {\left\langle \left\lVert \varphi_\mathrm{exact} \right\rVert_2^2 \right\rangle}\right)^{1/2},
\end{equation}
where $\varphi_\mathrm{exact}$ is the ground truth. The averaging operator $\langle \cdot \rangle$ may be taken over either the spatio-temporal domain,
\begin{equation}
        \label{equ: average}    
        \left\langle \varphi \right\rangle_{\mathcal{V} \times \mathcal{T}} = \frac{1}{\left| \mathcal{V} \times \mathcal{T} \right|} \int_\mathcal{T} \int_\mathcal{V} \varphi\mathopen{} \left(\boldsymbol{x}, t\right) \mathrm{d}\boldsymbol{x} \,\mathrm{d}t,
\end{equation}
or across the tracks,
\begin{equation}
        \label{equ: average: particle}    
        \left\langle \varphi \right\rangle_\mathrm{p} = \frac{1}{n_\mathrm{p}} \sum_{k=1}^{n_\mathrm{p}} \frac{1}{\left| \sdx{\mathcal{T}}[k] \right|} \int_{\sdx{\mathcal{T}}[k]} \varphi(t) \,\mathrm{d}t.
\end{equation}
In practice, these integrals are approximated by sums over DNS grid points and time steps for flow fields, and over discrete measurement instants in $\sdx{\mathcal{T}}[k]$ for tracks. In the unsteady flow cases, we report errors for the fluctuating component $\varphi^\prime$, defined by a Reynolds decomposition $\varphi = \overline{\varphi} + \varphi^\prime$. This yields a conservative estimate of reconstruction accuracy, since mean fields are easier to recover than turbulent fluctuations. In the steady cone flow case, errors are reported for the primitive fields without modification.\par

Spectral error analysis quantifies reconstruction accuracy across wavenumbers. The spherical averaging operator in Fourier space is
\begin{equation}
    \label{equ: spectral average}
    \left\langle \widetilde{\varphi} \right\rangle_{\kappa} = \int_{\mathcal{K}(\kappa)} \widetilde{\varphi}\mathopen{} \left(\boldsymbol{\kappa}\right) \mathrm{d}\boldsymbol{\kappa}
\end{equation}
where $\widetilde{\varphi}$ is the 3D Fourier transform of $\varphi$ and $\mathcal{K}(\kappa)$ is a shell of radius $\kappa$. In practice, the integration is approximated by averaging Fourier magnitudes within discrete shells of width $\Delta\kappa$. For velocity fields, the turbulent kinetic energy (TKE) spectrum is computed using \eqref{equ: spectral average} for $\varphi = \boldsymbol{u} \boldsymbol{\cdot} \boldsymbol{u}/2$. Normalized velocity error spectra are then given by
\begin{equation}
    \label{equ: spectral error}
    e_{\widetilde{\boldsymbol{u}}}(\kappa) =
    \left(\frac{\left\langle \left\lVert \widetilde{\boldsymbol{u}} - \widetilde{\boldsymbol{u}}_\mathrm{exact} \right\rVert_2^2 \right\rangle_\kappa} {\left\langle \left\lVert \widetilde{\boldsymbol{u}}_\mathrm{exact} \right\rVert_2^2 \right\rangle_\kappa}\right)^{1/2}.
\end{equation}
This spectrum measures the energy of velocity reconstruction errors relative to the true turbulent energy at each wavenumber. A key reference point for interpreting these spectra is the particle sampling Nyquist wavenumber,
\begin{equation}
    \label{equ: Nyquist}
    \kappa_\mathrm{Nyq} = \frac{\pi}{\delta},
\end{equation}
where $\delta$ is the mean inter-particle spacing. Note that $\kappa_\mathrm{Nyq}$ represents the highest resolvable wavenumber from interpolation of the particle velocities, alone. In the absence of physics-based constraints, error levels beyond this limit are expected to saturate at 100\%.\par

%%% Results %%%
\section{Joint estimation of flow fields and particle positions}
\label{sec: tracer}
We firstly examine the TBL case, in which the particle tracks correspond to ideal tracers. Because trajectory estimation is the first step in LPT workflows, we begin in \S~\ref{sec: tracer: track} by comparing several track processing methods applied directly to noisy LPT data, including finite differencing, B-splines with a fixed length, TrackFit, and KCTs trained without a physics loss. We then compare these results to the track estimates produced by our joint reconstruction procedure.\par

Next, we examine flow field reconstructions in \S~\ref{sec: tracer: flow}. Specifically, we compare baseline reconstructions, produced using the track-only kinematics, to the results of joint estimation. By varying the seeding density and magnitude of localization error, we assess how track density and track quality affect both stages of the reconstruction problem.\par

% Track-only optimization
\subsection{Particle track optimization}
\label{sec: tracer: track}
We process datasets with three seeding densities, corresponding to mean particle spacings of $\delta = 9\ell_\nu$, $18\ell_\nu$, and $36\ell_\nu$, where $\ell_\nu$ is the friction length. We also consider six localization error levels, with $\sigma_x \in \{0.09, 0.27, 0.45, 0.63, 0.81, 0.9\}\ell_\nu$. Five approaches are used to estimate the tracks. First, we present the raw tracks, with velocities obtained by finite differencing. Second, we apply a B-spline filter, described in appendix~\ref{app: B-spline}, with the segment length tuned in a supervised manner to minimize error across all the datasets. Third, we use TrackFit \citep{Gesemann2016, Gesemann2021}, which also represents trajectories with B-splines and approximates an optimal Wiener filter. Specifically, we use the advanced TrackFit variant of \citet{Buchwald2025}, wherein the weights are adapted from local track statistics to improve performance in inhomogeneous flows. Fourth, we train KCT models using the data loss in isolation. These four approaches rely solely on measurement data and do not incorporate any flow physics information. For our fifth and final approach, we train KCTs jointly with the flow model using the combined data, flow physics, particle physics, and boundary losses. This final setting embeds physical constraints and therefore indicates the extent to which such coupling improves track accuracy. Below, these five datasets are labeled \emph{raw}, \emph{filtered}, \emph{TrackFit}, \emph{KCT} (i.e., data only), and \emph{joint estimation}.\par

\Cref{tab: error x,tab: error z} report the error standard deviations in the $x$- and $z$-directions; statistics for the $x$- and $y$-directions are nearly identical. Finite differencing amplifies positional errors into very noisy velocity and acceleration estimates. B-spline filtering, by contrast, suppresses these errors by factors of two or more, consistent with the findings of \citet{Li2024b}. TrackFit further improves upon the B-spline results by adapting its filter parameters from the position amplitude spectra of local track statistics \citep{Buchwald2025}. Data-only KCTs perform worse than the raw tracks because the data loss merely enforces statistical consistency with the assumed localization uncertainty. The resulting optimization is therefore highly ill-posed because infinitely many non-physical trajectories are perfectly consistent with the target distribution of residuals. Therefore, we exclude data-only KCTs from the flow reconstructions in \S~\ref{sec: tracer: flow}.\par

\begin{table}[ht]
    \renewcommand{\arraystretch}{1}
    \caption{Standard deviations of track errors in the $x$-direction for varying noise levels. At each level, the three sub-rows correspond to the position, velocity, and acceleration errors, normalized by the friction length $\ell_\nu$, velocity $v_\nu$, and acceleration $v_\nu^2/\ell_\nu$, respectively. The lowest error in each row is shown in bold.}
    \centering
    \begin{tabular}{c c c c c c c c}
        \hline\hline
        \multirow{2}{*}{Noise level} &
        \multirow{2}{*}{Raw} &
        \multirow{2}{*}{Filtered} &
        \multirow{2}{*}{TrackFit} &
        \multirow{2}{*}{KCT} &
        \multicolumn{3}{c}{Joint estimation}\\
        & & & & & \multicolumn{1}{c}{$\delta = 9\ell_\nu$} & \multicolumn{1}{c}{$\delta = 18\ell_\nu$} & \multicolumn{1}{c} {$\delta = 36\ell_\nu$} \\
        \hline
        \multirow{3}{*}{0.05~px} & 0.09 & 0.08 & \textbf{0.04} & 0.13 & \textbf{0.04} & 0.05 & 0.05\\
        & 0.11 & 0.07 & 0.04 & 0.17 & \textbf{0.03} & 0.04 & 0.05\\
        & 0.14 & 0.07 & 0.04 & 0.30 & \textbf{0.03} & \textbf{0.03} & 0.05\\
        
        \multirow{3}{*}{0.15~px} & 0.27 & 0.16 & \textbf{0.12} & 0.38 & \textbf{0.12} & 0.13 & 0.14 \rule{0pt}{15pt}\\
        & 0.24 & 0.11 & 0.08 & 0.34 & \textbf{0.06} & 0.08 & 0.10\\
        & 0.32 & 0.10 & 0.05 & 0.35 & \textbf{0.03} & 0.04 & 0.07\\
        
        \multirow{3}{*}{0.25~px} & 0.45 & 0.24 & 0.18 & 0.63 & \textbf{0.17} & 0.20 & 0.21 \rule{0pt}{15pt}\\
        & 0.37 & 0.15 & 0.10 & 0.53 & \textbf{0.07} & 0.09 & 0.13\\
        & 0.50 & 0.14 & 0.06 & 0.65 & \textbf{0.03} & 0.04 & 0.07\\
        
        \multirow{3}{*}{0.35~px} & 0.63 & 0.33 & 0.24 & 0.88 & \textbf{0.23} & 0.26 & 0.29 \rule{0pt}{15pt}\\
        & 0.49 & 0.20 & 0.13 & 0.72 & \textbf{0.09} & 0.12 & 0.16\\
        & 0.67 & 0.19 & 0.07 & 0.82 & \textbf{0.04} & 0.05 & 0.09\\
        
        \multirow{3}{*}{0.45~px} & 0.81 & 0.42 & 0.31 & 1.14 & \textbf{0.28} & 0.33 & 0.36 \rule{0pt}{15pt}\\
        & 0.62 & 0.26 & 0.16 & 0.92 & \textbf{0.09} & 0.13 & 0.17\\
        & 0.84 & 0.24 & 0.08 & 1.20 & \textbf{0.03} & 0.06 & 0.11\\
        
        \multirow{3}{*}{0.5~px} & 0.90 & 0.47 & 0.34 & 1.26 & \textbf{0.30} & 0.38 & 0.40 \rule{0pt}{15pt}\\
        & 0.71 & 0.28 & 0.17 & 1.04 & \textbf{0.10} & 0.15 & 0.18\\
        & 0.97 & 0.26 & 0.09 & 1.29 & \textbf{0.04} & 0.05 & 0.12\\
        \hline\hline
    \end{tabular}
    \label{tab: error x}
\end{table}

\begin{figure}[htb!]
    \vspace{-0.5em}
    \centering
    \includegraphics[width=1\linewidth]{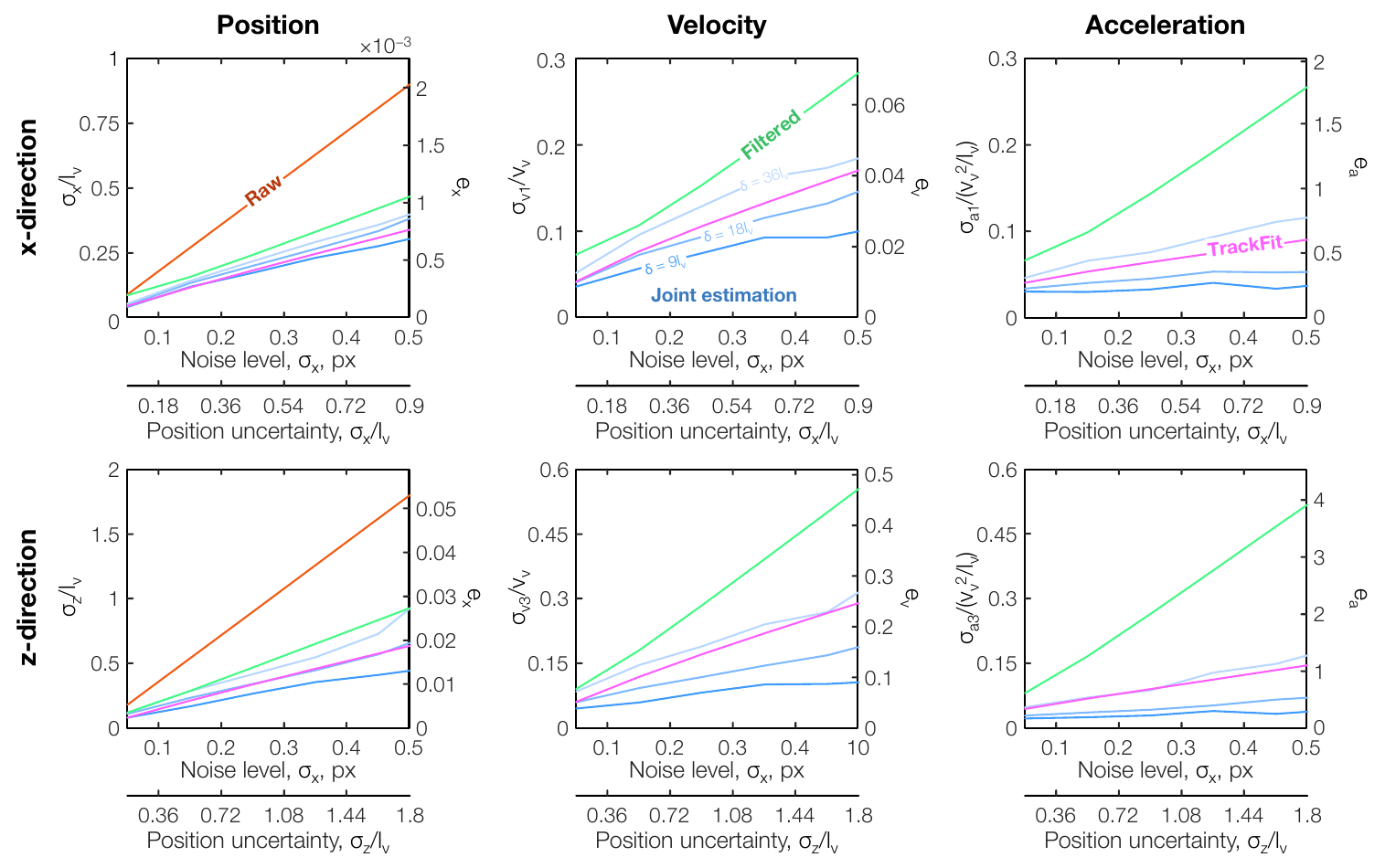}
    \caption{Standard deviations and NRMSEs of particle position, velocity, and acceleration errors in the $x$- and $z$-directions under varying noise levels. Joint flow--particle estimation is applied at three inter-particle spacings ($\delta = 9\ell_\nu$, $18\ell_\nu$, and $36\ell_\nu$). Standard deviations are normalized by viscous units. The corresponding position uncertainty at each noise level is labeled below. Joint estimation consistently yields the lowest errors, with some dependence on seeding density. TrackFit achieves accuracy comparable to joint estimation for $\delta = 18\ell_\nu$ to $36\ell_\nu$.}
    \label{fig: TBL: tracks error}
\end{figure}

In stark contrast, joint particle--flow estimation substantially improves the accuracy of the KCT estimates, yielding lower position, velocity, and acceleration errors than the filtering-based approaches across all the seeding densities considered here. Unlike the first four methods, however, performance of the jointly trained models depends upon the seeding density: as the number of particles decreases, the flow becomes progressively under-resolved, which in turn degrades the track estimates. \Cref{fig: TBL: tracks error} plots the error standard deviations for the raw, filtered, TrackFit, and jointly estimated tracks in the $x$- and $z$-directions against the noise level. At the lowest levels, all methods exhibit similar performance. At higher levels, however, joint estimation reduces the position, velocity, and acceleration errors by up to 60\% relative to B-spline filtering and TrackFit in the densely seeded cases. The benefit of joint estimation diminishes as the seeding density decreases, and TrackFit becomes competitive once the mean spacing $\delta$ lies between $18\ell_\nu$ and $36\ell_\nu$.\par

\begin{table}[ht]
    \renewcommand{\arraystretch}{1}
    \caption{Standard deviations of track errors in the $z$-direction for varying noise levels. At each level, the three sub-rows correspond to the position, velocity, and acceleration errors, normalized by the friction length $\ell_\nu$, velocity $v_\nu$, and acceleration $v_\nu^2/\ell_\nu$, respectively. The lowest error in each row is shown in bold.}
    \centering
    \begin{tabular}{c c c c c c c c}
        \hline\hline
        \multirow{2}{*}{Noise level} &
        \multirow{2}{*}{Raw} &
        \multirow{2}{*}{Filtered} &
        \multirow{2}{*}{TrackFit} &
        \multirow{2}{*}{KCT} &
        \multicolumn{3}{c}{Joint estimation}\\
        & & & & & \multicolumn{1}{c}{$\delta = 9\ell_\nu$} & \multicolumn{1}{c}{$\delta = 18\ell_\nu$} & \multicolumn{1}{c} {$\delta = 36\ell_\nu$} \\
        \hline
        \multirow{3}{*}{0.05~px} & 0.18 & 0.12 & \textbf{0.08} & 0.25 & \textbf{0.08} & 0.10 & 0.11\\
        & 0.21 & 0.09 & 0.06 & 0.35 & \textbf{0.04} & 0.06 & 0.08\\
        & 0.28 & 0.08 & 0.05 & 0.52 & \textbf{0.02} & 0.04 & 0.05\\
        
        \multirow{3}{*}{0.15~px} & 0.54 & 0.28 & 0.22 & 0.76 & \textbf{0.17} & 0.23 & 0.29 \rule{0pt}{15pt}\\
        & 0.45 & 0.18 & 0.12 & 0.71 & \textbf{0.06} & 0.09 & 0.15\\
        & 0.63 & 0.17 & 0.07 & 0.82 & \textbf{0.03} & 0.04 & 0.07\\
        
        \multirow{3}{*}{0.25~px} & 0.9 & 0.47 & 0.34 & 1.27 & \textbf{0.27} & 0.34 & 0.42 \rule{0pt}{15pt}\\
        & 0.69 & 0.28 & 0.17 & 1.08 & \textbf{0.08} & 0.12 & 0.19\\
        & 0.96 & 0.26 & 0.09 & 1.19 & \textbf{0.03} & 0.04 & 0.09\\
        
        \multirow{3}{*}{0.35~px} & 1.26 & 0.65 & 0.46 & 1.77 & \textbf{0.36} & 0.45 & 0.55 \rule{0pt}{15pt}\\
        & 0.94 & 0.39 & 0.22 & 1.81 & \textbf{0.10} & 0.14 & 0.24\\
        & 1.3 & 0.37 & 0.11 & 2.42 & \textbf{0.04} & 0.05 & 0.13\\
        
        \multirow{3}{*}{0.45~px} & 1.62 & 0.84 & 0.58 & 2.28 & \textbf{0.42} & 0.57 & 0.73 \rule{0pt}{15pt}\\
        & 1.17 & 0.50 & 0.27 & 1.99 & \textbf{0.10} & 0.17 & 0.27\\
        & 1.67 & 0.47 & 0.13 & 2.80 & \textbf{0.03} & 0.07 & 0.15\\
        
        \multirow{3}{*}{0.5~px} & 1.8 & 0.93 & 0.64 & 2.5 & \textbf{0.45} & 0.66 & 0.92 \rule{0pt}{15pt}\\
        & 1.36 & 0.55 & 0.29 & 2.11 & \textbf{0.11} & 0.19 & 0.31\\
        & 1.87 & 0.52 & 0.15 & 2.85 & \textbf{0.04} & 0.07 & 0.17\\
        \hline\hline
    \end{tabular}
    \label{tab: error z}
\end{table}

\begin{figure}[ht!]
    \vspace{-0.5em}
    \centering
    \includegraphics[width=1\linewidth]{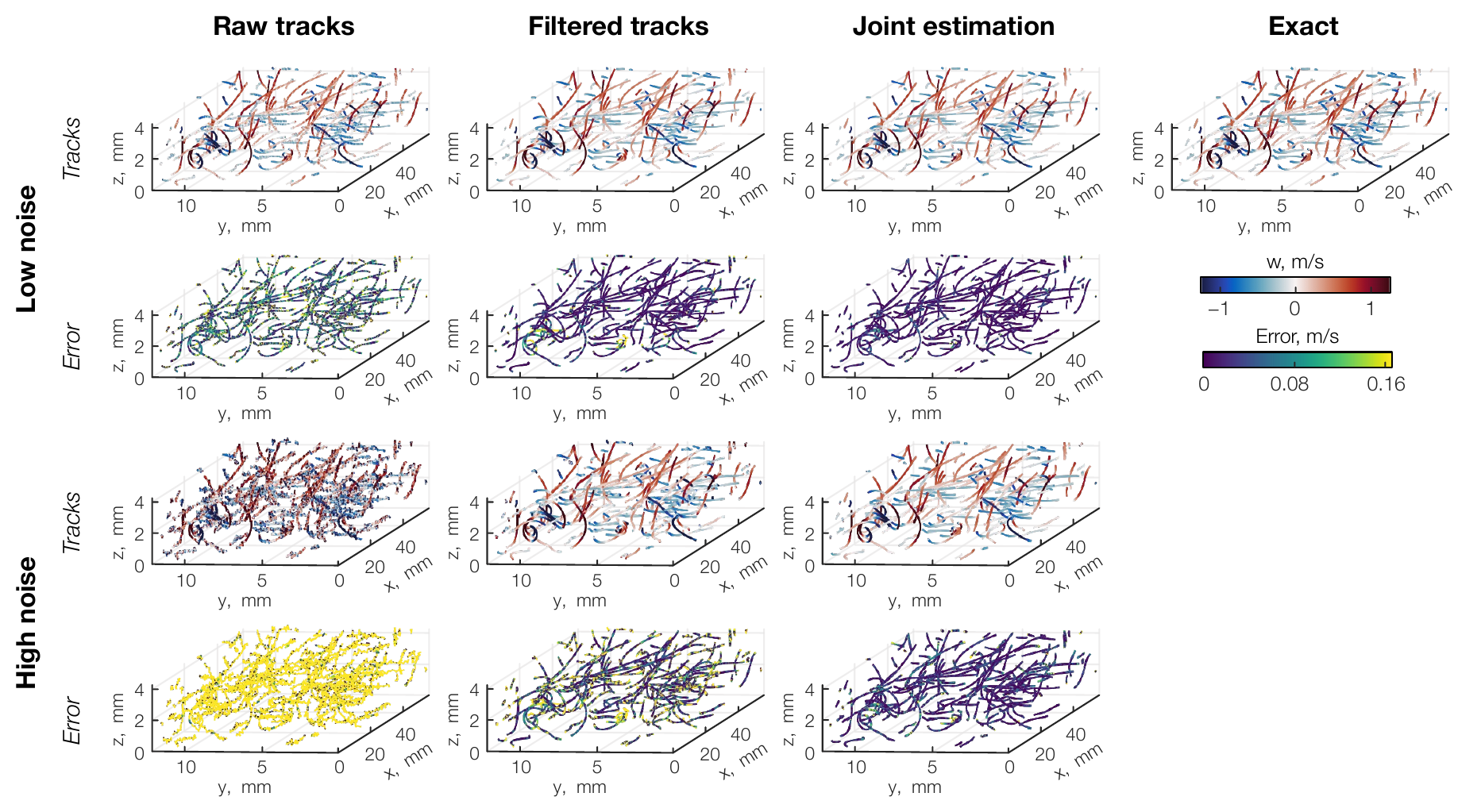}
    \caption{Comparison of exact, raw, filtered, and jointly estimated tracks, together with their pointwise velocity errors, at the lowest ($\sigma_x = 0.09\ell_\nu$) and highest ($\sigma_x = 0.9\ell_\nu$) noise levels for the densest seeding case ($\delta = 9\ell_\nu$). Only 200 tracks are shown for clarity. Colors indicate the $v_3$ velocity or its absolute error. Raw track velocity errors at high noise are downscaled by a factor of five for visualization. Joint estimation accurately reconstructs track geometries and velocities across noise levels, whereas finite differencing and filtering show large errors, especially near boundaries and in regions of high acceleration.}
    \label{fig: TBL: tracks}
\end{figure}

\begin{figure}[ht!]
    %\vspace{-0.5em}
    \centering
    \includegraphics[width=0.6\linewidth]{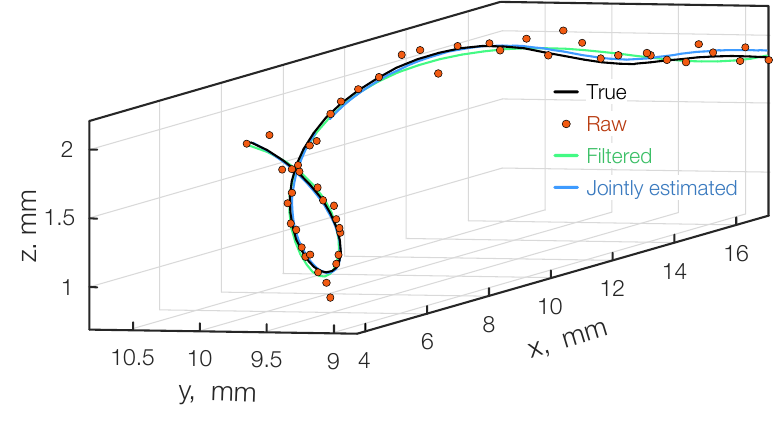}
    \caption{Close-up of a representative track at the noisiest condition. The true track is shown in black, the noisy data as red dots, the filtered track in green, and the jointly estimated track in blue. Although filtering improves upon the raw data, joint estimation more accurately recovers the true trajectory.}
    \label{fig: TBL: single track}
\end{figure}

\Cref{fig: TBL: tracks} shows representative raw, filtered, and jointly estimated tracks for the densest seeding case, $\delta = 9\ell_\nu$, at the lowest and highest noise levels. Tracks are colored by the $v_3$ component of velocity and by its absolute error. The $z$-direction coincides with the optical axis and therefore exhibits the largest localization errors. Finite difference velocities are highly sensitive to noise, largely obscuring the underlying flow. Track filtering suppresses this amplification and recovers tracks that qualitatively resemble the ground truth, but at high noise the filtered tracks still exhibit substantial errors. This is especially evident in segments of high curvature, where localization errors become comparable to the true particle motion, and near the first and last time steps, where the available temporal support is limited. Joint estimation ameliorates these issues through physical constraints, yielding low errors over the full extent of each track across all noise levels and seeding densities. A close-up in \cref{fig: TBL: single track} shows this contrast: the ``jointly estimated track'' captures subtle fluctuations of the true trajectory, whereas the filtered track remains visibly distorted. Quantitatively, at low noise, NRMSEs for the $v_3$ velocity component are 17.9\%, 7.7\%, and 3.9\% for finite differencing, B-spline filtering, and KCTs with joint estimation, respectively. At high noise, these errors increase to 174.4\%, 47.2\%, and 9.0\%. That is to say, noise amplifies the errors of the raw and filtered tracks by nearly an order of magnitude, whereas joint estimation limits the amplification to roughly a factor of two to three. This robustness carries directly into the flow reconstruction, as discussed next.\par

% Flow reconstructions
\subsection{Flow field reconstruction}
\label{sec: tracer: flow}
\begin{figure}[htb]
    \vspace{-0.5em}
    \centering
    \includegraphics[width=0.9\linewidth]{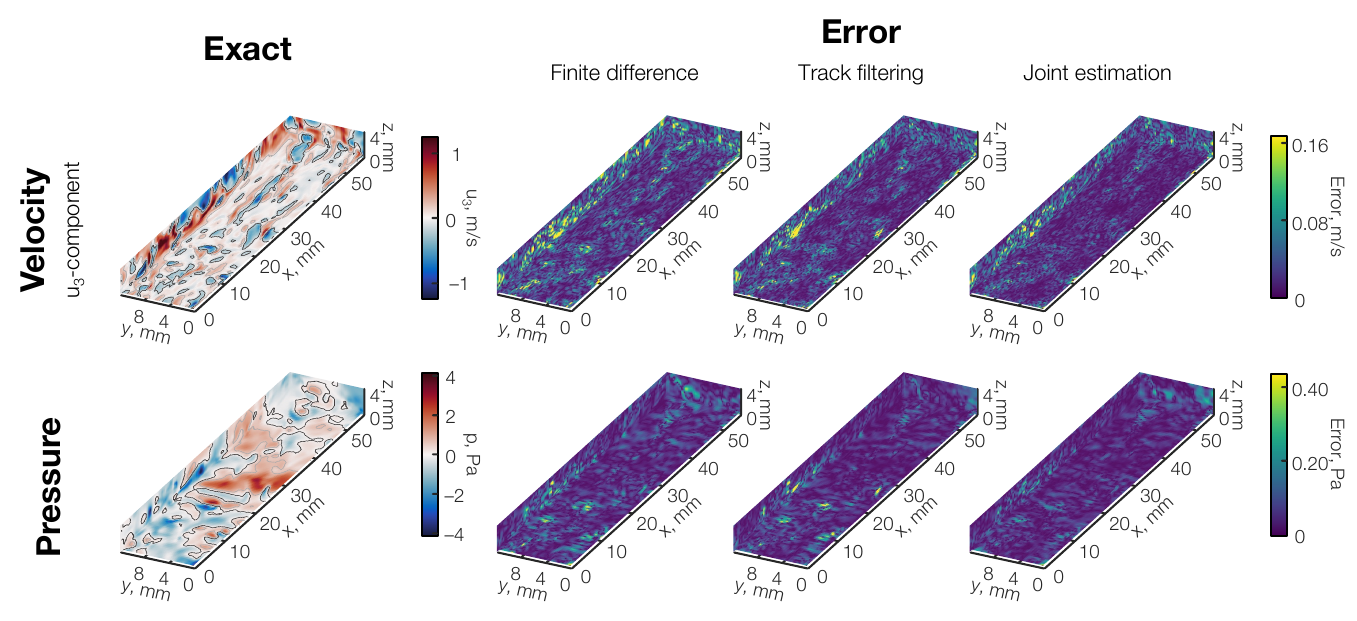}
    \caption{Exact flow fields (right) and absolute error fields (left) at the lowest noise level ($\sigma_x = 0.09\ell_\nu$) and highest seeding density ($\delta = 9\ell_\nu$). Dark and gray contour lines indicate low and high iso-values from the exact fields, overlaid on the reconstructions for comparison. The iso-values are $-0.17$ and 0.17~m/s for $u_3$, and 0 and 0.81~Pa for $p$, respectively. All methods recover the main flow features, but joint estimation yields the lowest errors.}
    \label{fig: TBL: low-noise}
\end{figure}

Flow states are reconstructed by two methods. First, in the baseline method, described in \S~\ref{sec: method: loss: baseline}, we train the flow model using velocities obtained either from raw tracks processed by finite differencing or using tracks filtered by B-spline smoothing. Second, in the joint estimation, we train the KCT and flow models together. \Cref{fig: TBL: low-noise} shows the ground truth and error fields of the $u_3$ velocity component and pressure at the lowest noise level, $\sigma_x = 0.09\ell_\nu$, using the tracks shown in \cref{fig: TBL: tracks}. The fields are rendered on three orthogonal planes at the central snapshot in time. All three methods recover the main flow structures with high fidelity, but finite differencing and filtering yield larger errors than joint estimation, with NRMSEs of 17.6\%, 8.0\%, and 4.2\%, respectively. The corresponding error fields show a clear reduction in reconstruction error from finite differencing to filtering to joint estimation, illustrating the effects of particle--flow feedbacks. For the $u_3$ field, low-noise NRMSEs are 11.6\%, 9.4\%, and 7.8\%, while for pressure they are 14.9\%, 13.4\%, and 12.9\%.\par

\begin{figure}[htb!]
    \vspace{-0.5em}
    \centering
    \includegraphics[width=0.9\linewidth]{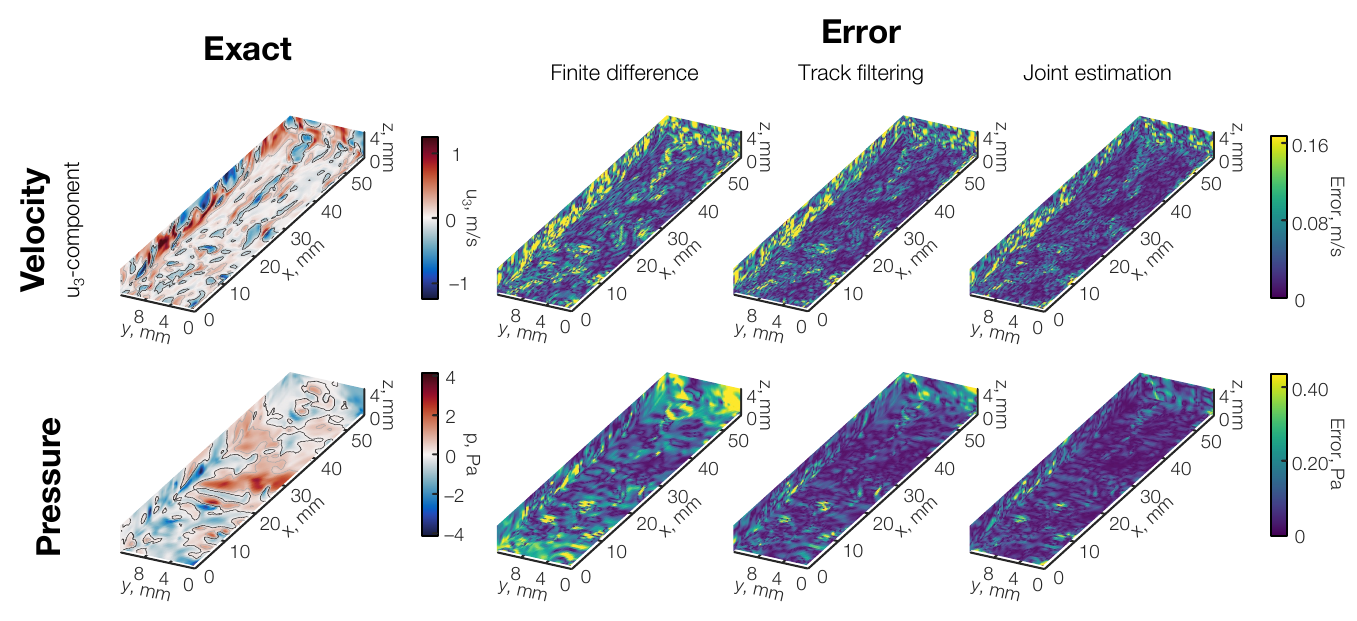}
    \caption{Exact flow fields (right) and absolute error fields (left) at the highest noise level ($\sigma_x = 0.9\ell_\nu$) and highest seeding density ($\delta = 9\ell_\nu$). Dark and gray contour lines indicate low and high iso-values from the exact fields, overlaid on the reconstructions for comparison. The iso-values are $-0.17$ and 0.17~m/s for $u_3$, and 0 and 0.81~Pa for $p$, respectively. Finite-difference and filtered reconstructions blur fine-scale features, whereas joint estimation more faithfully recovers the detailed flow structure.}
    \label{fig: TBL: high-noise}
\end{figure}

\begin{figure}[htb!]
    \vspace*{-0.5em}
    \centering
    \includegraphics[width=0.8\linewidth]{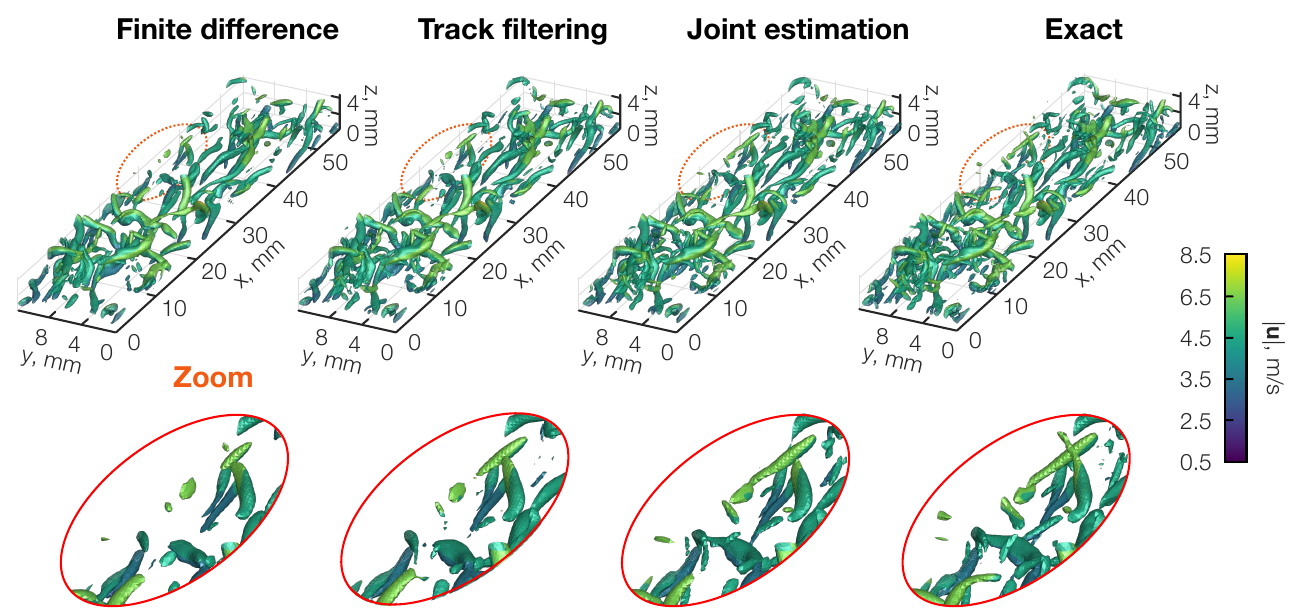}
    \vspace*{0.5em}
    \caption{Coherent structures in the TBL reconstructed at the highest noise level ($\sigma_x = 0.9\ell_\nu$) and densest seeding ($\delta = 9\ell_\nu$): finite differencing (far left), filtered tracks (left center), joint estimation (right center), and ground truth (far right). Structures are visualized as $Q$-criterion isosurfaces ($Q = 2.5\times10^6~\mathrm{s}^{-2}$) colored by velocity magnitude. Joint estimation best recovers the true structures.}
    \label{fig: TBL: Q}
\end{figure}

Real LPT experiments can be subject to large localization errors, especially when single-camera methods are used. \Cref{fig: TBL: high-noise} shows the ground truth flow fields and reconstruction errors at the highest noise level considered, which is representative of high-quality plenoptic and DIH LPT measurements. Baseline reconstructions obtained from finite difference velocities exhibit large errors throughout the probe volume, reflecting the strong blurring of fine-scale structures. Using velocities from filtered tracks improves the reconstruction markedly, while joint estimation yields the lowest errors and most faithfully recovers the highest-frequency turbulent features. High-noise NRMSEs for the $u_3$ velocity are 20.9\%, 15.1\%, and 12.1\% for finite differencing, filtering, and joint estimation, respectively, with corresponding pressure errors of 18.9\%, 13.5\%, and 12.2\%. The same trend is evident in the coherent structures extracted from the reconstructions. Specifically, \cref{fig: TBL: Q} shows that, whereas the baseline methods yield flow fields with varying degrees of smoothing, joint estimation resolves more bona fide structures across a broad range of scales.\par

\begin{figure}[htb!]
    \vspace*{-0.5em}
    \centering
    \includegraphics[width=\linewidth]{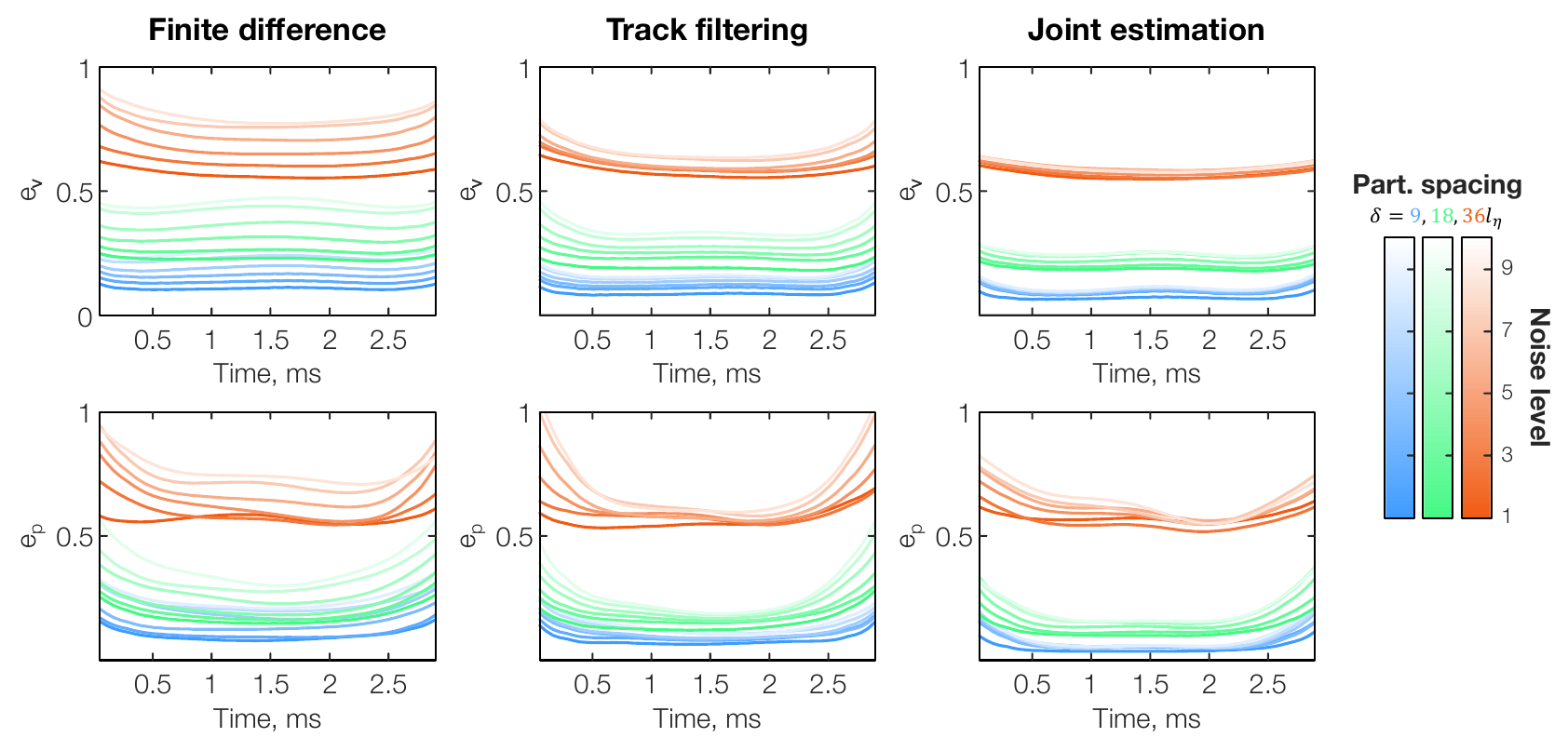}
    \caption{Velocity and pressure errors of the reconstructed flows over the full time window. Colors denote inter-particle spacings of $9\ell_\nu$ (blue), $18\ell_\nu$ (green), and $36\ell_\nu$ (red), with darker shades indicating less noise and brighter shades indicating more noise. Errors increase with both noise level and particle spacing. Finite difference reconstructions are the most sensitive; filtering provides modest noise suppression; and joint flow--particle estimation consistently maintains the lowest errors.}
    \label{fig: TBL: error}
\end{figure}

Estimates shown in \cref{fig: TBL: low-noise,fig: TBL: high-noise,fig: TBL: Q} correspond to the densest seeding case, which is highly favorable, with particles spaced about 10 viscous units apart. To examine how seeding density and noise interact, we applied both the baseline and joint reconstruction methods to the full matrix of TBL cases described in \S~\ref{sec: cases: TBL}. The resulting NRMSEs of the velocity and pressure fields are plotted over time in \cref{fig: TBL: error}. Naturally, both larger inter-particle spacings and higher noise increase the errors for all methods. Baseline reconstructions with velocities computed by finite differencing are especially sensitive. Reconstructions based on filtered velocities benefit from smoothing, but these estimates still degrade substantially as the level of noise increases.\par

A notable feature of the error traces is the high errors near the beginning and end of the assimilation window. They reflect the limited temporal support available for differentiation near the track ends. In contrast, joint estimation keeps the errors within a relatively narrow band across all the seeding densities and noise levels that we tested, consistent with the trends in track error shown in \cref{fig: TBL: tracks error}. The accuracy of joint reconstructions is also fairly uniform over time. Although more advanced optimizers, such as SOAP \citep{Wang2025}, may mitigate this effect, the underlying difficulty is tied to the limited temporal support of noisy, sparse, and truncated tracks near the observation boundaries. More broadly, this boundary behavior suggests that reliable flow reconstruction could require a minimum of temporal support. How such a requirement might depend on the flow time scales, measurement rate, noise level, and imposed physical constraints remains a topic for future study.\par

Taken together, the track and flow results reported in \S\S~\ref{sec: tracer: track} and \ref{sec: tracer: flow} show that incorporating uncertainty information and physics into a joint reconstruction can improve the recovery of particle trajectories as well as flow fields. The governing equations constrain the track geometries; when combined with a prescribed uncertainty model, those geometries permit better localization of the particles, and the resulting kinematics support more accurate flow reconstruction: a virtuous circle! In this demonstration, we have assumed that the localization uncertainty distribution is known. In practice, it may be estimated a priori using model-based methods \citep{Bhattacharya2020} or a posteriori by estimating the high-frequency noise floor of the position spectrum \citep{Buchwald2025}.\par

%%% Inertial particles %%%
\section{Joint estimation of flow fields and inertial particle properties}
\label{sec: inertial}
We next study the joint estimation of flow fields and inertial particle properties from particle tracks. Unlike ideal tracers, inertial particles deviate from local flow motion and exhibit dissipative dynamics \citep{Bec2003}. It is therefore not obvious whether inertial track data contain sufficient information to uniquely identify both flow states and the unknown particle properties that affect slip. Flow reconstruction from inertial tracks amounts to a parameterized PDE-constrained inverse problem, with the Navier--Stokes equations coupled to one parameterized Maxey--Riley equation per particle (here parameterized by particle diameter $d_\mathrm{p}$ and density $\rho_\mathrm{p}$). The framework presented in this text allows us to empirically test whether flow fields and particle properties can be inferred together from the Lagrangian data, alone. We show that such reconstructions are in fact possible. Two representative cases are considered: incompressible turbulence seeded with bidisperse particles that follow the Schiller--Naumann drag law \citep{Schiller1933}, and a supersonic, shock-dominated flow with particle motion described by the compressible Loth drag law \citep{Loth2008}. These examples feature one-way coupling with $St \sim 1$--5 in the former case and nontrivial compressible dynamics with shock–particle interactions in the latter.\par

% HIT with bidisperse particles
\subsection{Homogeneous isotropic turbulence with bidisperse particles}
\label{sec: inertial: HIT}
\begin{figure}[htb]
\vspace*{-0.5em}
    \centering
    \includegraphics[width=0.9\textwidth]{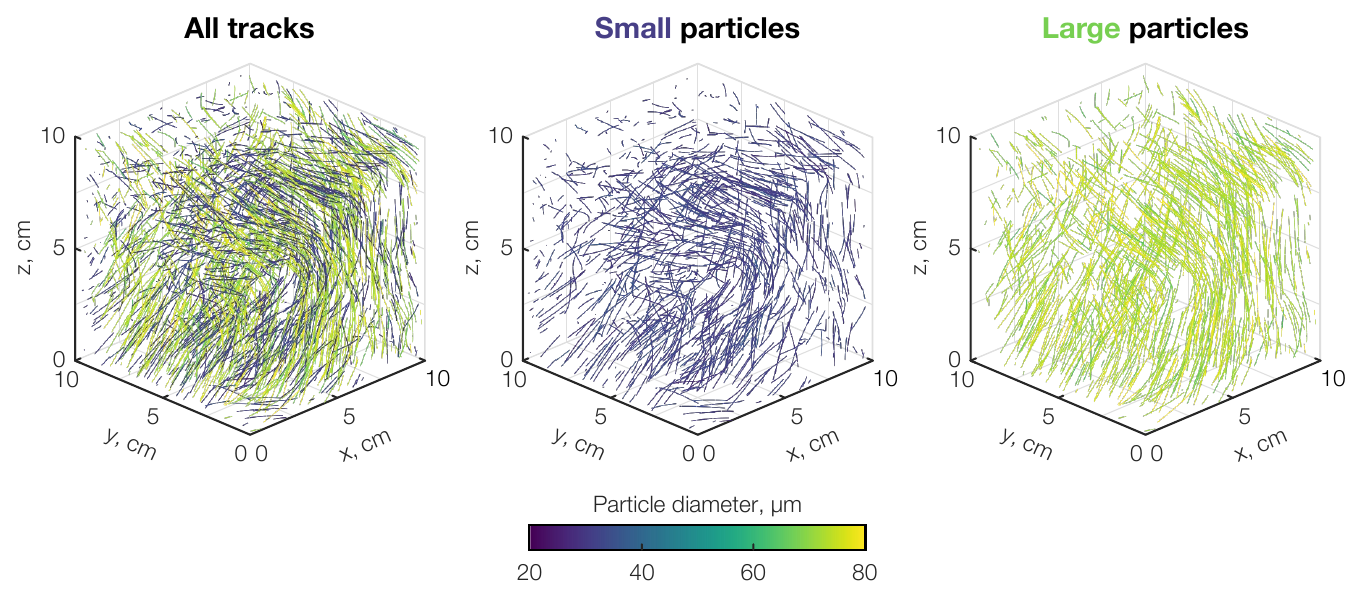}
    \caption{Random selection of bidisperse particle tracks in isotropic turbulence: (left) 3100 tracks; (middle) small particles; (right) large particles. Tracks are colored by particle diameter. Small particles meander in all directions more uniformly, whereas larger particles exhibit pronounced gravitational settling. Tracks are unlabeled in the reconstruction, since $d_\mathrm{p}$ and hence $St$ are unknown.}
    \label{fig: HIT tracks}
\end{figure}

\Cref{fig: HIT tracks} shows a random subset of 3100 inertial tracks in the HIT flow, colored by particle diameter. Tracks from small and large particles are intertwined in a dense cluster, with the middle and right subplots isolating each group. The comparison highlights qualitative differences between $St \sim 1$ and $St \sim 5$ transport. Smaller particles (purple) whirl around in all directions with slight velocity lag, while larger particles (chartreuse) bear the mark of gravitational settling, drifting downward in the negative $z$-direction over time.  The mean and maximum slip velocity normalized by the Kolmogorov velocity scale, i.e., $|\boldsymbol{u}-\boldsymbol{v}|/u_\eta$, are 2.2 and 13.3 for the small particles ($St \sim 1$) and 9.3 and 27.6 for the large particles ($St \sim 5$). Unlike ideal tracers, which remain strongly correlated to the flow, inertial tracks begin to decorrelate at finite $St$ and become effectively uncorrelated at high $St$. As a result, both sets of inertial tracks ``mask'' the underlying flow in distinct ways, and the particles are unlabeled in the reconstruction (i.e., with no knowledge of their diameter available in the optimization). This poses a challenge for reconstructing coupled flow states and particle trajectories.\par

\begin{figure}[htb!]
\vspace*{-0.5em}
    \centering
    \includegraphics[width=\textwidth]{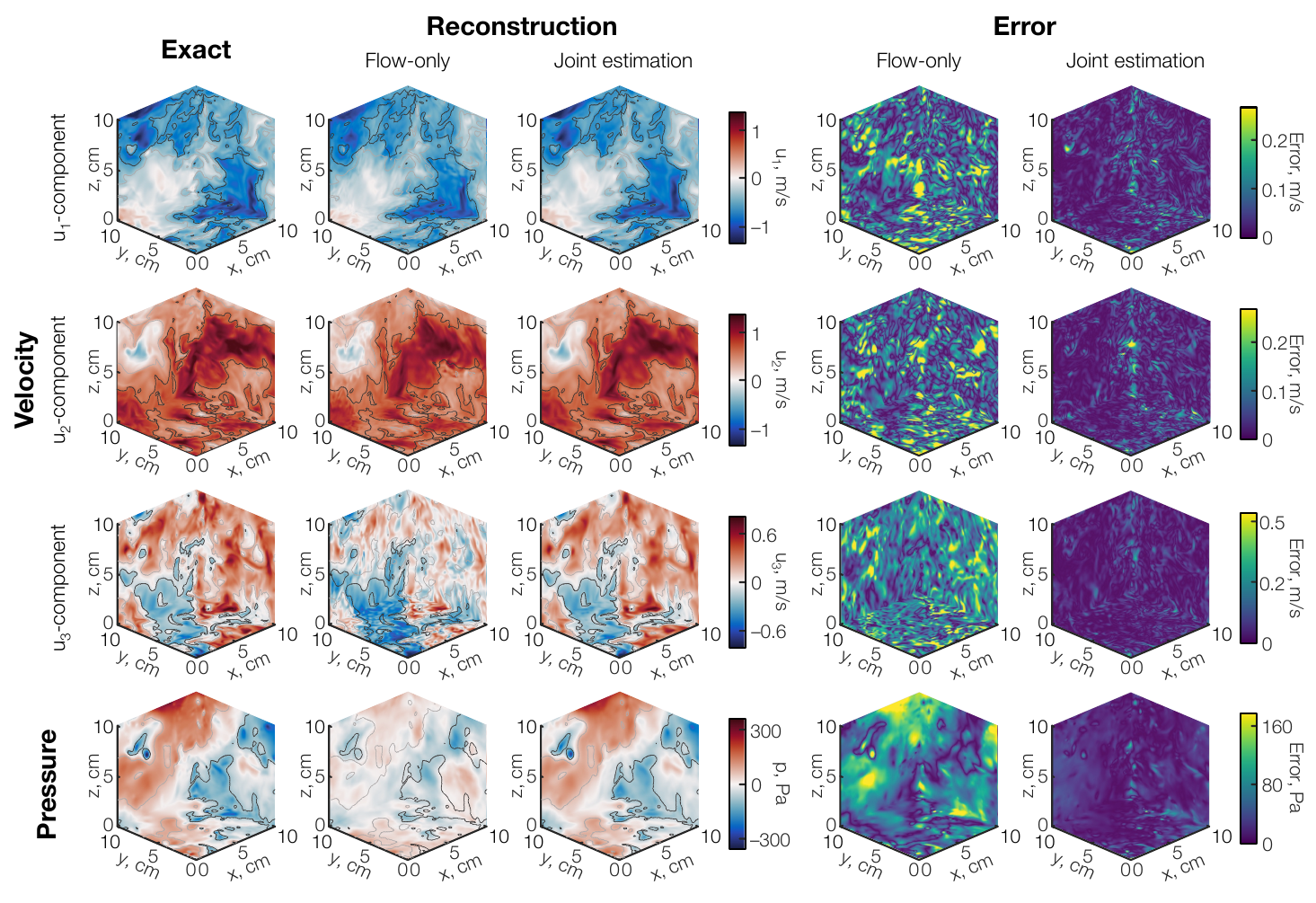}
    \caption{Isotropic turbulent flow fields: (left) exact and reconstructed velocity fields; (right) absolute error fields. Dark and gray contour lines indicate low and high iso-values from the exact fields, overlaid on the reconstructions for comparison. The iso-values are $-636.8$ and $-318.4$~m/s for $u_1$, 318.4 and 636.8~m/s for $u_2$, $-127.3$ and 127.3~m/s for $u_3$, and $-81.1$ and 81.1~Pa for $p$, respectively. Accounting for particle dynamics enables accurate recovery of turbulent flow from inertial particle tracks.}
    \label{fig: HIT reconstructions}
\end{figure}

Inertial tracks from the bidisperse particles are used in both the baseline flow-only reconstruction from \S~\ref{sec: method: loss: baseline} and the joint particle--flow reconstruction. In the baseline case, particles are treated as ideal tracers, such that $\boldsymbol{u} = \boldsymbol{v}$, leading to the velocity-based data loss in \eqref{equ: baseline: data loss}. Particle velocities are obtained from the track data using the B-spline filter. In the joint reconstruction, each particle diameter $d_\mathrm{p}$ is a trainable parameter that determine's the particle's relaxation time $\tau_\mathrm{p}$. For inertial particles, $\tau_\mathrm{p}$ can vary dynamically with the slip velocity through $Re_\mathrm{p}$ \eqref{equ: Reynolds} and $C_\mathrm{D}$ \eqref{equ: Schiller-Naumann}, so the particle dynamics depend on both intrinsic particle properties and the local instantaneous flow velocity. For initialization, $d_\mathrm{p}$ values are drawn from a single Gaussian distribution with mean $52.5~\upmu$m and standard deviation $4~\upmu$m, chosen to have minimal overlap with the true size distributions.\par

\begin{figure}[htb]
\vspace*{-0.5em}
    \centering
    \includegraphics[width=0.9\textwidth]{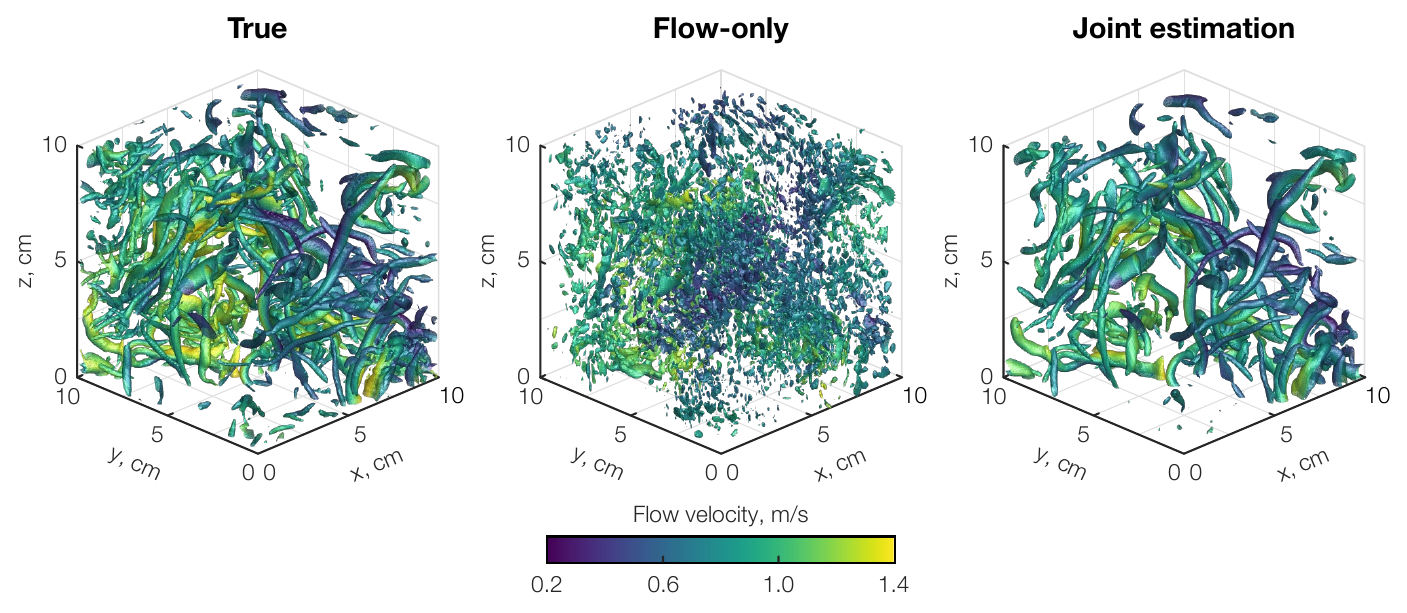}
    \caption{Coherent structures in HIT: (left) exact structures; (middle) flow-only reconstruction under the ideal-tracer assumption ($\boldsymbol{u} = \boldsymbol{v}$); (right) joint estimation with inferred $St$ (through $d_\mathrm{p}$). Structures are shown as $Q$-criterion isosurfaces ($Q = 5000~\mathrm{s}^{-2}$) colored by velocity magnitude. Flow-only reconstructions are dominated by artifacts, whereas joint estimation recovers physically consistent structures with some filtering.}
    \label{fig: HIT Q-criterion}
\end{figure}

\Cref{fig: HIT reconstructions} shows velocity and pressure cut plots from the ground truth DNS and both reconstructions. Cuts are taken at the bottom ($z = 0$~cm), rear ($y = 10$~cm), and right ($x = 10$~cm) faces of the domain, which represent locations of high error due to the lack of boundary conditions in the reconstructions. While there is qualitative agreement between the DNS and flow-only reconstructions in the $u_1$ and $u_2$ components, significant errors appear in $u_3$ and pressure. These $z$-direction errors arise because the flow-only method cannot separate gravitational settling from advection, and inaccurate velocities prevent pressure recovery \citep{Pan2016, Faiella2021, Nie2022}. Even for the \emph{apparently} accurate $u_1$ and $u_2$ fields, the corresponding error fields reveal large deviations. In contrast, the joint reconstructions are highly accurate across the board, as seen in the dark purple (null) error maps. Time-averaged NRMSEs are 4.2\%, 3.6\%, and 11.0\% for the velocity components and 17.2\% for pressure, compared to 18.9\%, 18.3\%, 92.6\%, and 72.3\% for the flow-only method. The reconstructions are further examined through coherent structures identified by isosurfaces of the $Q$-criterion field. \Cref{fig: HIT Q-criterion} shows that flow-only reconstruction produces entirely spurious high-frequency structures, despite the superficial similarity in velocity fields. Meanwhile, joint estimation recovers rich, physically consistent structures, albeit with some noticeable spatial filtering. These results establish the possibility of flow field reconstruction through the veil of inertial particle dynamics.\par

\begin{figure}[htb]
\vspace*{-0.5em}
    \centering
    \includegraphics[width=0.8\textwidth]{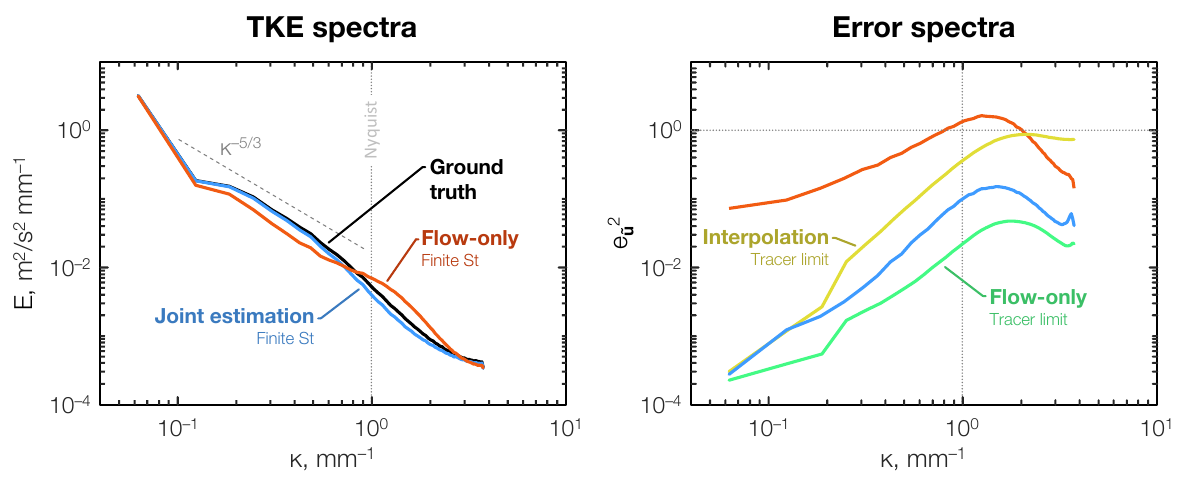}
    \caption{Comparison of TKE spectra. Left: DNS reference (black), flow-only reconstruction (red), and joint reconstruction (blue). Right: normalized error spectra relative to the true energy at each wavenumber, including errors for the flow-only reconstruction (red), joint estimation (blue), and tracer-data baselines under identical conditions: interpolation (yellow) and flow-only reconstruction with tracer data (green). The Nyquist wavenumber $\kappa_\mathrm{Nyq} = 1.007$ is indicated for reference and coincides with the peak error. Neural reconstructions show decaying error beyond $\kappa_\mathrm{Nyq}$ because of the networks' inductive bias. Finite-$St$ data produce larger errors, and the flow-only model shows a spurious peak in the super-Nyquist range because of the $St \to 0$ assumption.}
    \label{fig: HIT spectra}
\end{figure}

The left panel of \cref{fig: HIT spectra} compares the TKE spectra \eqref{equ: spectral average} from flow-only and joint reconstructions to the DNS reference. For context, we compute the Nyquist wavenumber $\kappa_\mathrm{Nyq}$ of particle sampling via \eqref{equ: Nyquist}, which sets the maximum recoverable wavenumber for ideal tracers by interpolation alone. Inertial particles, however, alter this picture. Flow-only reconstructions exhibit abnormal TKE behavior: underestimating energy at low wavenumbers and giving rise to spurious amplification at high wavenumbers, with the low and high wavenumber regions demarcated by $\kappa_\mathrm{Nyq}$. The low-wavenumber deficit occurs because inertial particles respond to turbulent fluctuations with a delay, collectively acting as a low-pass filter on the carrier-phase velocity field \citep{Mei1996}. At high wavenumbers, weakly inertial particles ($St \sim 1$) continue to track the flow, but heavier particles ($St \sim 5$) detach from fluid parcels that are rapidly accelerating, such as in vortices and shear layers \citep{Bewley2013, Vosskuhle2014}. This detachment leads to multi-valued velocities in an Eulerian description, which cannot be reconciled without accounting for slip velocities, thus producing spurious fluctuations at small scales in a flow-only approach.\par

By contrast, joint estimation reproduces the DNS spectra well across the wavenumbers. A slight underestimation of TKE appears before $\kappa_\mathrm{Nyq}$, consistent with the spatial filtering effect seen in \cref{fig: HIT Q-criterion}. This filtering reflects the additional challenge that comes with using inertial tracks, since particle properties (e.g., $d_\mathrm{p}$) must be inferred simultaneously with the flow states. Because inertial particles encode only indirect information about the carrier phase, they seem to sample the flow less efficiently than ideal tracers. In \S~\ref{sec: sensitivity: seeding}, we show that increasing seeding density mitigates this issue.\par

Strictly speaking, matching the true TKE spectrum does not guarantee recovery of the target flow fields. For example, distinct snapshots of statistically stationary HIT may share the same TKE spectrum but differ in spatial structure. We therefore compute normalized velocity error spectra, defined in \eqref{equ: spectral error}, and plot them in the right panel of \cref{fig: HIT spectra}. These spectra quantify reconstruction errors relative to TKE across wavenumbers. For reference, we also include results from ideal tracer tracks that were reconstructed by two conventional methods: adaptive Gaussian windowing, a na{\"i}ve interpolation approach \citep{Agui1987}, and a flow-only reconstruction. As expected, interpolation without physics asymptotes to 100\% error beyond the Nyquist wavenumber, while flow-only reconstruction at the tracer limit achieves the lowest errors, particularly in the super-Nyquist region, demonstrating the ability of DA methods to recover under-resolved dynamics. A systematic study of neural DA's performance under varying seeding densities of \emph{ideal} tracers is presented in appendix~\ref{app: perfect tracers}. By comparison, joint estimation with inertial tracks yields errors lower than na{\"i}ve interpolation but higher than the tracer-based reference, supporting the notion that the information content of inertial tracks that can be leveraged for flow reconstruction is diminished relative to ideal tracers. Finally, flow-only reconstruction with inertial tracks performs worst in the sub-Nyquist region due to inconsistent physical assumptions, but it appears to outperform interpolation at high wavenumbers. This apparent advantage is simply a byproduct of implicit filtering by the network (viz., its low-frequency inductive bias), which coincides with the natural decay of turbulent energy at small scales.\par

\begin{figure}[htb]
\vspace*{-0.5em}
    \centering
    \includegraphics[width=0.9\textwidth]{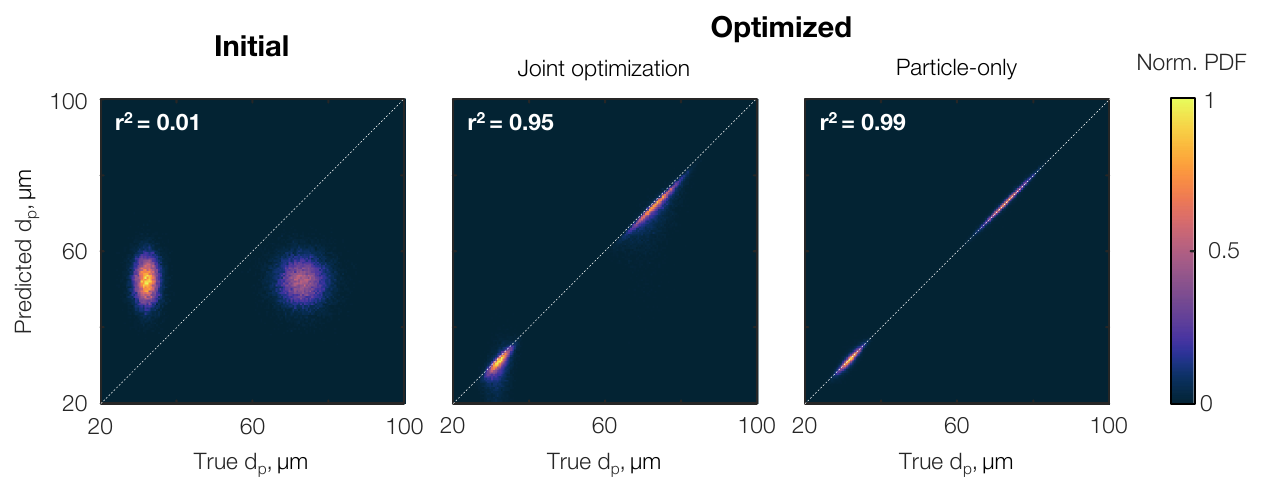}
    \caption{Normalized joint PDFs of estimated and true particle diameters: (left) initialization; (middle) estimates from joint reconstruction; (right) KCT-only estimates computed using the true velocity fields. Joint estimation recovers $d_\mathrm{p}$ accurately, with slight underestimation due to filtered slip velocities; reconstructions based on the exact flow are more accurate and essentially unbiased.}
    \label{fig: joint particle PDFs}
\end{figure}

By training the inertial particle and flow models in tandem, we also recover each particle's diameter. \Cref{fig: joint particle PDFs} shows normalized joint PDFs of the true and inferred values of $d_\mathrm{p}$. The left panel depicts the random initialization, drawn from a Gaussian distribution that is centered between the true size distributions. Following joint estimation, the inferred diameters separate into two clusters (middle panel), raising the Pearson correlation between true and estimated values of $d_\mathrm{p}$ from 0.01 to 0.95. A slight bias toward underestimation is visible, with density bowing ever so slightly below the 45$^\circ$ line. We attribute this to spatial filtering in the reconstructed velocity fields. That is, high-acceleration events are smoothed out, reducing the apparent slip velocities $|\boldsymbol{u}-\boldsymbol{v}|$ and thus lowering the inferred particle sizes. To test this, we pretrained a high-fidelity neural flow model on DNS velcity data and froze it during the particle inference. With the flow known, the estimated diameters are unbiased (right panel), and the correlation rises to 0.99.\par

Before leaving this section, we emphasize that the $d_\mathrm{p}$ classification does not rely on prior knowledge of the flow fields nor on analysis of particle images, cf. \citet{Zhang2008} and \citet{Khalitov2002}. Rather, the results are obtained solely from the tracks and governing physics. When additional sizing information is available, as in \citet{Huang2021} or \citet{delaTorre2023}, it can be incorporated into the estimation of $d_\mathrm{p}$ to further improve accuracy of both the flow states and particle properties. We confirmed this through supplementary tests (not shown), wherein exact values of $d_\mathrm{p}$ were prescribed, which yielded more accurate reconstructions of $\boldsymbol{u}$ and $p$. Statistical priors on $d_\mathrm{p}$ from calibration measurements can be similarly be beneficial.\par

% Cone-cylinder flow
\subsection{Supersonic flow over a cone--cylinder body}
\label{sec: inertial: cone}
We next turn to joint flow--particle estimation in a compressible, shock-dominated flow with inertial particle transport. In high-speed PIV/LPT experiments, seed particles such as \ce{TiO2} or \ce{Al2O4} are subject to agglomeration due to electrostatic forces, leading to variability in the size and density of the aggregates \citep{Williams2015}. These unknown properties, combined with inertial tracks, obscure the flow field. Prior work has shown that PIV measurements of seed traversing an oblique shock wave can be used to calibrate the particle property distributions, yielding values close to manufacturer specifications \citep{Ragni2011, Williams2015}. Such calibration data have been employed to assign a constant particle relaxation time $\tau_\mathrm{p}$, which could then be used to correct the apparent velocity fields from a cross-correlation analysis of the image pairs \citep{Koike2007, Boiko2015}. In practice, however, $\tau_\mathrm{p}$ usually varies between particles, due to size and density differences, and along individual trajectories, due to local changes in temperature and viscosity. These variations introduce uncertainty into particle properties and flow fields. This issue can be addressed via a joint reconstruction with trainable particle properties.\par

\begin{figure}[htb]
    \vspace*{-0.5em}
    \centering
    \includegraphics[width=0.9\linewidth]{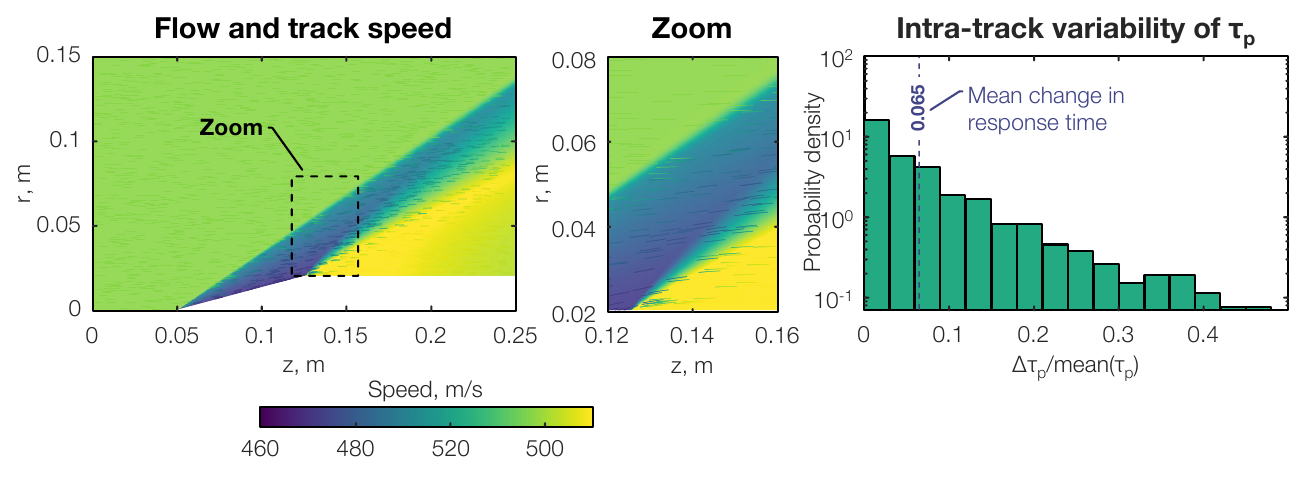}
    \caption{(left) Tracks of agglomerated \ce{TiO2} particles in supersonic flow, colored by particle speed and overlaid on a flow-speed map. Slip is visible in the aft-shock region and expansion fan, where it appears as streaking. (right) PDFs of the normalized intra-track variation of $\tau_\mathrm{p}$, highlighting transient changes in particle response across shocks and expansions.}
    \label{fig: cone tracks}
\end{figure}

\begin{figure}[htb]
\vspace*{-1em}
    \centering
    \includegraphics[width=1\linewidth]{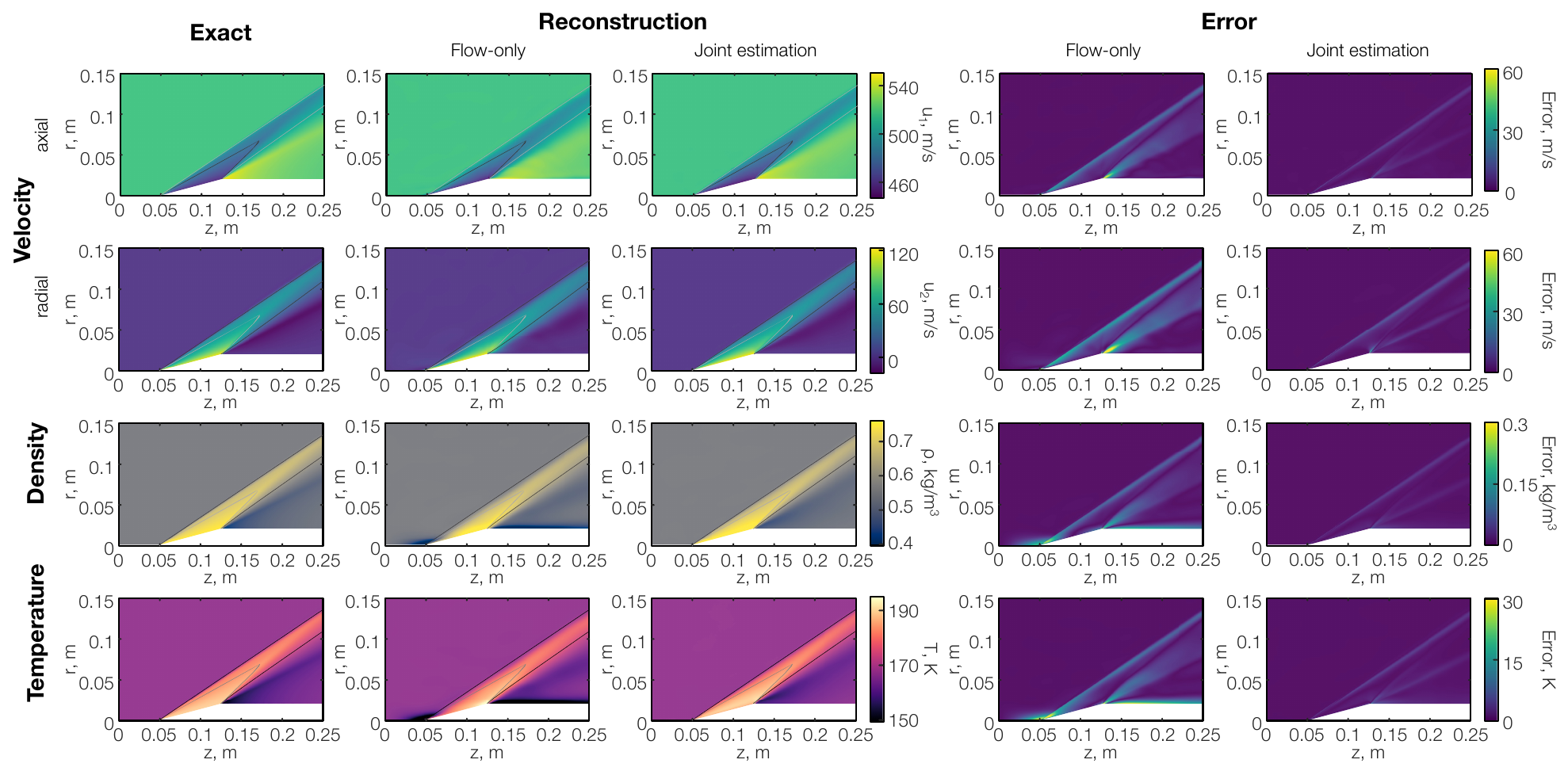}
    \caption{(left) Reconstructed cone--cylinder flow fields compared with the CFD reference. (right) Absolute error fields, showing sharper shocks and fewer artifacts with joint estimation than with the flow-only baseline. Dark and gray contour lines indicate low and high iso-values from the CFD reference, overlaid on the reconstructions for comparison. The iso-values are 480 and 506~m/s for the axial velocity component, 22 and 57~m/s for the radial velocity component, 0.59 and 0.70~kg/m$^3$ for density, and 172 and 183.5~K for temperature, respectively.}
    \label{fig: cone panel}
\end{figure}

The left side of \cref{fig: cone tracks} shows simulated particle tracks colored by the local particle speed, with background shading indicating the flow speed. Regions of slip, especially in the aft-shock and expansion fan regions, are clearly visible, manifesting as streaks in the continuous flow speed field. Steep gradients in carrier-phase viscosity, density, and speed of sound strongly influence the particle dynamics across shocks and expansions, causing the particle response time to vary substantially throughout the domain \citep{Williams2014, Williams2015}. The right side of \cref{fig: cone tracks} presents PDFs of the normalized intra-track range of $\tau_\mathrm{p}$ for particles punching through the shock wave or lurching forward in the expansion fan. Particles upstream of the shock are excluded, as they are initialized without slip and retain a constant $\tau_\mathrm{p}$. On average, $\tau_\mathrm{p}$ varies by 5.5\% along the tracks in this case, with changes up to 40\% across the shock. These variations underscore the transient character of particle relaxation times in supersonic flows, which has been neglected in some correction schemes for high-speed PIV \citep{Koike2007, Boiko2015}.\par

\begin{figure}[htb!]
    \vspace*{-0.5em}
    \centering
    \includegraphics[width=0.7\linewidth]{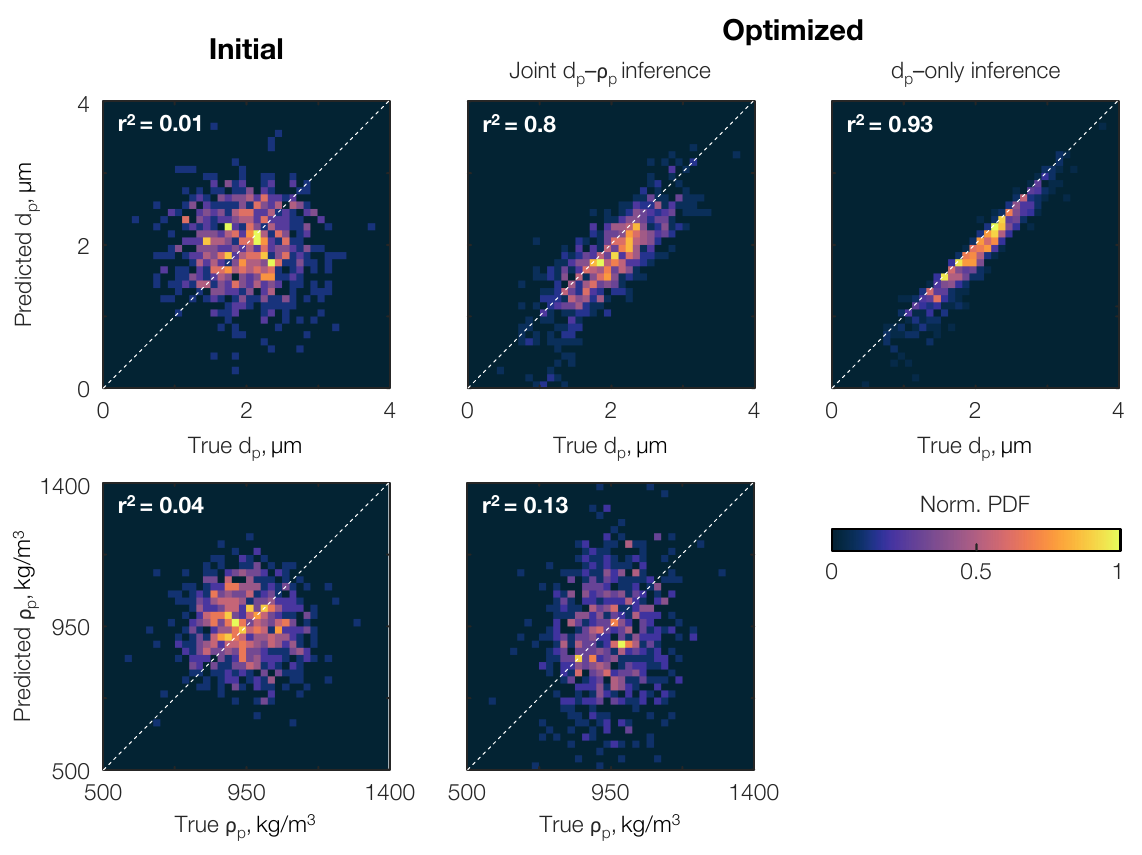}
    \caption{Implicit particle characterization in the cone--cylinder flow. Initial $d_\mathrm{p}$ and $\rho_\mathrm{p}$ values (left), jointly optimized estimates (middle), and $d_\mathrm{p}$-only estimates (right) are compared with the ground truth. As expected, fixing $\rho_\mathrm{p}$ improves the accuracy of the inferred $d_\mathrm{p}$ values.}
    \label{fig: cone PDFs}
\end{figure}

Inertial tracks from the cone--cylinder case are pre-processed for joint estimation. Since all tracks have a uniform length of eight positions, no track splitting is required. Initial particle diameters are sampled from a Gaussian distribution with mean 2~$\upmu$m and standard deviation 0.5~$\upmu$m, while densities are drawn from a Gaussian distribution with mean 950~kg/m$^3$ and standard deviation 100~kg/m$^3$. These distributions provide coarse estimates of particle property statistics, assumed to be available in practice from calibration experiments or manufacturer specifications. \Cref{fig: cone panel} presents reconstructed axial and radial velocity, density, and temperature fields for the cone--cylinder flow. Results are compared against the baseline flow-only reconstruction, which assumes ideal tracers. In the baseline estimates, the shock interface is smeared and density\slash temperature artifacts appear near the surface, consistent with prior observations \citep{Samimy1991, Ragni2011}. The joint estimation more accurately resolves the shock structure. Absolute error maps, presented on the right side of \cref{fig: cone panel}, quantify this improvement: joint estimation achieves NRMSEs of 1\% (axial velocity), 6\% (radial velocity), 3\% (density), and 2\% (temperature), compared to 2\%, 25\%, 18\%, and 22\% for the flow-only mode.\par

\Cref{fig: cone PDFs} shows the inferred particle properties from the joint estimation in the middle column. The top row plots normalized joint PDFs of the estimated and true particle diameters. Although initialized randomly, $d_\mathrm{p}$ is effectively optimized to align with the ground truth, with the correlation coefficient improving from 0.01 to 0.8. As in results from the incompressible HIT case, $d_\mathrm{p}$ tends to be underestimated due to implicit filtering, particularly across the shock. The bottom row shows the joint PDFs of the estimated and true particle densities $\rho_\mathrm{p}$, which are poorly estimated compared to $d_\mathrm{p}$ since $\tau_\mathrm{p}$ depends quadratically on $d_\mathrm{p}$ and linearly on $\rho_\mathrm{p}$, that is: $\tau_\mathrm{p} \sim \rho_\mathrm{p} d_\mathrm{p}^2 / C_\mathrm{D} Re_\mathrm{p}$ per \eqref{equ: response time: general}. We numerically confirmed that the product $C_\mathrm{D} Re_\mathrm{p}$ remains of order 0.01 for this flow, with minimal variation across particles. Importantly, even with divergent $\rho_\mathrm{p}$ estimates, the flow fields are reconstructed with high accuracy, as per \cref{fig: cone panel}, indicating that the flow model is relatively insensitive to uncertainties in particle density at these conditions.\par

The joint reconstruction above assumed both $d_\mathrm{p}$ and $\rho_\mathrm{p}$ to be unknown, representing the most challenging scenario in supersonic PIV\slash LPT. To assess how prior knowledge of particle properties affects the reconstruction, we reduce the degrees of freedom by fixing $\rho_\mathrm{p}$ to its true values and allowing only $d_\mathrm{p}$ to vary. The resulting $d_\mathrm{p}$ estimates are shown on the right side of \cref{fig: cone PDFs}. Compared to the fully unconstrained case, shown in the middle of \cref{fig: cone PDFs}, the alignment between estimated and true $d_\mathrm{p}$ values is substantially improved, with the Pearson correlation coefficient rising from 0.8 to 0.93. This enhanced particle characterization translates directly into better flow reconstruction, lowering the NRMSEs to 0.5\% (axial velocity), 4.8\% (radial velocity), 2.3\% (density), and 1.3\% (temperature). These results underscore the complementary nature of the two phases: even partial knowledge of particle properties (here $\rho_\mathrm{p}$) improves the accuracy of both flow fields and particle properties.\par

%%% Interaction effects %%%
\section{Interaction of noise and inertia in flow--particle reconstruction}
\label{sec: sensitivity}
Section~\ref{sec: inertial} establishes the possibility of jointly estimating flow fields and inertial particle properties, albeit under idealized conditions with dense, noise-free tracks. In real LPT experiments, one must contend with sparse and noisy tracks that could admit many possible solutions. When the number of particles is small, the inversion problem becomes ill-posed: the available data eventually become insufficient to recover both particle properties and flow states. Even when the problem remains formally well-posed, reconstructions may be ill-conditioned in the presence of localization errors or high-$St$ particles, which behave ballistically and interact only weakly with the carrier flow, heightening the sensitivity to noise. In this section, we examine how seeding density, localization uncertainty, and Stokes number interact to govern the feasibility and robustness of joint reconstructions with inertial tracks.\par

% Seeding density
\subsection{Seeding density effects}
\label{sec: sensitivity: seeding}
\begin{figure}[htb]
    \vspace*{-0.5em}
    \centering
    \includegraphics[width=0.8\linewidth]{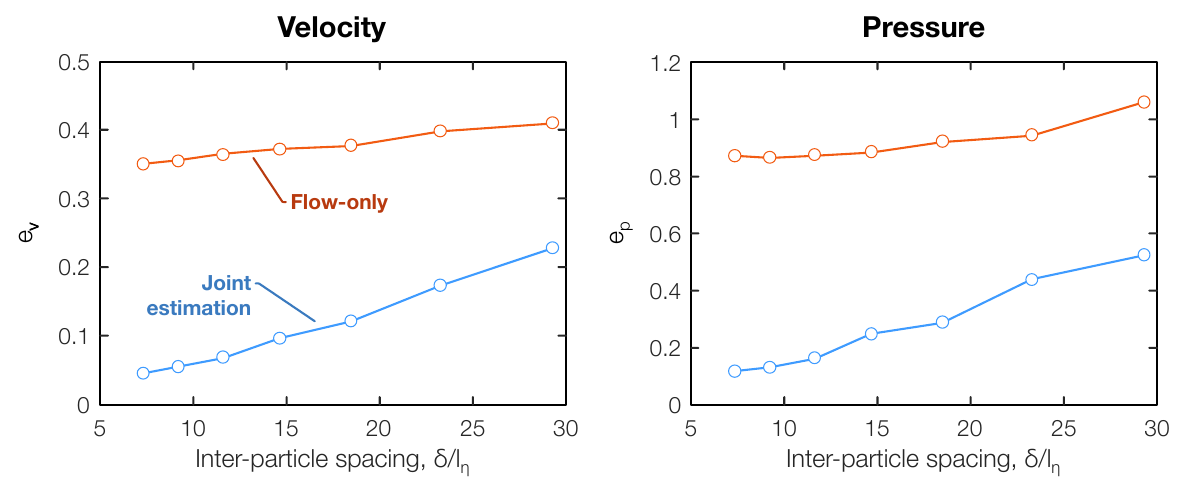}
    \caption{NRMSEs of reconstructed HIT flow fields versus seeding density: (left) velocity; (right) pressure. Flow-only reconstructions are shown in red and joint estimation in blue. Errors from joint estimation decrease as the seeding density increases, indicating a transition from a severely ill-posed regime to a much better constrained one.}
    \label{fig: flow error seeding}
\end{figure}

To start, we examine the influence of particle seeding density. A suite of test cases is generated in a reduced $64^3$ HIT domain, with seeding density varied as described in \S~\ref{sec: cases: sensitivity}. Both the flow-only and joint estimation methods are tested for comparison. \Cref{fig: flow error seeding} reports NRMSEs of the reconstructed velocity and pressure fields as functions of the normalized inter-particle spacing $\delta/\ell_\eta$. Because the flow-only method neglects particle--fluid coupling, reducing the particle spacing does not substantially improve the conditioning of the inverse problem, and the reconstruction remains poor, with velocity and pressure NRMSEs hovering near 40\% and 100\%, respectively. The velocity field is not entirely out of reach, however, since inertial particles ($St = 3$) retain some correlation with large-scale structures. By contrast, joint estimation improves markedly as the particle spacing decreases, with $e_{\boldsymbol{u}}$ and $e_p$ dropping from roughly 20\% and 50\% to 5\% and 10\%, respectively. This major reduction in error indicates a transition from a severely ill-posed regime to a well-posed one; interestingly, the transition is very gradual.\par

\begin{figure}[htb]
    \vspace*{-0.5em}
    \centering
    \includegraphics[width=0.9\linewidth]{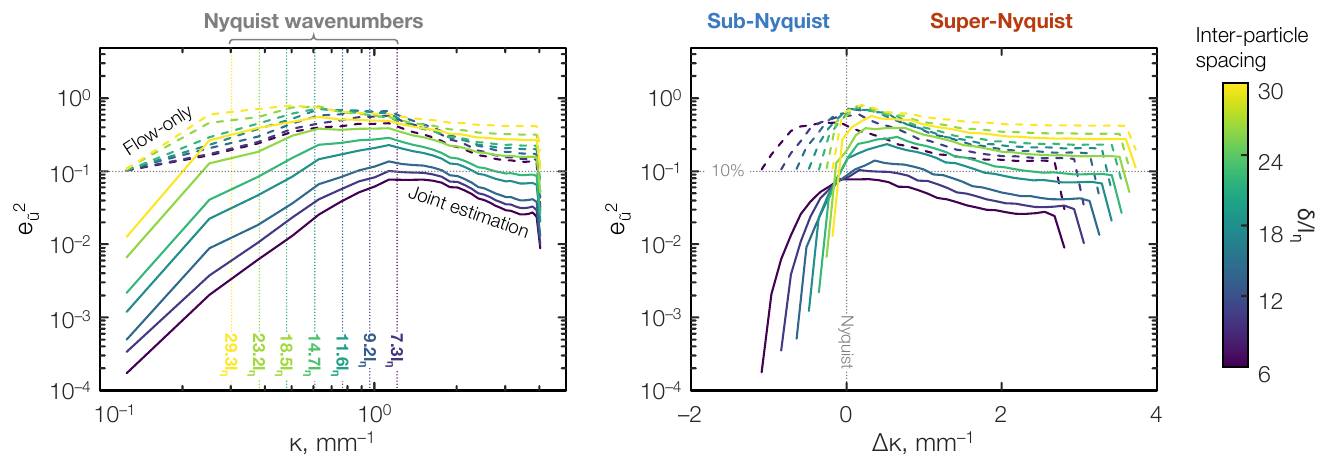}
    \caption{Error spectra of HIT reconstructions under varying inter-particle spacings. Spectra are shown versus wavenumber (left) and normalized by the Nyquist wavenumber (right). Colors denote particle spacing. Dashed lines indicate flow-only reconstructions, and solid lines indicate joint estimation. Errors in joint estimation decrease rapidly as the seeding density increases, whereas flow-only reconstructions retain large errors across all scales.}
    \label{fig: spectra seeding}
\end{figure}

\begin{figure}[htb!]
    \vspace*{-0.5em}
    \centering
    \includegraphics[width=\linewidth]{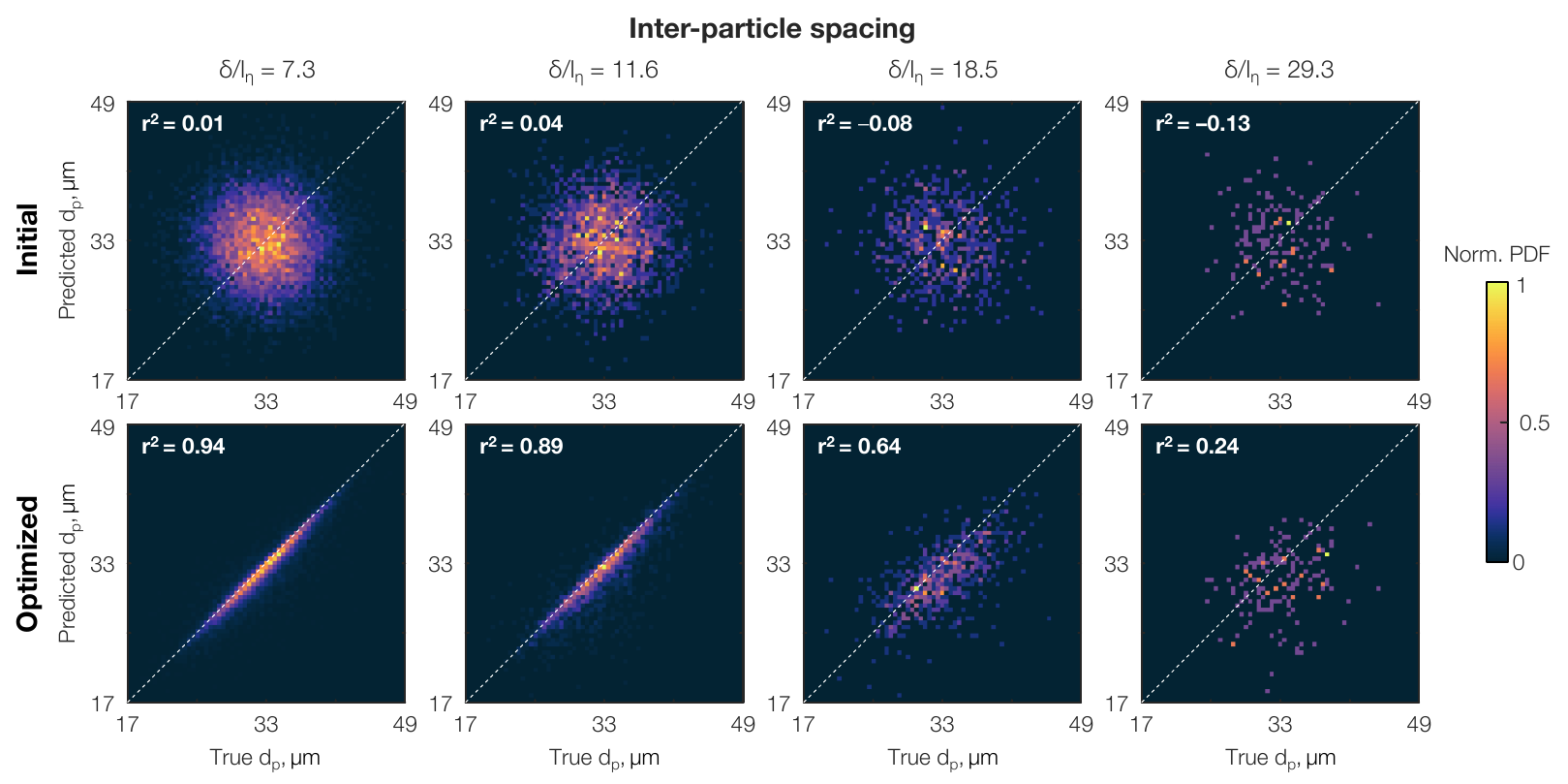}
    \caption{Implicit particle characterization in a reduced HIT domain under varying seeding densities. Initial (top) and optimized (bottom) estimates of $d_\mathrm{p}$ are compared with the ground truth. At sparse seeding, the optimized $d_\mathrm{p}$ values diverge from the truth, consistent with poor flow reconstruction.}
    \label{fig: PDF seeding}
\end{figure}

The left panel of \cref{fig: spectra seeding} shows the normalized spectral errors, per \eqref{equ: spectral error}, for the flow-only and joint reconstructions. Nyquist wavenumbers are plotted to aid interpretation. Errors in the flow-only reconstruction stagnate at a high level, exceeding 10\% across most wavenumbers, whereas the joint estimation method achieves much smaller errors as the particle spacing decreases. To better visualize these trends, the right panel of \cref{fig: spectra seeding} plots the error spectra relative to the corresponding Nyquist wavenumbers and separates the sub- and super-Nyquist regions. In the sub-Nyquist region, joint estimation increasingly outperforms the flow-only approach toward lower wavenumbers, in some cases by up to three orders of magnitude. This highlights an important limitation of the flow-only inversion: even the large-scale structures do not improve much as more particles are added. In the super-Nyquist region, the flow-only reconstruction plateaus at high error, with NRMSEs exceeding 30\%, whereas joint estimation achieves super-resolution at high seeding densities, with errors dropping below 8\% beyond the Nyquist wavenumber in the densest case.\par

\begin{figure}[htb]
    \vspace*{-0.5em}
    \centering
    \includegraphics[width=0.8\linewidth]{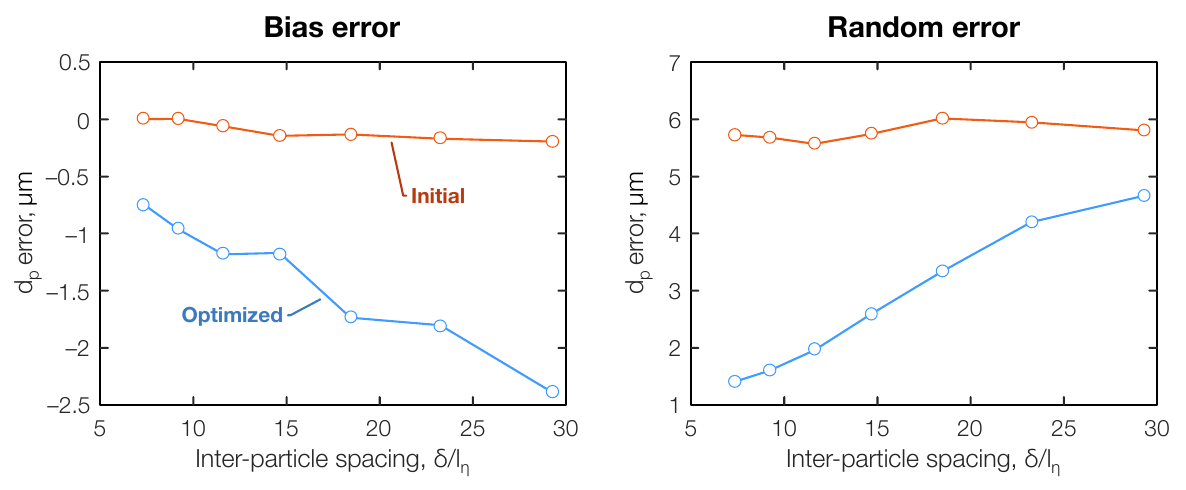}
    \caption{Bias (left) and random (right) errors in $d_\mathrm{p}$ before (red) and after (blue) optimization across seeding densities. Both decrease sharply at high density because the flow reconstruction improves.}
    \label{fig: dp error seeding}
\end{figure}

As before, particle sizes are inferred jointly through the inertial model. \Cref{fig: PDF seeding} shows normalized PDFs of the initial and optimized $d_\mathrm{p}$ values against the ground truth for four representative particle spacings, labeled at the top of each sub-figure. At small spacings, the optimized values of $d_\mathrm{p}$ are closely aligned with the true distribution, whereas at large spacings, the optimization fails to converge, reflecting the poor flow fields recovered at sparse seeding conditions. The correlation coefficient drops from 0.94 to 0.24 between the smallest and largest spacings. As in \cref{fig: joint particle PDFs,fig: cone PDFs}, $d_\mathrm{p}$ is systematically underestimated, consistent with the filtering of slip velocities in the reconstructed flow fields. \Cref{fig: dp error seeding} shows the bias and random error in $d_\mathrm{p}$ before and after optimization, where the bias is the mean error across particles and the random error is its standard deviation. As expected from the random initialization, the initial $d_\mathrm{p}$ values are unbiased, whereas the optimized values develop an increasingly negative bias as the particle spacing grows, again reflecting the degradation of the reconstructed velocity field (see \cref{fig: spectra seeding}). Random error, by contrast, is sharply reduced with denser seeding, falling from about 5~$\upmu$m in the sparsest case to 1.5~$\upmu$m at the highest seeding density.\par

Taken together, these results show that joint reconstructions with inertial particles depends strongly on seeding density. Dense track data provide enough information for the coupled models to accurately recover the carrier flow and particle properties, whereas sparse seeding leaves the inverse problem too poorly constrained for this task, leading to filtered flow fields and degraded particle property estimates.\par

% Noise and Stokes effects
\subsection{Noise and Stokes number effects}
\label{sec: sensitivity: noise}
\begin{figure}[htb]
    \vspace*{-0.5em}
    \centering
    \includegraphics[width=0.9\linewidth]{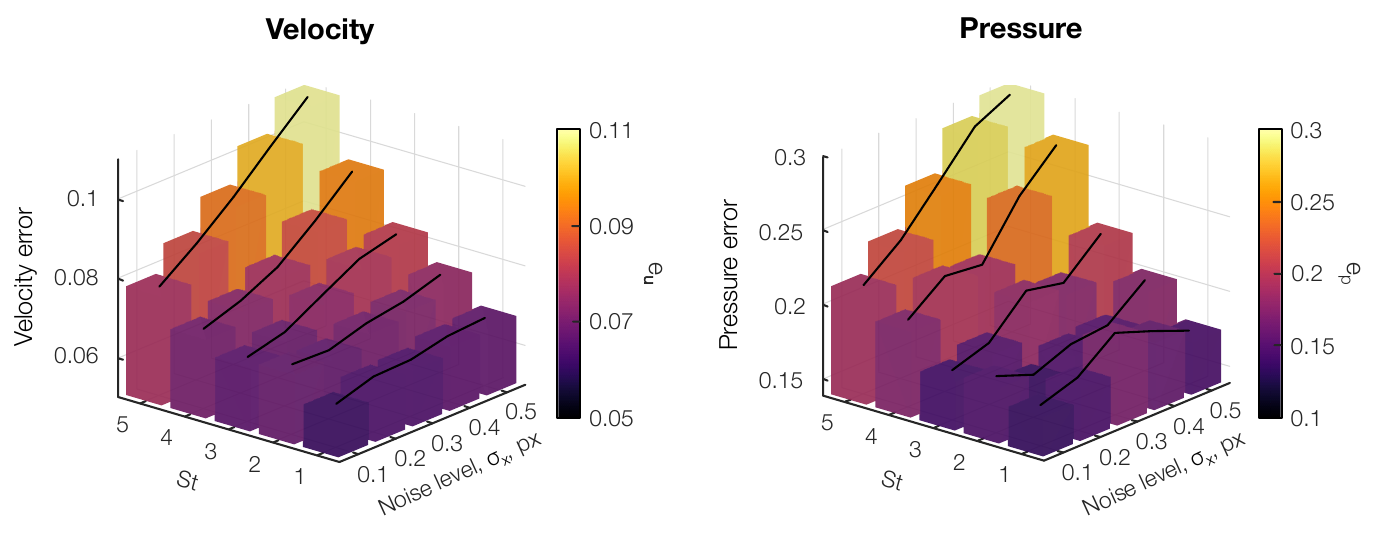}
    \caption{NRMSEs of reconstructed HIT flow fields: (left) velocity; (right) pressure, under varying noise levels and Stokes numbers. Accuracy deteriorates with increasing noise and $St$. In the $St \to \infty$ limit, accurate flow recovery becomes extremely difficult because of strong sensitivity to errors in particle acceleration.}
    \label{fig: St flow error}
\end{figure}

Lastly, we examine how measurement uncertainty and particle inertia interact in joint estimation. Inertial tracks are reconstructed over a range of noise levels and Stokes numbers. The resulting NRMSEs for velocity and pressure are shown in the 3D bar plots of \cref{fig: St flow error}. Both quantities have similar effects on error, which increases with both noise levels and $St$. At low $St$, the reconstruction is \emph{relatively} insensitive to noise, whereas at high $St$, the errors increase rapidly with noise. This behavior is expected. To see why, we rearrange the Maxey--Riley equation \eqref{equ: Maxey-Riley} to isolate the carrier velocity,
\begin{equation}
    \label{equ: up}
    \boldsymbol{u} = \mathopen{}\left(\frac{\mathrm{d} \boldsymbol{v}}{\mathrm{d} t} - \boldsymbol{g} \right) \tau_\mathrm{p} + \boldsymbol{v}.
\end{equation}
Here, terms II--IV of \eqref{equ: Maxey-Riley} are neglected because of the large the density ratio, $\rho_\mathrm{p}/\rho \sim O(10^3)$. In the low-$St$ limit, with $\tau_\mathrm{p} \to 0$, this reduces to $\boldsymbol{u} = \boldsymbol{v}$, so the particle velocity directly tracks the fluid velocity and the reconstruction remains robust, as already seen in \S~\ref{sec: tracer}. At high $St$, however, the acceleration term $(\mathrm{d}\boldsymbol{v}/\mathrm{d}t - \boldsymbol{g})\,\tau_\mathrm{p}$ becomes increasingly important, making $\boldsymbol{u}$ highly sensitive to errors in particle acceleration. Since acceleration estimates degrade rapidly with noise \citep{Berk2024a}, the reconstruction becomes much less reliable. In the ballistic limit, where $\tau_\mathrm{p} \to \infty$, even small acceleration errors can be amplified without bound. In that regime, the tracks carry too little usable imprint of the surrounding flow to permit accurate reconstruction.\par

\begin{figure}[htb]
    \vspace*{-0.5em}
    \centering
    \includegraphics[width=0.6\linewidth]{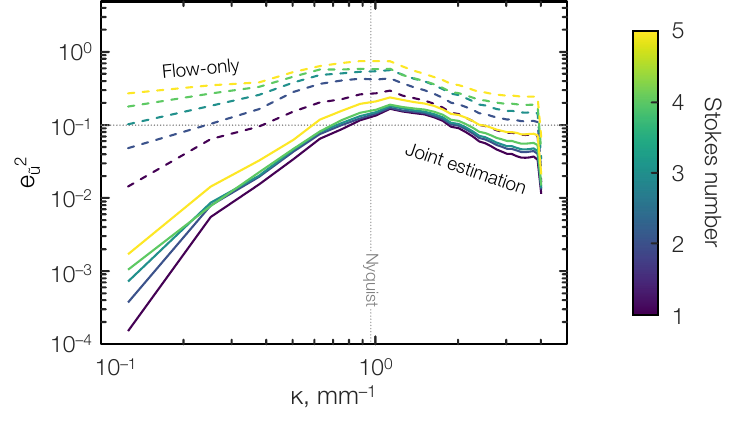}
    \caption{Error spectra of HIT reconstructions under varying Stokes numbers at the medium noise level, $\sigma_x = 0.3$~px. Color denotes $St$. Dashed lines indicate flow-only reconstructions, and solid lines indicate joint estimation. Errors from joint estimation collapse into a relatively narrow band, whereas flow-only reconstructions deteriorate rapidly as $St$ increases.}
    \label{fig: St error spectra}
\end{figure}

To illustrate what information is practically lost from the inertial tracks and what can be recovered by our joint estimation procedure, \cref{fig: St error spectra} shows the normalized error spectra of both the flow-only and joint reconstructions over a range of $St$. A medium noise level, $\sigma_x = 0.3$~px, is used, corresponding to the upper end of STB uncertainty. The particle-sampling Nyquist wavenumber is plotted for reference. As expected, the flow-only reconstruction yields large errors in both the sub- and super-Nyquist regions, and these increase sharply with $St$. This arises mainly because the particle velocity field becomes multivalued at finite $St$, which cannot be reconciled without explicitly accounting for slip velocities. In contrast, the error spectra from joint estimation collapse into a much narrower band and show only a weak dependence on $St$. The residual spread across $St$ reflects the noise sensitivity described above: reconstructions based on higher-$St$ particles are more vulnerable to acceleration errors through \eqref{equ: up}. Quantitatively, the cut-off wavenumbers of the flow-only reconstruction at $e_{\widetilde{\boldsymbol{u}}}^2 = 0.1$ are 0.37~mm$^{-1}$ and 0.24~mm$^{-1}$ for $St = 1$ and 2, respectively; beyond $St = 2$, the spectral errors are uniformly large. Conversely, joint estimation yields cut-off wavenumbers near 0.75~mm$^{-1}$ across all cases, showing that it can recover substantially smaller flow scales than are available under an erroneous assumption of tracer behavior.\par

\begin{figure}[htb]
    \vspace*{-0.5em}
    \centering
    \includegraphics[width=0.9\linewidth]{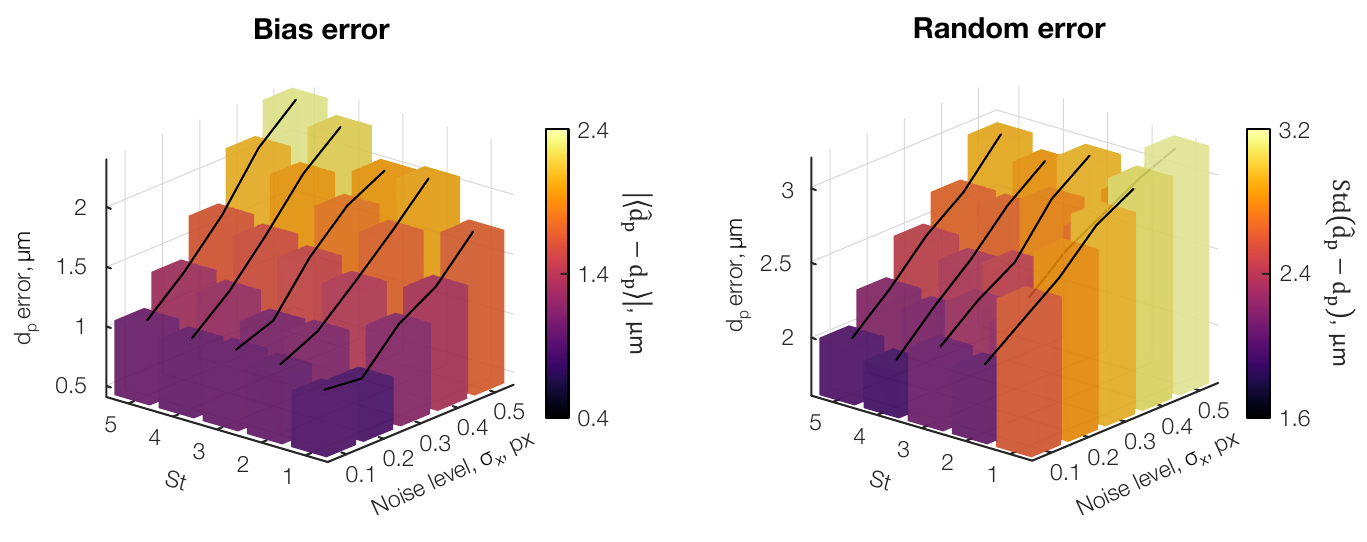}
    \caption{Absolute bias (left) and random (right) errors of the inferred particle diameters under varying noise levels and Stokes numbers. The bias increases with both noise and $St$ because the reconstructed velocity fields become more strongly filtered. The random error decreases with increasing $St$ because the physics loss becomes more sensitive to $d_\mathrm{p}$. In the tracer limit, $d_\mathrm{p}$ is nearly unidentifiable.}
    \label{fig: St dp error}
\end{figure}

\Cref{fig: St dp error} shows the absolute bias and random errors in the inferred particle diameters from joint reconstruction. The bias increases with both noise level and Stokes number, mirroring the trends in the velocity and pressure errors in \cref{fig: St flow error}. This likely reflects stronger filtering of the reconstructed flow field at larger noise and $St$, which then propagates into the inferred particle sizes. The random error, however, shows a more nuanced trend. As expected, it grows with noise, but it decreases as $St$ increases. Thus, particles with some degree of inertia make the diameter estimate more repeatable, even though the associated flow reconstruction worsens. This behavior is consistent with the particle physics loss \eqref{equ: particle physics loss}, which becomes more sensitive to $d_\mathrm{p}$ at larger Stokes number. During training, $d_\mathrm{p}$ is optimized together with the flow model to estimate $\tau_\mathrm{p}$, i.e., through \eqref{equ: Reynolds}--\eqref{equ: Schiller-Naumann}, such that the residuals of \eqref{equ: up} are minimized. According to \eqref{equ: response time: general} and \eqref{equ: Schiller-Naumann}, $\tau_\mathrm{p}$ scales as $\tau_\mathrm{p} \sim \rho_\mathrm{p} d_\mathrm{p}^{1.313}$, so the sensitivity of $\tau_\mathrm{p}$ to $d_\mathrm{p}$ increases with particle inertia. In the tracer limit, by contrast, perturbations in $d_\mathrm{p}$ have little effect on the dynamics, such the true particle size cannot be determined from track data. It should be noted that this lack of information about $d_\mathrm{p}$ does not compromise recovery of the flow field, however, since near-tracer particles closely track the carrier phase.\par

Several practical (and largely intuitive) guidelines follow from these results. First, seeding density should be increased whenever possible, provided that one-way coupling remains valid and that particle tracking is not degraded by overlapping particle images. Second, low-inertia particles are generally preferable for accurate flow reconstruction. When only high-inertia particles are available, high-resolution LPT systems become especially important because they reduce noise and improve acceleration estimates during joint reconstruction. Third, when particle properties are also of interest, lower measurement noise reduces both the bias and the random error in the inferred quantities. High seeding densities are again beneficial because they reduces implicit filtering of the flow and thereby limit bias in the recovered particle properties.\par

%%% Conclusions %%%
\section{Concluding remarks}
\label{sec: conclusions}
This work numerically investigates the joint estimation of flow fields and particle properties using Lagrangian track data. The problem is motivated by two central challenges in LPT experiments: noisy particle tracks and inertial transport effects. Existing DA algorithms rely on error-prone velocity estimates derived from fitted tracks and assume ideal tracer particles with zero slip, both of which can compromise the fidelity of Eulerian reconstructions. To overcome these limitations, we propose to reconstruct the flow and particle states jointly from track data, under the combined constraints of disperse multiphase flow physics and known\slash estimated localization uncertainties. The resulting framework, termed NIPA, is built on a PINN architecture that couples a neural flow model with a set of kinematics-constrained track models, with one for each particle. Joint training of these models yields flow fields and particle properties that at once satisfy governing physics and match the LPT data.\par

We test NIPA across a range of particle-laden flows: incompressible turbulence with ideal tracers in \S~\ref{sec: tracer}, incompressible turbulence with inertial particles in \S~\ref{sec: inertial: HIT}, and compressible, shock-dominated flow with inertial particles in \S~\ref{sec: inertial: cone}. For each case, we ask whether the available track data, taken together with the governing physics, suffices to recover both flow states and unknown particle properties. We also examine how seeding density, noise magnitude, and Stokes number influence the robustness of reconstructions in \S~\ref{sec: sensitivity}. From these studies, four main conclusions emerge, as summarized below.\par

\begin{enumerate}[topsep=.5ex, itemsep=-.5ex, partopsep=.5ex, parsep=.5ex]
    \item For noisy tracks of ideal tracers, joint estimation can recover both the flow field and the underlying particle trajectories in the TBL case, provided a sufficient number of tracks are available (\S~\ref{sec: tracer}). Across a broad range of seeding densities, jointly estimated tracks and flow fields are more accurate than filtered tracks or baseline flow-only reconstructions, highlighting the value of coupled flow--particle learning. As seeding density decreases, accuracy degrades, and conventional filtering may become preferable once the Eulerian field can no longer be inferred reliably from the available tracks. At large localization uncertainties, however, joint estimation remains robust and outperforms conventional filtering by up to 50\%, owing to the integration of physics-informed tracking with the flow reconstruction.
    
    \item Inertial particle properties can be inferred jointly with incompressible (\S~\ref{sec: inertial: HIT}) and compressible (\S~\ref{sec: inertial: cone}) flow fields. Reconstructions that account for particle--fluid interactions, e.g., through the Maxey--Riley equation, recover the flow accurately from inertial tracks. Particle properties such as their diameter can be inferred from the tracks and governing physics. In the compressible case, particle densities do not reliably converge, reflecting the weak sensitivity of the reconstruction to $\rho_\mathrm{p}$, although this does not noticeably degrade the flow field. When particle properties are known and supplied during training, the flow reconstruction improves substantially.
    
    \item In inertial-particle cases, reconstruction quality depends strongly on the information content of the track data (\S~\ref{sec: sensitivity: seeding}). Dense seeding enables scale-resolving reconstructions, in some cases beyond the particle Nyquist wavenumber, whereas sparse seeding makes the inversion severely ill-posed and yields heavily filtered flow fields that miss much of the high-wavenumber content. These filtered fields in turn degrade the particle inference, leading to systematic underestimation of particle diameters and increased random errors.
    
    \item Both flow reconstruction and particle inference depend strongly on localization uncertainties and Stokes numbers (\S~\ref{sec: sensitivity: noise}). As expected, reconstruction accuracy declines with increasing noise and $St$. At high $St$, ballistic particle motion amplifies even small acceleration errors through the Maxey--Riley dynamics, making accurate flow recovery increasingly difficult. Bias in the inferred particle size also grows with noise and $St$ because the flow field becomes more strongly filtered. At the same time, random errors in particle size decrease with increasing $St$, since higher-inertia particles make the physics loss more sensitive to changes in $d_\mathrm{p}$. In the tracer limit, particle size is effectively unidentifiable, though this does not hinder recovery of the flow field.
\end{enumerate}

Lastly, we note that the ideas developed here are not tied specifically to NIPA. The central point is that Lagrangian information can be incorporated directly into a coupled, physics-constrained flow--particle reconstruction. Similar ideas could be pursued with other DA frameworks, including adjoint--variational approaches \citep{Gutierrez2026, Ke2026}. Continued advances in solvers, track models, and optimization methods may further improve performance, especially in the inertial regime, thereby broadening the range of LPT conditions under which such reconstructions are useful.\par

% Appendices
\appendix

%%% Flow Equations %%%
\section{Carrier-phase governing equations}
\label{app: carrier}
Lagrangian particle tracking experiments involve disperse multiphase flows, wherein tracer particles constitute the disperse phase and the fluid of interest is the carrier phase. Depending on the particle mass loading and volume fraction, particle--fluid interactions are modeled using one-way, two-way, or four-way coupling schemes \citep{Subramaniam2022}. Analysis in this work applies to the one-way coupled regime, where momentum transfer from particles to the carrier flow is negligible. This assumption is appropriate for most PIV and LPT experiments, which typically operate in the dilute limit. Governing equations for flows in the test cases introduced in \S~\ref{sec: cases} are summarized below.\par

% Incompressible
\subsection{Equations for unsteady 3D incompressible flow}
\label{app: carrier: incompressible}
Flows in the TBL (\S~\ref{sec: cases: TBL}) and HIT (\S~\ref{sec: cases: HIT}) cases are governed by the 3D continuity and momentum equations for incompressible flow,
\begin{subequations}
    \label{equ: incompressible}
    \begin{align}
        \boldsymbol{\nabla} \boldsymbol{\cdot} \boldsymbol{u} &= 0, \label{equ: incompressible:continuity}\\
        \frac{\partial \boldsymbol{u}}{\partial t} +  \boldsymbol{u} \boldsymbol{\cdot} \boldsymbol{\nabla} \boldsymbol{u} &= -\frac{1}{\rho}\boldsymbol{\nabla} p + \nu\nabla^2\boldsymbol{u} + \boldsymbol{F}, \label{equ: incompressible:momentum}
    \end{align}
\end{subequations}
where $\boldsymbol{u}$ is the 3D velocity vector and $\boldsymbol{\nabla}$ denotes the del operator in Cartesian coordinates. For isotropic turbulence, a forcing term is introduced to sustain stationary turbulence \citep{Rosales2005},
\begin{equation}
    \boldsymbol{F} = \frac{\varepsilon}{3 \,u^{2}_\mathrm{rms}} \boldsymbol{u},
    \label{equ: incompressible:forcing}
\end{equation}
with $\varepsilon$ being the mean energy dissipation rate and $u_\mathrm{rms}$ the root-mean-square velocity. For the TBL case, we set $\boldsymbol{F} = \mathbf{0}$. In both the TBL and HIT cases, the residual vector $\boldsymbol{e}_\mathrm{f}$ in \eqref{equ: flow physics loss} is formed from the components of \eqref{equ: incompressible}.\par

% Compressible
\subsection{Equations for steady axisymmetric compressible flow}
\label{app: carrier: compressible}
The cone–cylinder flow (\S~\ref{sec: cases: cone}) is governed by the steady, axisymmetric compressible Navier--Stokes equations for the conservation of mass, momentum, and energy,
\begin{subequations}
    \label{equ: compressible}
    \begin{align}
        \boldsymbol{\nabla} \boldsymbol{\cdot} \left(\rho \,\boldsymbol{u}\right) &= 0, \label{equ: compressible:continuity} \\
        \boldsymbol{\nabla} \boldsymbol{\cdot} \left(\rho \,\boldsymbol{u} \,\boldsymbol{u}^\top\right) &= -\boldsymbol{\nabla} p +
        \boldsymbol{\nabla} \boldsymbol{\cdot} \left[\mu \left(\boldsymbol{\nabla}\boldsymbol{u} + \boldsymbol{\nabla}\boldsymbol{u}^\top \right) -
        \frac{2}{3} \mu \left(\boldsymbol{\nabla} \boldsymbol{\cdot} \boldsymbol{u}\right) \mathsfbi{I}\right], \label{equ: compressible:momentum} \\
        \boldsymbol{\nabla} \boldsymbol{\cdot} \left[\left(\rho E + p\right)\boldsymbol{u}\right] &= \boldsymbol{\nabla} \boldsymbol{\cdot} \left(k \boldsymbol{\nabla} T\right) +
        \boldsymbol{\nabla} \boldsymbol{\cdot} \left\{\left[\mu \left(\boldsymbol{\nabla}\boldsymbol{u} + \boldsymbol{\nabla}\boldsymbol{u}^\top \right) -
        \frac{2}{3} \mu \left(\boldsymbol{\nabla} \boldsymbol{\cdot} \boldsymbol{u}\right) \mathsfbi{I}\right] \boldsymbol{\cdot} \boldsymbol{u}\right\}. \label{equ: compressible:energy}
    \end{align}
\end{subequations}
In these expressions, $\boldsymbol{u}$ denotes the velocity vector with axial and radial components and $\boldsymbol{\nabla}$ is the two-dimensional (2D) del operator in cylindrical coordinates (axial--radial). The thermodynamic variables are the density $\rho$, pressure $p$, and temperature $T$, while $E$ is the specific total energy and $k$ the thermal conductivity of the carrier phase. The temperature $T$ is obtained from the local total energy and velocity magnitude, and transport properties $\mu$ and $k$ are evaluated via Sutherland's law \citep{Anderson1990}.\par

Equation~\eqref{equ: compressible} comprises four governing equations with five unknowns and must therefore be closed with an equation of state. We adopt the calorically perfect gas law,
\begin{equation}
    p = \left(\gamma - 1\right) \rho \underbrace{\left(E - \frac{1}{2} \boldsymbol{u} \boldsymbol{\cdot} \boldsymbol{u}\right)}_{C_\mathrm{v} T},
\end{equation}
where $\gamma = C_\mathrm{p}/C_\mathrm{v}$ is the ratio of specific heat at constant pressure $C_\mathrm{p}$ to that at constant volume $C_\mathrm{v}$. For the cone--cylinder test case, the residual vector $\boldsymbol{e}_\mathrm{f}$ in \eqref{equ: flow physics loss} collects the contributions from each of the conservation laws in \eqref{equ: compressible}.\par

%%% Particle Equations %%%
\section{Disperse-phase governing equations}
\label{app: disperse}
This appendix summarizes the particle dynamics models employed for the bidispersed HIT case (\S~\ref{sec: cases: HIT}) and the supersonic cone--cylinder flow case (\S~\ref{sec: cases: cone}). We begin with the full Maxey--Riley equation, which governs the motion of small spherical particles in incompressible fluids. We then outline its modification for compressible flows, relevant to tracer particle dynamics in high-speed PIV\slash LPT applications. Next, we introduce key dimensionless numbers that quantify the relative importance of viscous, compressibility, and rarefaction effects on particle motion. Finally, we summarize a drag law suitable for supersonic conditions, which incorporates corrections for these effects.\par

% MR equation
\subsection{Maxey--Riley equation}
\label{app: disperse: MR}
Small spherical particles moving in a locally uniform flow are subject to both inertial and viscous forces. The inertial transport regime is commonly characterized by the particle Reynolds number, defined in terms of a characteristic particle length (diameter $d_\mathrm{p}$), slip velocity, and fluid density and viscosity,
\begin{equation}
    Re_\mathrm{p} = \frac{\rho d_\mathrm{p} \overbrace{\left|\boldsymbol{u} - \boldsymbol{v}\right|}^\text{slip}}{\mu}.
    \label{equ: Reynolds}
\end{equation}
Here, ``slip'' refers to the ballistic motion of the particle relative to the carrier fluid. A related measure is the particle response time, which quantifies the time scale at which a particle relaxes toward the local fluid velocity,
\begin{equation}
    \tau_\mathrm{p} = \frac{4}{3 C_\mathrm{D} Re_\mathrm{p}} \frac{\rho_\mathrm{p} d_\mathrm{p}^2}{\mu} = \frac{4}{3 C_\mathrm{D}} 
    \frac{\rho_\mathrm{p}}{\rho}
    \frac{d_\mathrm{p}}{\left|\boldsymbol{u} - \boldsymbol{v}\right|},
    \label{equ: response time: general}
\end{equation}
where $C_\mathrm{D}$ is the drag coefficient. In the creeping-flow limit, where $Re_\mathrm{p} \ll 1$, Stokes' law applies, with $C_\mathrm{D} = 24/Re_\mathrm{p}$. For finite $Re_\mathrm{p}$, however, inertial effects necessitate modification of the drag law. For our HIT case, we adopt the Schiller--Naumann correlation,
\begin{equation}
    C_\mathrm{D} = \frac{24}{Re_\mathrm{p}} \mathopen{}\left(1+0.15Re_\mathrm{p}^{0.687}\right), \quad Re_\mathrm{p} < 800, 
    \label{equ: Schiller-Naumann}
\end{equation}
which has been validated over a broad range of $Re_\mathrm{p}$. More general drag laws applicable at higher $Re_\mathrm{p}$ are reviewed in Chapter~8 of Subramaniam and Balachandar \citep{Subramaniam2022} and can be incorporated into the DA framework where applicable.\par

When particles are much smaller than the relevant hydrodynamic length scale (i.e., quasi-point particles), their dynamics are governed by the version of the Maxey--Riley equation \citep{Maxey1997} modified by \citet{Mei1996},
\begin{align}
    \frac{\mathrm{d} \boldsymbol{v}}{\mathrm{d} t} =
    \underbrace{\frac{\boldsymbol{u} - \boldsymbol{v}}{\tau_\mathrm{p}}}_\text{I} +
    \underbrace{\frac{\rho}{\rho_\mathrm{p}} \frac{ \mathrm{D}\boldsymbol{u}}{\mathrm{D}t}}_\text{II} +
    \underbrace{\frac{1}{2} \frac{\rho}{\rho_\mathrm{p}} \left(\frac{ \mathrm{D}\boldsymbol{u}}{\mathrm{D}t} - \frac{\mathrm{d}\boldsymbol{v}}{\mathrm{d}t} \right)}_\text{III} +
    \underbrace{\sqrt{\frac{9}{2\pi} \frac{\rho}{\rho_\mathrm{p} \tau_\mathrm{p}}} \int_{-\infty}^t \frac{1}{\sqrt{t-\tau}} \frac{\mathrm{d} (\boldsymbol{u} - \boldsymbol{v})}{\mathrm{d} \tau} \mathrm{d}\tau}_\text{IV} +
    \underbrace{\boldsymbol{g}}_\text{V},
    \label{equ: Maxey-Riley}
\end{align}
where $\mathrm{D}/\mathrm{D}t$ and $\mathrm{d}/\mathrm{d}t$ denote total derivatives following a fluid parcel and a particle, respectively. The five terms on the right-hand side correspond to: (I)~quasi-steady drag, (II)~pressure gradient force, (III)~added mass effect, (IV)~Basset history term (unsteady vorticity diffusion), and (V)~gravitational force. Since the carrier-phase momentum equation, i.e., \eqref{equ: incompressible:momentum}, neglects gravity, buoyancy is not included in \eqref{equ: Maxey-Riley}. If gravity were retained, the last term would instead appear as $(1 - \rho/\rho_\mathrm{p}) \boldsymbol{g}$, consistent with the formulation of \citet{Mei1996}. The relative magnitude of each contribution depends on flow conditions and particle properties, e.g., density, diameter \citep{Thomas1992, Ling2013}.\par

Two simplifications are made based on the properties of the small, dense particles assumed in our simulations. First, owing to the large particle-to-fluid density ratio, $\rho_\mathrm{p}/\rho \sim O(10^3)$, we neglect the Basset history force in the forward simulations \citep{Eaton2009, Ling2013}. In the reconstructions, we additionally omit the pressure gradient and added mass forces, which are indeed included in the forward simulations, since their magnitudes are roughly three orders of magnitude smaller than Stokes drag. This deliberate mismatch between the forward and reconstruction force models highlights the robustness of NIPA to imperfect particle dynamics, which is important in practice. Second, given the minute particle size, we neglect finite-size corrections such as the Fax{\'e}n term and Saffman lift \citep{Maxey1997}. The Kolmogorov length scale in \S~\ref{sec: cases: HIT} is about 350~$\upmu$m: substantially larger than the maximum particle diameter of ${\sim}70$~$\upmu$m, thereby justifying the quasi-point-particle assumption. Consequently, for the HIT case, the residual vector $\sdx{\boldsymbol{e}_\mathrm{p}}[k]$ in \eqref{equ: particle physics loss} includes only the contributions from Stokes drag and gravity, i.e., terms (I) and (V) in \eqref{equ: Maxey-Riley}, for the $k$th particle.\par

% Particle dynamics in compressible flows
\subsection{Particle dynamics in compressible flows}
\label{app: disperse: compressible}
Tracer particles in high-speed flow are often modeled as solid spheres immersed in an unbounded fluid and subject only to quasi-steady drag \citep{Williams2015}. For typical tracers in PIV or LPT, i.e., very small ($d_\mathrm{p} \sim 1$~$\upmu$m) and with large density ratios ($\rho_\mathrm{p}/\rho \gg 1$), the contributions of pressure gradient, added mass, Basset history, and body forces are negligible in high-speed conditions \citep{Melling1997, Ragni2011}. Under these assumptions, the Maxey--Riley equation reduces to a balance between slip velocity and quasi-steady drag,
\begin{equation}
    \label{equ: particle motion}
    \frac{\mathrm{d} \boldsymbol{v}}{\mathrm{d} t} = \frac{\boldsymbol{u} - \boldsymbol{v}}{\tau_\mathrm{p}},
\end{equation}
where $\tau_\mathrm{p}$ is the particle response time given in \eqref{equ: response time: general}. In our supersonic cone flow case, $\boldsymbol{u}$ and $\boldsymbol{v}$ are 2D vectors (axial and radial), and the residual $\sdx{\boldsymbol{e}_\mathrm{p}}[k]$ in \eqref{equ: particle physics loss} comprises both components of \eqref{equ: particle motion} for the $k$th particle. Although pressure gradient, added mass, and Basset history forces can momentarily exceed Stokes drag as particles traverse a shock wave, their cumulative effect on particle trajectories is negligible in the high-density-ratio limit \citep{Thomas1992, Parmar2009, Capecelatro2023}, making them safe to neglect for our purposes.\par

% Compressible drag
\subsection{Compressible drag law}
\label{app: disperse: compressible drag}
Particle drag in high-speed flow depends not only on viscous forces from the carrier phase but also on compressibility and rarefaction effects. Compressibility effects scale with the particle Mach number,
\begin{equation}
    Ma_\mathrm{p} = \frac{\left|\boldsymbol{u} - \boldsymbol{v}\right|}{\sqrt{\gamma R T}}, 
    \label{equ: Mach}
\end{equation}
where $R$ is the specific gas constant of the carrier phase. Rarefaction effects are governed by the ratio of the mean free path of the carrier fluid, $\lambda$, to a characteristic length scale, typically the particle diameter at low $Re_\mathrm{p}$. This ratio defines the particle Knudsen number,
\begin{equation}
    Kn_\mathrm{p} = \frac{\lambda}{d_\mathrm{p}} = \frac{Ma_\mathrm{p}}{Re_\mathrm{p}} \sqrt{\frac{\pi \gamma}{2}},
    \label{equ: Knudsen}
\end{equation}
where the expression on the right-hand side follows from the ideal gas law. Thus, drag correlations for compressible particle-laden flows can be expressed in terms of any two of the three nondimensional groups: $Re_\mathrm{p}$, $Ma_\mathrm{p}$, and $Kn_\mathrm{p}$. In practice, this formulation allows models such as Loth's drag law, discussed next, to bridge viscous, compressibility-dominated, and rarefaction-dominated regimes.\par

\begin{figure}[t]
\vspace*{-0.5em}
    \centering
    \includegraphics[height=5cm]{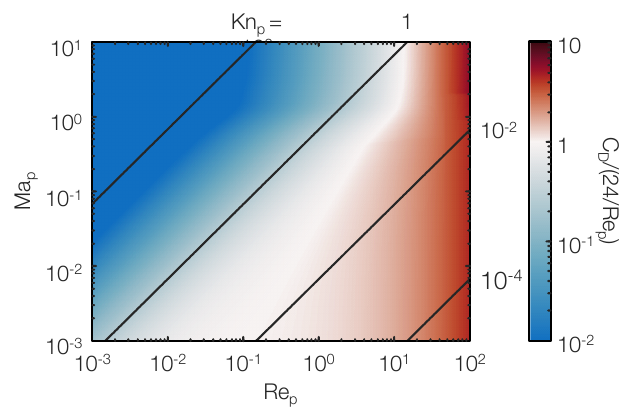}
    \caption{Loth drag model normalized by Stokes drag, showing its dependence on $Re_\mathrm{p}$ and $Ma_\mathrm{p}$. Black lines indicate contours of constant $Kn_\mathrm{p}$.}
    \label{fig: Loth}
\end{figure}

\citet{Loth2008} put forth a comprehensive drag correlation for compressible particle-laden flows. The model expresses the drag coefficient $C_\mathrm{D}$ as a function of particle Reynolds and Mach numbers, thereby defining the particle response time $\tau_\mathrm{p}$ through \eqref{equ: response time: general}. The general form is
\begin{equation}
    C_\mathrm{D} = \left\{\begin{array}{ll}\dfrac{C_{\mathrm{D}, Kn, Re}}{1 + Ma_\mathrm{p}^4} + \dfrac{Ma_\mathrm{p}^4 C_{\mathrm{D}, f_\mathrm{M}, Re}}{1+Ma_\mathrm{p}^4}, &\quad Re_\mathrm{p} < 45 \\
    \frac{24}{Re_\mathrm{p}} \left[1 + 0.15 Re_\mathrm{p}^{0.687}\right] H_\mathrm{M} + \dfrac{0.42 C_\mathrm{M}}{1 + \frac{42~500 \,G_\mathrm{M}}{Re_\mathrm{p}^{1.16}}}, &\quad Re_\mathrm{p} > 45\end{array}\right..
    \label{equ: Loth}
\end{equation}
The first branch of \eqref{equ: Loth} applies in the rarefaction-dominated regime ($Re_\mathrm{p} < 45$) and the second branch in the compression-dominated regime ($Re_\mathrm{p} > 45$). Although we implement the full model in forward simulations, the cone--cylinder case has $Re_\mathrm{p} < 23$ throughout, so only the rarefaction branch is used in our reconstructions. \Cref{fig: Loth} shows $C_\mathrm{D}$ normalized by Stokes drag as a function of $Re_\mathrm{p}$ and $Ma_\mathrm{p}$. At low $Re_\mathrm{p}$, contours of $C_\mathrm{D}$ align with those of $Kn_\mathrm{p}$, indicating rarefaction control. As $Re_\mathrm{p}$ increases, the gradient of $C_\mathrm{D} Re_\mathrm{p}/24$ bends toward the $Ma_\mathrm{p}$ axis, marking the onset of compressibility effects. In our case, however, the flow remains entirely within the aforementioned rarefaction regime.\par

Returning to \eqref{equ: Loth}, the rarefaction-specific terms are
\begin{subequations}
    \begin{align}
        C_{\mathrm{D}, Kn, Re} &= \frac{24}{Re_\mathrm{p}} \left(1 + 0.15 Re_\mathrm{p}^{0.687}\right) f_{Kn}, \\
        f_{Kn} &= \frac{1}{1+Kn_\mathrm{p} \left[2.514 + 0.8 \,\exp\mathopen{}\left(-\frac{0.55}{Kn_\mathrm{p}}\right)\right]}, \\
        C_{\mathrm{D}, f_\mathrm{M}, Re} &= \frac{C_\mathrm{D, f_M}}{1+\left(\frac{C_\mathrm{D, f_M}}{1.63}-1\right) \sqrt{\frac{Re_\mathrm{p}}{45}}}, \\
        C_\mathrm{D, f_M} &= \frac{\left(1 + 2s_\mathrm{M}^2\right) \mathrm{erf}\mathopen{}\left(-s_\mathrm{M}^2\right)}{s_\mathrm{M}^3 \sqrt{\pi}} + \frac{\left(4s_\mathrm{M}^4 + 4s_\mathrm{M}^2 - 1\right) \mathrm{erf}\mathopen{}\left(s\right)}{2s_\mathrm{M}^4} + \frac{2}{3s_\mathrm{M}} \sqrt{\frac{\pi T_\mathrm{p}}{T}}, \\
        s_\mathrm{M} &\equiv Ma_\mathrm{p} \sqrt{\gamma/2},
    \end{align}
\end{subequations}
where $T_\mathrm{p}$ is the particle temperature. The compression-specific contributions to \eqref{equ: Loth} are
\begin{subequations}
    \begin{align}
        H_\mathrm{M} &= 1 - \frac{0.258 C_\mathrm{M}}{1 + 514 \,G_\mathrm{M}}, \\
        G_\mathrm{M} &= \left\{\begin{array}{ll} 1 - 1.525 \,Ma_\mathrm{p}^4, &\quad Ma_\mathrm{p} < 0.89 \\ 0.0002 + 0.0008 \,\mathrm{tanh}\mathopen{}\left[12.77 \left(Ma_\mathrm{p} - 2.02\right) \right], &\quad Ma_\mathrm{p} > 0.89\end{array}\right., \\
        C_\mathrm{M} &= \left\{\begin{array}{ll} \frac{5}{3} + \frac{2}{3} \,\mathrm{tanh}\mathopen{}\left[3 \,\log\mathopen{}\left(Ma_\mathrm{p} - 0.1\right) \right], &\quad Ma_\mathrm{p} < 1.45 \\ 2.044 + 0.2 \,\exp\mathopen{}\left[-1.8 \,\log\mathopen{}\left(\frac{Ma_\mathrm{p}}{1.5} \right)^2 \right], &\quad Ma_\mathrm{p} > 1.45\end{array}\right..
    \end{align}
\end{subequations}
The resultant drag law has been extensively benchmarked using experimental data and employed for many simulations of high-speed particle-laden flow. Recently, \citet{Loth2021} published a comprehensive review of relevant results, obtained from particle-resolved DNSs, rarefied-gas simulations, and wind tunnel experiments. The authors found that Loth's original model was not empirically supported near $Re_\mathrm{p} = 45$. They reported an updated model that corrects for these discrepancies in \citet{Loth2021}. However, updates in that paper do not meaningfully affect the cone--cylinder simulation in \S~\ref{sec: cases: cone} because, again, it has a maximum $Re_\mathrm{p}$ of about 23. We thus employ the original formulation of \citet{Loth2008}, as presented above.\par

%%% B-spline %%%
\section{B-spline filtering}
\label{app: B-spline}
For the baseline flow-only reconstruction defined in \S~\ref{sec: method: loss: baseline}, particle velocities are estimated from measured tracks via finite differencing as well as smoothing. Among available options, cubic B-splines provide a standard filtering approach because they yield $C^2$-continuous fits with good stability while relying only on low-order piecewise polynomials \citep{Gesemann2016, Li2024a}. A representative cubic B-spline approximation for the $x$-coordinate of a particle trajectory is
\begin{equation}
    \label{equ: B-spline}
    \widehat{x}(t) = \sum_{j=0}^{K} c_j \,\alpha\mathopen{} \left(\frac{t-t_j^\prime}{\Delta t}\right),
\end{equation}
where $c_j$ are spline coefficients, $\{t_j^\prime \mid j = 0, \dots, K\}$ are uniformly spaced knots with spacing $\Delta t$, and $\alpha$ denotes the cubic basis function \citep{Skare2005},
\begin{equation}
    \label{equ: B-spline basis}
    \alpha \mathopen{}\left(t\right) = \left\{\begin{array}{lll}\frac{2}{3} - \left(1-\frac{\lvert t \rvert}{2}\right)t^2, &\quad 0 < \lvert t \rvert < 1 \\
    \frac{\left(2-\lvert t \rvert^3\right)}{6}, &\quad 1 < \lvert t \rvert < 2 \\
    0, &\quad \lvert t \rvert > 2\end{array}\right..
\end{equation}
Velocity and acceleration are then obtained analytically as the first and second derivatives of $\widehat{x}$. Naturally, the same construction applies to the other spatial components.\par

The number of knots $K$ controls the trade-off between smoothness and fidelity. Increasing $K$ improves the ability of the spline to follow fine-scale fluctuations but risks fitting to measurement noise. Conversely, small-$K$ splines over-smooth the trajectories. In practice, we set $K$ adaptively such that each spline segment spans about ten measured points, a choice tuned to ensure sufficient expressivity while effectively suppressing measurement noise. This choice of segment length is robust across the noise levels and achieves the optimal accuracy through supervision against the ground truth. This setup represents an idealized, best-case scenario filtering that is not available in real experimental data. The spline coefficients are obtained by minimizing the residual between measured and filtered positions in a least-squares sense, i.e., by minimizing
\begin{equation*}
    \sum_{j=1}^{n_k} \left\lVert \sdx{x_j}[k] - \widehat{x}(t_j) \right\rVert_2^2,
\end{equation*}
where $\sdx{x_j}[k]$ is the measured position at the $j$th time step $t_j$, $\widehat{x}$ is the spline estimate from \eqref{equ: B-spline} evaluated at the same time, and $n_k$ is the total number of points in the track.\par

In our implementation, filtering is performed via MATLAB's \texttt{spap2} and \texttt{fnder} routines. While details of software usage are secondary, the essential point here is that B-spline filtering yields smoothed particle tracks from which velocities and accelerations can be determined in a consistent manner. These filtered quantities serve as inputs to the baseline reconstruction method introduced in \S~\ref{sec: method: loss: baseline}, used for comparison throughout our tests.\par

%%% Low-Inertial Particles %%%
\section{Cone--cylinder flow with low-inertia particles}
\label{app: low-inertia particles}
\begin{figure}[htb]
\vspace*{-1em}
    \centering
    \includegraphics[width=1\linewidth]{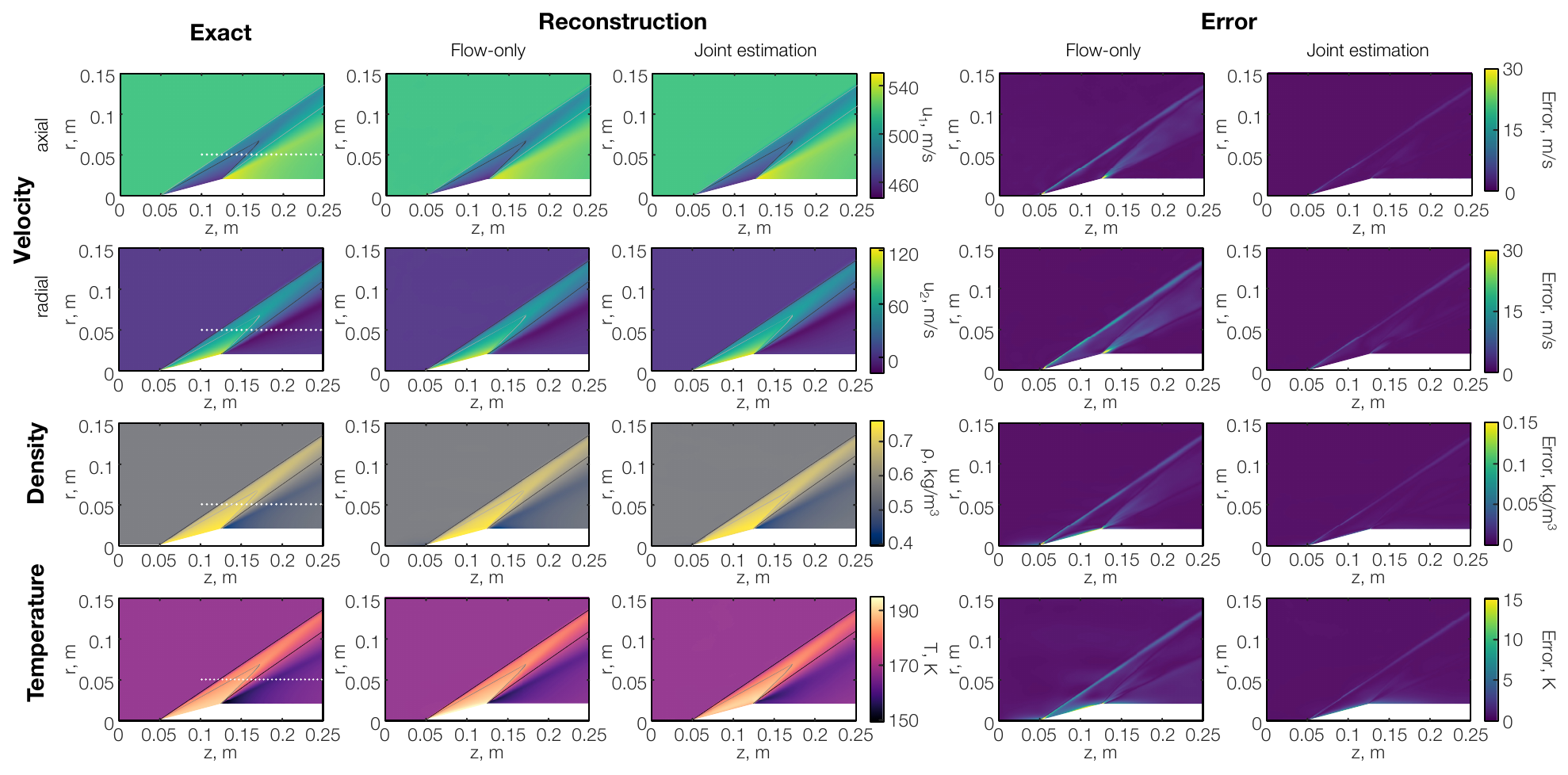}
    \caption{Cone--cylinder flow reconstruction with low-inertia particles at $\tau_\mathrm{p} \approx 3~\upmu$s. (left) Reconstructed flow fields compared with the CFD reference. (right) Absolute error fields, showing substantial error reduction with joint estimation even at low particle inertia. Dark and gray contour lines indicate low and high iso-values from the CFD reference, overlaid on the reconstructions for comparison. The iso-values are 480 and 506~m/s for the axial velocity component, 22 and 57~m/s for the radial velocity component, 0.59 and 0.70~kg/m$^3$ for density, and 172 and 183.5~K for temperature, respectively. The white dashed lines indicate the locations at which the flow profiles in \cref{fig: cone line-low} are extracted.}
    \label{fig: cone panel-low}
\end{figure}

Particles of lower inertia are simulated in the cone--cylinder flow, following the procedure in \S~\ref{sec: cases: cone}. This case helps to assess the advantage of joint estimation over the flow-only mode much closer the tracer limit, as might be expected in a highly controlled experiment. To reduce inertia, the particle diameters are drawn from a Gaussian distribution with mean $d_\mathrm{p} = 0.8$~$\upmu$m and standard deviation 0.2~$\upmu$m. The particle density and all other computational settings, including the number of particles and frames, advection scheme, and boundary treatment, are kept the same as in \S~\ref{sec: cases: cone}. The resulting response time decreases to $\tau_\mathrm{p} \approx 3.1$~$\upmu$s, comparable to the \ce{TiO2} response times reported by \citet{Ragni2011}. \Cref{fig: cone panel-low} shows the reconstructed flow fields for the flow-only and joint estimation modes. Because these lower-inertia particles have better traceability, both reconstructions resemble the DNS reference closely, with only subtle visual differences.\par

\begin{figure}[htb]
    \centering
    \includegraphics[width=1\linewidth]{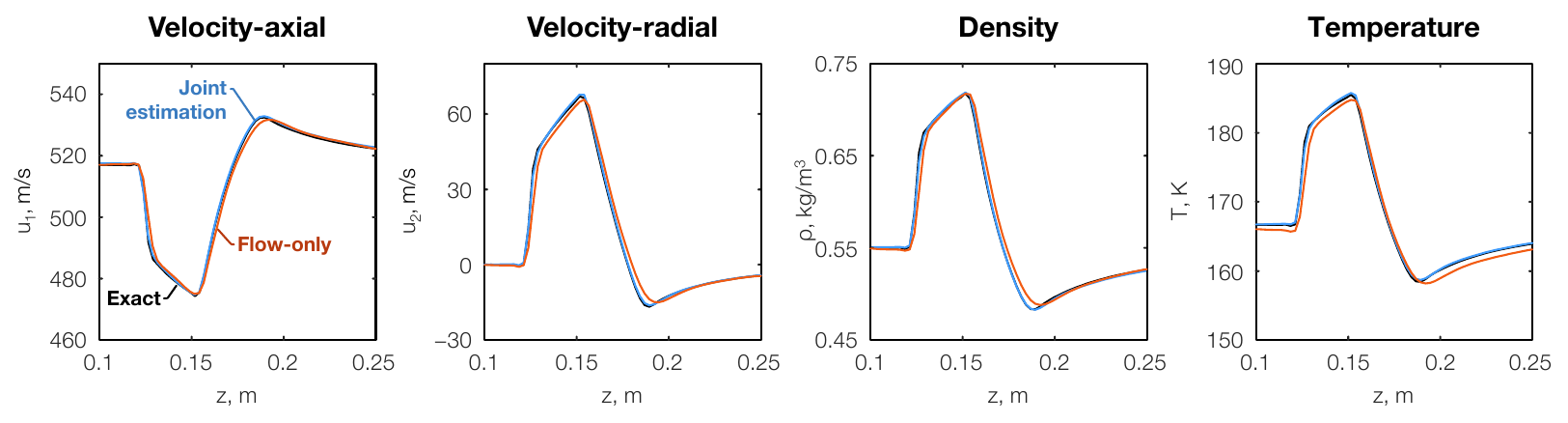}
    \caption{Flow field cuts along the horizontal line at $r = 0.05$~m with $0.1~\mathrm{m} < z < 0.25~\mathrm{m}$. The exact simulation is shown in black, joint estimation in blue, and flow-only reconstruction in red. Joint estimation nearly overlaps with the CFD data, whereas the flow-only result is shifted toward the post-shock region because of delayed particle response.}
    \label{fig: cone line-low}
\end{figure}

To resolve those differences, \cref{fig: cone line-low} plots velocity, density, and temperature profiles along a horizontal line at $r = 0.05$~m for $0.1~\mathrm{m} < z < 0.25~\mathrm{m}$. These profiles cross both the shock interface ($z \approx 0.12$~m) and the expansion fan ($z \approx 0.18$~m). The joint estimation results nearly overlap with the ground truth across the full line. The flow-only results capture the overall trends, but are shifted slightly toward the post-shock region because of the delayed particle response. Such discrepancies are expected to be much more significant for fluctuations in high-speed flows. The corresponding absolute-error fields are shown on the right side of \cref{fig: cone panel-low}. As expected, the flow-only mode exhibits its largest errors across the shock and expansion regions, much as in the larger-inertia case of \S~\ref{sec: inertial: cone}. Joint estimation substantially reduces these errors, yielding NRMSEs of 0.4\% for the axial velocity, 5.2\% for the radial velocity, 2.0\% for the density, and 2.0\% for the temperature, compared with 1.0\%, 10.7\%, 6.8\%, and 3.4\% for the flow-only mode.\par

\begin{figure}[htb!]
    \vspace*{-0.5em}
    \centering
    \includegraphics[width=0.6\linewidth]{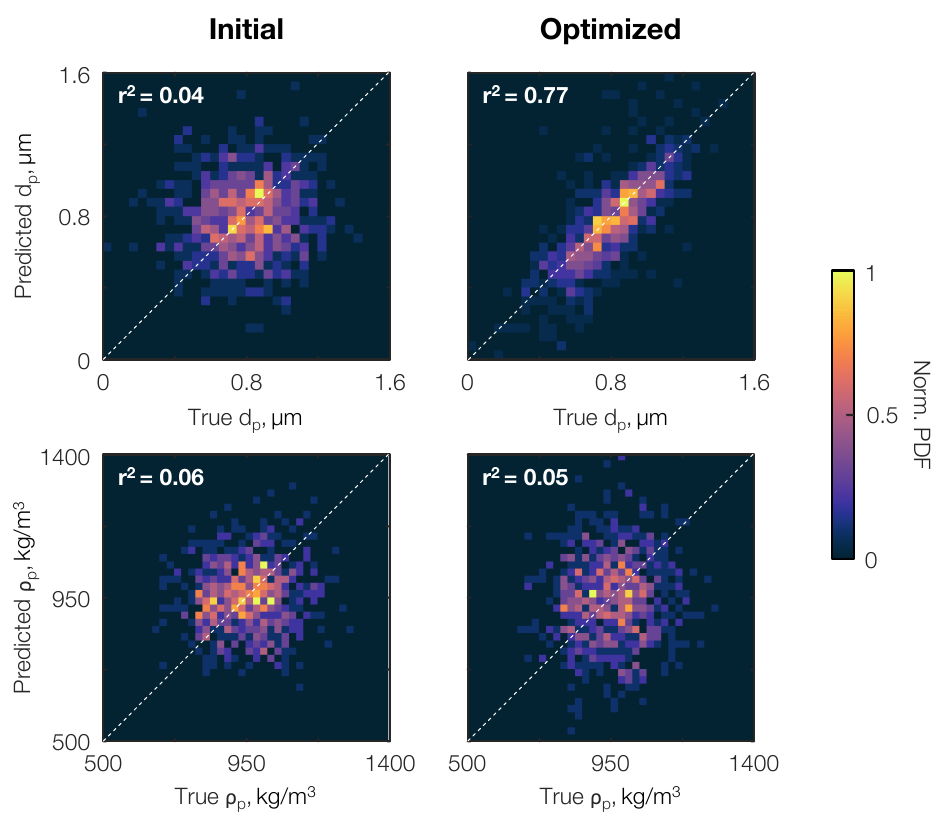}
    \caption{Implicit particle characterization in the cone--cylinder flow with low-inertia particles at $\tau_\mathrm{p} \approx 3~\upmu$s. Initial $d_\mathrm{p}$ and $\rho_\mathrm{p}$ values (left) and jointly optimized estimates (right) are compared with the ground truth.}
    \label{fig: cone PDFs-low}
\end{figure}

\Cref{fig: cone PDFs-low} shows the normalized PDFs of the inferred particle diameter and density from joint estimation, compared with their true distributions. As in the case with greater inertia case, shown in \cref{fig: cone PDFs}, $d_\mathrm{p}$ is recovered successfully from a random initialization, whereas $\rho_\mathrm{p}$ does not converge because the reconstruction is only weakly sensitive to the particle density, as discussed in \S~\ref{sec: inertial: cone}. Taken together, these results indicate that joint particle--flow reconstruction remains sensitive to particle size even for sub-micron particles with relatively weak inertial effects.\par

%%% Neural Expressivity %%%
\section{Characterization of neural expressivity}
\label{app: expressivity}
The HIT architecture summarized in \cref{tab: architecture} is evaluated here, using DNS data to assess neural expressivity. Three sub-volumes are extracted from the full HIT dataset described in \S~\ref{sec: cases: HIT}, with domain sizes of $128^3$, $256^3$, and $512^3$ voxels. After dimensionalization, these correspond to physical domains of $10^3$, $20^3$, and $40^3$~cm$^3$, respectively. Five training datasets are constructed at two temporal resolutions, each spanning 100 frames over 0.04~s. Two high-resolution cases use the $128^3$ and $256^3$ domains at a temporal resolution of $0.05\tau_\eta$. Owing to memory limits, the $512^3$ case is only tested at a coarser temporal resolution of $0.25\tau_\eta$, for which 21 frames are sampled uniformly from the corresponding high-resolution data. The $128^3$ high-resolution case matches the conditions used for the inertial HIT study in \S~\ref{sec: cases: HIT}, allowing for direct comparison.\par

\begin{figure}[t]
\vspace*{-0.5em}
    \centering
    \includegraphics[height=5cm]{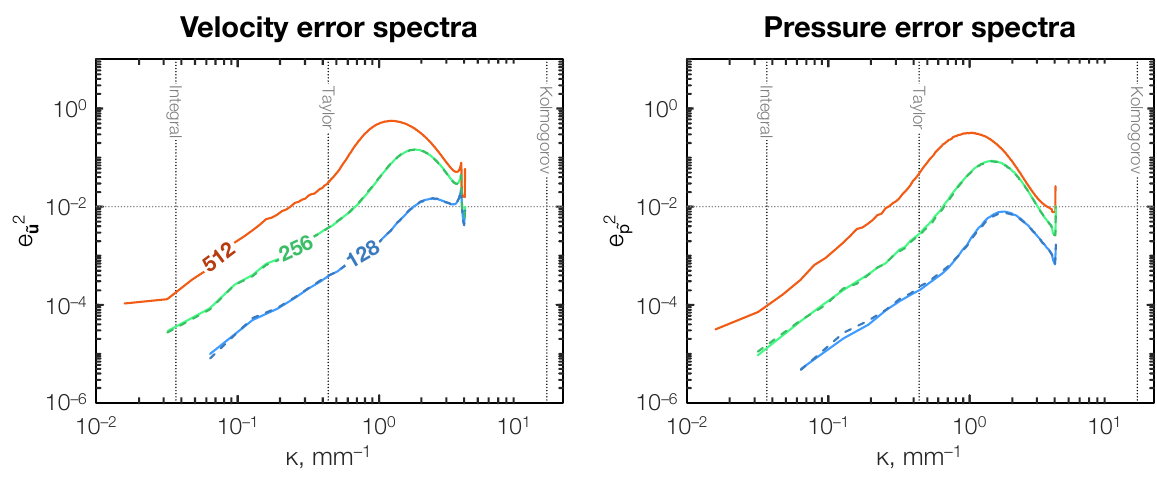}
    \caption{Velocity (left) and pressure (right) error spectra from idealized training on DNS data of homogeneous isotropic turbulence over three domain sizes: $128^3$, $256^3$, and $512^3$ voxels. Dashed and solid lines denote high- and low-temporal-resolution cases, respectively. Vertical lines indicate the wavenumbers associated with the three turbulence length scales. The horizontal line marks an error threshold of 0.01, used to define the cutoff wavenumber for each domain size. The selected neural network resolves scales beyond the Taylor scale for the $128^3$ and $256^3$ domains, but only the large integral scales for the $512^3$ domain.}
    \label{fig: HIT expressivity}
\end{figure}

To isolate representation error, the networks are trained to learn the DNS data with the idealized loss functions for velocity and pressure:
\begin{equation}
    \label{equ: DNS data loss}
    \mathscr{J}_\mathrm{ideal}^{\boldsymbol{u}} =
    \frac{\dim(\boldsymbol{e}_\mathrm{f})^{-1}}{\left|\mathcal{V} \times \mathcal{T}\right|} \int_\mathcal{T} \int_\mathcal{V} \left\lVert \boldsymbol{u} - \boldsymbol{u}_\mathrm{exact} \right\rVert_2^2 \mathrm{d}\boldsymbol{x} \,\mathrm{d}t
    \quad\text{and}\quad
    \mathscr{J}_\mathrm{ideal}^p =
    \frac{\dim(\boldsymbol{e}_\mathrm{f})^{-1}}{\left|\mathcal{V} \times \mathcal{T}\right|} \int_\mathcal{T} \int_\mathcal{V} \left\lVert p - p_\mathrm{exact} \right\rVert_2^2 \mathrm{d}\boldsymbol{x} \,\mathrm{d}t,
\end{equation}
where $(\cdot)_\mathrm{exact}$ denotes a DNS quantity. The integrals are approximated by Monte Carlo sampling of the voxels and time steps, using batches of 5000 randomly sampled points. Both the velocity and pressure networks are trained using the procedure described in \S~\ref{sec: implementation: training}, and all errors are evaluated at the higher temporal resolution. Hence, reconstruction errors serves as a proxy for network expressivity.\par

\Cref{fig: HIT expressivity} shows normalized velocity and pressure error spectra for the different domain sizes and temporal resolutions. Results for low and high temporal resolutions overlap nearly perfectly, indicating that temporal resolution is not the limiting factor in these tests. At low temporal resolution, the mean velocity NRMSEs are 1.1\%, 3.6\%, and 9.6\% for the $128^3$, $256^3$, and $512^3$ domains, respectively, while the corresponding pressure NRMSEs are 1.6\%, 3.6\%, and 7.1\%. In each case, the difference relative to the corresponding high-resolution result is below 0.1\%. Naturally, errors systematically increase with domain size, due to the increased spectral content and constant neural expressivity. By comparison, the inertial HIT reconstruction in \S~\ref{sec: cases: HIT} yields velocity and pressure NRMSEs of 5.0\% and 17.2\%, both being far larger than the ideal-training errors of 1.1\% and 1.6\% for the matched $128^3$ case. This indicates that the inertial HIT results are not limited by the neural representation, but rather by the conditioning of the inverse problem, viz. the finite seeding density and particle inertia.\par

%%% KCT Implementation %%%
\section{Initialization of the KCT models}
\label{app: KCT}

% Track splitting
\subsection{Track splitting}
\label{app: KCT: split}
Particle tracks in LPT vary in length, ranging from only a few positions to several hundred. Since TensorFlow requires fixed tensor shapes for computational graphs, we divide each trajectory into fixed-length segments of 20 positions. Shorter segments are zero-padded. When a single trajectory spans multiple segments, its particle properties (e.g., size, density) are shared across all segments; these properties may be trainable in cases with inertial particles. Although longer segments are possible, they increase the number and dimension of coefficient matrices in \eqref{equ: track position matrix}--\eqref{equ: track acceleration matrix}, thereby raising the computational cost and memory demand of KCTs. Conversely, very short segments disrupt the continuity of tracks and introduce boundary-related artifacts. In practice, therefore, segment lengths of 15--30 points strike a good balance between efficiency and fidelity.\par

% Warm starting
\subsection{Warm-starting KCTs with filtered tracks}
\label{app: KCT: warm start}
For inertial particles, joint optimization of flow states, particle positions, and per-particle properties from noisy data is ill-posed and prone to divergence. To improve stability, we ``warm-start'' each KCT model using filtered tracks obtained from a conventional smoothing technique, such as polynomial regression or kernel convolution \citep{Berk2024a}. Initialization is posed as a per-track optimization that identifies the displacement vector $\boldsymbol{\delta}$ and velocity parameters $\boldsymbol{\theta}$ which minimize
\begin{equation}
    \label{equ: KCT init}
    \mathscr{J}_\mathrm{warm} = \sum_{j=1}^{n_\mathrm{k}}
    \left(\lVert\boldsymbol{v}_j-\widehat{\boldsymbol{v}}_j\rVert_2^2 + \chi \lVert\boldsymbol{a}_j-\widehat{\boldsymbol{a}}_j\rVert_2^2\right),
\end{equation}
where $\chi$ balances velocity and acceleration residuals and is assigned as
\begin{equation}
    \chi = \frac{\langle \lVert \widehat{\boldsymbol{v}} \rVert_2 \rangle_\mathrm{p}}{\langle \lVert \widehat{\boldsymbol{a}} \rVert_2 \rangle_\mathrm{p}},
\end{equation}
with $\langle \cdot \rangle_\mathrm{p}$ indicating an average over the current track. Velocities $\boldsymbol{v}$ and accelerations $\boldsymbol{a}$ are the outputs of the particle model $\sdx{\mathsf{P}}[k]$ from \eqref{equ: track velocity matrix} and \eqref{equ: track acceleration matrix}.\par

We minimize \eqref{equ: KCT init} using MATLAB's implementation of the Levenberg--Marquardt algorithm, initializing $\boldsymbol{\delta}$ with raw (noisy, observed) positions and $\boldsymbol{\theta}$ with zeros. In this study, fifth-order polynomials provide the filtered velocity and acceleration vectors $\widehat{\boldsymbol{v}}$ and $\widehat{\boldsymbol{a}}$, though more advanced methods such as TrackFit \citep{Gesemann2016} could also be employed. This warm-start is only used for cases with inertial particles. For ideal tracers, initializing with filtered tracks offers no improvement and risks biasing the reconstruction, so we simply initialize $\boldsymbol{\delta}$ with raw data and $\boldsymbol{\theta}$ with zeros.\par

% Particle property transformation
\subsection{Particle property transform}
\label{app: KCT: transform}
In inertial cases, particle properties such as their size and density can differ in units and span several orders of magnitude, limiting the precision of optimization. Therefore, we map each positive property $\varphi > 0$ to a dimensionless variable $\xi \sim O(1)$, with the latter variable being optimized. The forward transformation is
\begin{equation}
    \label{equ: forward transform}
    \xi = \frac{\log\mathopen{} \left[\exp\mathopen{} \left(c_1 \varphi\right) - 1\right]}{c_2} - c_3,
\end{equation}
with parameters $c_1$, $c_2$, and $c_3$ chosen so that $\xi$ is order one. The inverse transform recovers $\varphi$,
\begin{equation}
    \label{equ: backward transform}
    \varphi = \frac{\log\mathopen{} \left[\exp\mathopen{} \left(c_2 \xi + c_2c_3\right)+1 \right]}{c_1}
    = \frac{\mathrm{softplus}\mathopen{} \left(c_2 \xi + c_2c_3\right)}{c_1},
\end{equation}
where the softplus function ensures that $\varphi > 0$ for $c_1 > 0$.\par

Parameters are chosen as follows. First, $c_1$ is set to $\varphi^{-1}$, so $c_1 \varphi = 1$. Second, given an expected range $\varphi \in [\varphi_\mathrm{min}, \varphi_\mathrm{max}]$, we determine $c_2$ and $c_3$ by enforcing $\xi(\varphi_\mathrm{min}) = -1$ and $\xi(\varphi_\mathrm{max}) = 1$, yielding
\begin{subequations}
    \begin{equation}
        c_2 = \frac{\log\mathopen{} \left[\exp\mathopen{} \left(c_1\varphi_\mathrm{min}\right)-1\right]}{c_3-1}, \qquad
        c_3 = \frac{2}{1-\zeta}-1,
    \end{equation}
    with
    \begin{equation}
        \zeta = \frac{\log\mathopen{} \left[\exp\mathopen{} \left(c_1\varphi_\mathrm{min}\right) - 1\right]}{\log\mathopen{} \left[\exp\mathopen{} \left(c_1\varphi_\mathrm{max}\right) - 1\right]}.
    \end{equation}
\end{subequations}
This mapping is bijective and has bounded gradients, ensuring one-to-one correspondence between $\psi$ and $\xi$ and stable optimization during backpropagation. The formulation is valid for strictly positive (or negative) properties such as size and density; properties that may cross zero, such as electrical charge, require a different mapping.\par

%%% Noise-Free Tracer DA %%%
\section{Reconstruction accuracy for noise-free tracers}
\label{app: perfect tracers}
The capability of joint estimation to recover under-resolved dynamics from ideal tracers is numerically assessed and compared to na{\"i}ve interpolation by adaptive Gaussian windowing \citep{Agui1987}. HIT is used as a representative flow case. Noise-free tracks of perfect tracers are simulated using the same numerical settings as in \S~\ref{sec: cases: HIT}, including the domain size, temporal resolution, and advection scheme. The number of tracks is varied to produce inter-particle spacings from $6\ell_\eta$ to $24\ell_\eta$, corresponding to particle image densities from 0.1~ppp to 0.0016~ppp for a 1~MP camera.\par

\begin{figure}[ht]
\vspace*{-0.5em}
    \centering
    \includegraphics[height=5cm]{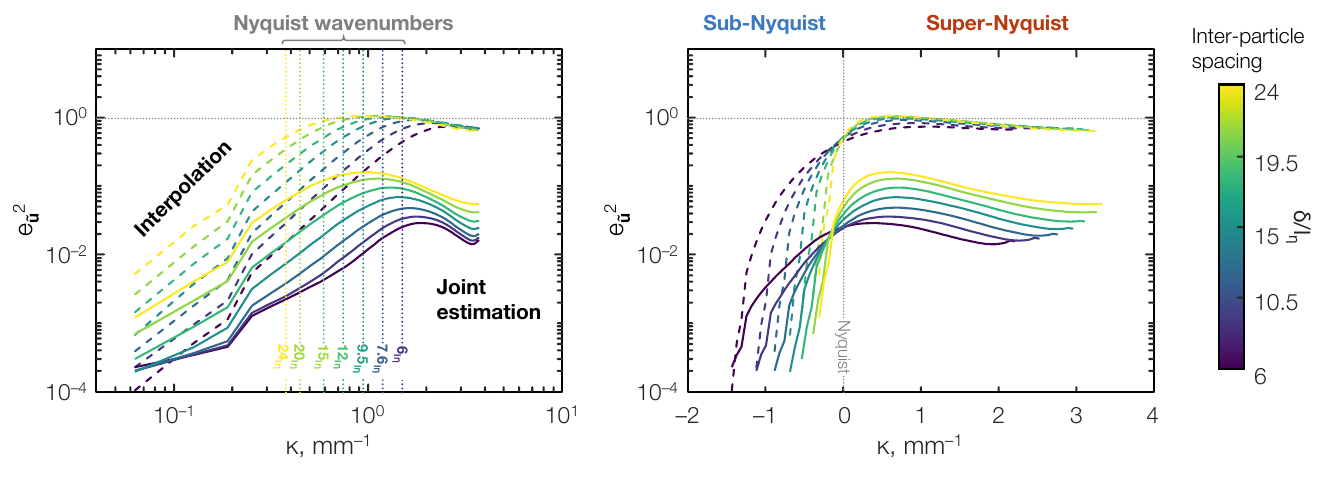}
    \caption{Error spectra of HIT reconstructions in the tracer limit ($St \to 0$) under varying inter-particle spacings. Spectra are shown versus wavenumber (left) and normalized by the Nyquist wavenumber (right). Colors denote particle spacing. Dashed lines indicate interpolation, and solid lines indicate joint estimation. Errors from joint estimation decrease rapidly in the super-Nyquist region because of the physical constraints, whereas na{\"i}ve interpolation asymptotes to 100\% error.}
    \label{fig: DA performance}
\end{figure}

\Cref{fig: DA performance} compares the velocity error spectra obtained by joint estimation and by interpolation. The left panel shows the spectra in absolute wavenumbers, with the particle-sampling Nyquist wavenumbers indicated for reference. Joint estimation reduces the error by one to two orders of magnitude compared to interpolation, especially at high wavenumbers. The right panel shows the error spectra versus the relative wavenumbers, i.e., relative to $\kappa_\mathrm{Nyq}$. Whereas the interpolation method saturates near 100\% error across all cases, joint estimation yields progressively smaller errors as the particle spacing decreases. This result illustrates the possibility for DA to recover flow information beyond the nominal sampling limit. However, this capability has not yet been definitively established for experimental data, which motivates further exploration.\par

%%% Acknowledgments %%%
\subsection*{Acknowledgments}
The authors acknowledge the generous support of Prof.~Andreas Schr{\"o}der, Dr.~Daniel Schanz, Mr.~Tom Buchwald, and Mr.~Philipp Godbersen of the German Aerospace Center (DLR) G{\"o}ttingen and Brandenburg University of Technology Cottbus-Senftenberg for their assistance in generating the TrackFit results and for helpful discussions regarding their interpretation.\par

%%% References %%%

\end{document}